\documentclass[twocolumn,epjc3]{svjour3}  
\usepackage{amsmath,amssymb,amsfonts}
\usepackage{graphicx,graphics}
\usepackage{graphics}
\usepackage{url}
\usepackage{slashed}
\usepackage{color}
\usepackage[normalem]{ulem}
\usepackage{mathtools}

\journalname{Eur. Phys. J. C}

\begin{document}

\title{EPPS21: A global QCD analysis of nuclear PDFs}

\author{Kari~J.~Eskola$^{1,2,}$\thanksref{e2},\,Petja~Paakkinen$^{1,2,3,}$\thanksref{e3},\,Hannu~Paukkunen$^{1,2,}$\thanksref{e1},\,Carlos~A.~Salgado$^{3,}$\thanksref{e4}}

\thankstext{e2}{e-mail: kari.eskola@jyu.fi}
\thankstext{e3}{e-mail: petja.k.m.paakkinen@jyu.fi}
\thankstext{e1}{e-mail: hannu.paukkunen@jyu.fi}
\thankstext{e4}{e-mail: carlos.salgado@usc.es}

\institute{University of Jyvaskyla, Department of Physics, P.O. Box 35, FI-40014 University of Jyvaskyla, Finland \and
Helsinki Institute of Physics, P.O. Box 64, FI-00014 University of Helsinki, Finland \and
Instituto Galego de F\'\i sica de Altas Enerx\'\i as (IGFAE), Universidade de Santiago de Compostela, E-15782 Galicia, Spain
}

\maketitle

\abstract{
We present an updated global analysis of collinearly factorized nuclear parton distribution functions (PDFs) at next-to-leading order in perturbative QCD. In comparison to our previous fit, EPPS16, the present analysis includes more data from proton-lead collisions at the Large Hadron Collider: 5\,TeV double-differential CMS dijet and LHCb D-meson data, as well as 8\,TeV CMS W$^\pm$ data. These new data lead to significantly better-constrained gluon distributions at small and intermediate values of the momentum fraction $x$, confirming the presence of shadowing and antishadowing for gluons in large nuclei. In addition, we include Jefferson Lab measurements of deeply inelastic scattering which probe nuclear PDFs at large $x$ and low virtualities. For the first time within the Hessian framework, we now also explore the uncertainties of nuclear PDFs due to the errors in the baseline proton PDFs. We release the results of our analysis as a new public parametrization of nuclear PDFs called EPPS21.
}

\section{Introduction}
\label{Introduction}

Within the past decade, the proton-nucleus (pA) program at the Large Hadron Collider (LHC) has evolved from the first exploratory steps to a level where the measurements are precise enough to constrain and thereby seriously test the universality of nuclear parton distributions (PDFs). While in our previous analysis, EPPS16 \cite{Eskola:2016oht}, the impact of the available LHC data was still somewhat limited (see also Ref.~\cite{Armesto:2015lrg}) the situation has undergone a significant change: For example, it is already known that the double differential CMS dijet \cite{Sirunyan:2018qel} and LHCb D-meson \cite{Aaij:2017gcy} data will both have a huge impact on the nuclear gluon PDFs \cite{Eskola:2019dui,Eskola:2019bgf}. Other recent heavy-flavour measurements such as B-meson and vector-meson production have been identified as potential gluon constraints as well  \cite{Kusina:2017gkz,Kusina:2020dki}. The recent W$^\pm$ measurements \cite{Sirunyan:2019dox}, in turn, have been argued to carry further sensitivity to the flavour separation \cite{AbdulKhalek:2020yuc} and strangeness~\cite{Kusina:2020lyz} of the nuclear quark distributions. These new LHC pPb data are the main motivation for our updated global analysis of nuclear PDFs.

The fixed-target program at Jefferson Laboratory (JLab) has recently included new measurements of deeply inelastic scattering (DIS) \cite{Seely:2009gt,Schmookler:2019nvf,Arrington:2021vuu}. These data probe nucleons at large momentum fractions $x \gtrsim 0.1$, the interaction scale $Q^2$ being typically lower than in the older fixed-target experiments. Out of these, the latest measurements by the JLab CLAS Collaboration \cite{Seely:2009gt} have been studied in the context of nuclear PDFs first in Ref.~\cite{Paukkunen:2020rnb}. No visible evidence of large, nuclear-size dependent $1/Q^2$-type higher-twist corrections was found, which also justifies the inclusion of these data in global fits of nuclear PDFs. These data were found to be helpful in disentangling the flavour dependence of the valence-quark nuclear effects. For subsequent further studies on the JLab data, see Refs.~\cite{Segarra:2020gtj,Khanpour:2020zyu}. 

At large interaction scales $Q \gg \Lambda_{\rm QCD}$, the nuclear effects in PDFs do not typically exceed some tens of percents and they are often regarded as ``corrections'' to the free-proton PDFs. Most of the free-proton fits indeed use data on nuclear targets with varying prescriptions in correcting for the nuclear effects. On the other hand, nuclear-PDF analyses also require a proton PDF as an input and the outcome is bound to carry some sensitivity on how the baseline is chosen. Such intertwining of the nuclear and free-proton PDFs has already been considered within the NNPDF framework: In the NNPDF4.0 \cite{Ball:2021leu} analysis of the free-proton PDFs the nuclear-PDF uncertainties were considered as correlated uncertainties following Ref.~\cite{Ball:2018twp}. In the nNNPDF2.0 analysis \cite{AbdulKhalek:2020yuc}, on the other hand, the uncertainties from the free-proton PDFs were propagated into nuclear PDFs.\footnote{After the preprint of the present article was submitted, also an updated nNNPDF analysis appeared \cite{Khalek:2022zqe}.} In the present work, we will now carry out the latter within a Hessian prescription. Eventually, in a complete analysis, both the free- and bound-proton PDFs should be fitted simultaneously and the first steps towards this direction have also recently been taken \cite{Walt:2019slu,Helenius:2021tof}. 

\section{Nuclear PDFs and proton baseline}
\label{TheNuclearPDFs}

\subsection{Parametrization of nuclear modifications}

We write the bound-proton PDF $f^{{\rm p}/A}_i(x,Q^2)$ as a product of the nuclear modification $R^{{\rm p}/A}_i(x,Q^2)$ and the free proton PDF $f^{\rm p}_i(x,Q^2)$, 
\begin{align}
f^{{\rm p}/A}_i(x,Q^2) = R^{{\rm p}/A}_i(x,Q^2) f^{\rm p}_i(x,Q^2) \,. \label{eq:defnPDF}
\end{align}
Here $A$ denotes the mass number of the nucleus and $i$ indexes the parton flavour. Our proton baseline here is the recent set CT18ANLO \cite{Hou:2019efy}. 
The CT18A differs from the default CT18 in that it includes also the ATLAS $7 \,{\rm TeV}$ data on W$^\pm$- and Z-boson production \cite{Aaboud:2016btc}. The inclusion of these data was found to impact primarily the strange-quark PDF and to worsen the description of the neutrino-iron dimuon data \cite{Goncharov:2001qe} in which the strange-quark PDF plays a central role. By adopting the version ``A'' our strange-quark baseline PDF is thus less sensitive to the data on heavy nuclei.

The PDFs of a bound neutron $f_i^{{\rm n}/A}(x,Q^2)$ follow from the bound-proton PDFs by virtue of the approximate isospin symmetry, 
\begin{align}
f_{u}^{{\rm n}/A}(x,Q^2) & =  f_{d}^{{\rm p}/A}(x,Q^2), \nonumber \\
f_{d}^{{\rm n}/A}(x,Q^2) & =  f_{u}^{{\rm p}/A}(x,Q^2), \nonumber \\
f_{\overline{u}}^{{\rm n}/A}(x,Q^2) & =  f_{\overline{d}}^{{\rm p}/A}(x,Q^2), \\
f_{\overline{d}}^{{\rm n}/A}(x,Q^2) & =  f_{\overline{u}}^{{\rm p}/A}(x,Q^2), \nonumber \\
f_{i}^{{\rm n}/A}(x,Q^2) & =  f_{i}^{{\rm p}/A}(x,Q^2) \quad {\rm for} \, {\rm other} \, {\rm flavours}. \nonumber
\end{align}
The full nuclear PDFs that enter the cross-section calculations are always linear combinations that depend on the number of protons $Z$ and number of neutrons $N=A-Z$, 
\begin{align}
f_i^A(x,Q^2) = Z f^{{\rm p}/A}_i(x,Q^2) + N f^{{\rm n}/A}_i(x,Q^2) \,. \label{eq:FullPDF}
\end{align}
We define the nuclear modifications of the full nuclear PDFs by 
\begin{align}
R_i^A(x,Q^2) = \frac{Z f^{{\rm p}/A}_i(x,Q^2) + N f^{{\rm n}/A}_i(x,Q^2)}{Z f^{{\rm p}}_i(x,Q^2) + N f^{{\rm n}}_i(x,Q^2)} \,. \label{eq:FullR}
\end{align}

As in our earlier fits, we prefer to parametrize the nuclear modifications $R^{{\rm p}/A}_i(x,Q_0^2)$ instead of the absolute PDFs $f^{{\rm p}/A}_i(x,Q_0^2)$. The two options are of course fully equivalent but since most of the observables in the analysis are normalized to measurements involving either the free proton or deuteron (whose nuclear effects we neglect -- see the last paragraph of this subsection), the relative differences with respect to the free proton PDF are what truly matter. 

The nuclear modifications are parametrized at the charm pole-mass threshold $Q_0=m_{\rm charm}=1.3 \, {\rm GeV}$. The value of $m_{\rm charm}$ is set here by the value adopted in the CT18A analysis \cite{Hou:2019efy} to retain a full consistency with the baseline proton PDFs. Coming up with a decent functional form for the parametrization and deciding which parameters can be free is among the biggest challenges in the entire global analysis of nuclear PDFs. On one hand the parametrization should be flexible enough in regions where there are data constraints. On the other hand, the outcome of the fit should be physically feasible. For example, it is reasonable to expect that the nuclear effects are broadly larger in heavy nuclei like lead than what they are in a light nucleus like carbon. Coming up with the functional form finally used in the present analysis is a combination of experience from a entire chain of global fits we have performed in the past \cite{Eskola:1998iy,Eskola:2007my,Eskola:2008ca,Eskola:2009uj,Eskola:2016oht}, and trial and error. Our parametrization is a piecewise-smooth function defined as, 
\begin{align}
& R_i^A(x,Q^2_0) = \label{eq:FitForm} \\[5pt]
& \left\{
\begin{array}{lr}
a_0 + a_1\big(x-x_a\big) 
\Big[e^{-xa_2/x_a}-e^{-a_2} \Big]
\,, &   x \leq x_a \\[5pt]
b_0x^{b_1}\big(1-x\big)^{b_2}
e^{xb_3} \,, 
& \hspace{-0.0cm} x_a \leq x \leq x_e \\[5pt]
c_0 + c_1\left(c_2-x \right) \left(1-x\right)^{-\beta} \,, & \hspace{-0.0cm} x_e \leq x \leq 1. 
\end{array}
\right.  
\nonumber
\end{align}

\begin{figure}[htb!]
\centering
\includegraphics[width=\linewidth]{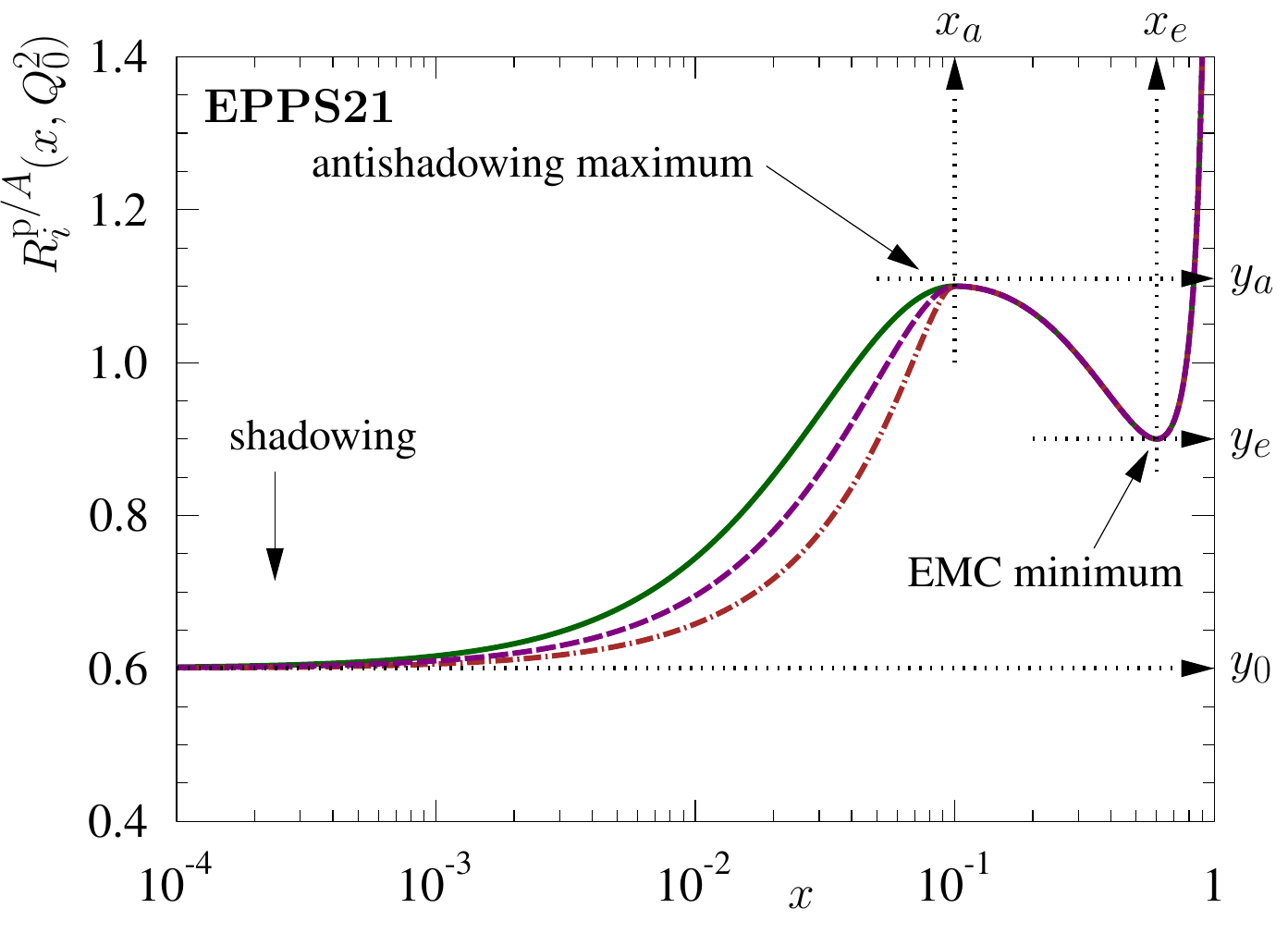}
\caption{Prototype of the EPPS21 fit functions $R_i^A(x,Q^2_0)$. The solid green line corresponds to $a_2=2$, the dashed purple line to $a_2=0$, and the brown dotted-dashed line to $a_2=-3$.}
\label{fig:FitForm}
\end{figure} 

In comparison to EPPS16, we have made some adjustments to the parametrization. First, the small-$x$ part involves the additional factor $e^{-xa_2/x_a} - e^{-a_2}$, which increases the flexibility at small $x$ \cite{Eskola:2007my}. Second, at intermediate values of $x$ we use a functional form that is often used to parametrize the absolute PDF. The first derivatives are taken to be zero at the matching points $x_a$ and $x_e$ corresponding to the locations of the anticipated antishadowing maximum and EMC minimum. This fixes four parameters. Apart from the new small-$x$ parameter $a_2$ and the large-$x$ parameter $c_0$, the rest of the parameters $a_i, b_i, c_i$ are expressed in terms of $y_a$, $y_e$ and $y_0$ which correspond to the values of the function at $x=x_a$, $x=x_e$ and $x=0$. The parametrization is illustrated in Fig.~\ref{fig:FitForm}, where the variation induced by the parameter $a_2$ is also demonstrated. 

For the gluons and valence quarks the $y_0$ parameters are determined separately for each nucleus by imposing the sum rules, 
\begin{align}
& \int_0^1 dx f_{u_{\rm V}}^{{\rm p}/A}(x,Q_0^2) = 2 \,, \label{eq:summ1} \\
& \int_0^1 dx f_{d_{\rm V}}^{{\rm p}/A}(x,Q_0^2) = 1 \,, \label{eq:summ2} \\
& \int_0^1 dx x \sum_i f_i^{{\rm p}/A}(x,Q_0^2) = 1 \,.  \label{eq:summ3}
\end{align}
The rest of the $A$ dependence is encoded into the height parameters $y_i$ as, 
\begin{equation}
y_i(A) = 1 + \Big[y_i(A_{\rm ref}) - 1 \Big] \left(\frac{A}{A_{\rm ref}} \right)^{\gamma_i}\,,  \label{eq:Adep}
\end{equation}
where $A_{\rm ref} = 12$, following our earlier analyses \cite{Eskola:1998iy,Eskola:2007my,Eskola:2008ca,Eskola:2009uj,Eskola:2016oht}. In other words the nuclear effect -- the distance from unity -- is assumed to scale as a power law. For strange quarks the small-$x$ exponent $\gamma_{y_0}$ is modified by
\begin{align}
\gamma_{y_0} \longrightarrow \gamma_{y_0} y_0\theta(1-y_0) \,, \label{eq:Adep2}
\end{align}
so that the $A$ dependence becomes weaker as $y_0 \rightarrow 0$. This is to keep the strange-quark PDFs from becoming overly negative which easily leads to negative charm-production cross sections in neutrino-nucleus DIS. 

The values of the strong coupling and heavy-quark pole masses are taken to be the same as in the CT18ANLO analysis \cite{Hou:2019efy}: the charm mass is set to $m_c= 1.3\,{\rm GeV}$,  the bottom-quark mass to $m_b= 4.75\,{\rm GeV}$, and the strong coupling is fixed to $\alpha_{\rm s}(M_{\rm Z}) = 0.118$, where $M_{\rm Z}$ is the $Z$ boson mass. At higher scales $Q^2>Q^2_0$ the nuclear PDFs are obtained through solving the 2-loop \cite{Furmanski:1980cm,Curci:1980uw} DGLAP evolution equations \cite{Dokshitzer:1977sg,Gribov:1972ri,Gribov:1972rt,Altarelli:1977zs} for which we use the method introduced in Ref.~\cite{Santorelli:1998yt}.

In the course of the analysis we also noticed that the DIS data for Li-6 and He-3 are not optimally reproduced by the monotonic power-law ansatz of Eq.~(\ref{eq:Adep}). Therefore we have introduced extra parameters, $f_3$ and $f_6$, and replace the nuclear modifications $R^{{\rm p}/3}_i(x,Q_0^2)$ and $R^{{\rm p}/6}_i(x,Q_0^2)$ by 
\begin{align}
R^{{\rm p}/3}_i(x,Q_0^2) & \longrightarrow 1 + f_3 \Big[R^{{\rm p}/3}_i(x,Q_0^2) - 1 \Big] \,, \\
R^{{\rm p}/6}_i(x,Q_0^2) & \longrightarrow 1 + f_6 \Big[R^{{\rm p}/6}_i(x,Q_0^2) - 1 \Big] \,, \label{eq:f3f6}
\end{align}
for all parton flavours $i$. The effect is larger for He-3 and keeping $f_3=1$ would lead to a completely incorrect EMC slope in the case of JLab He-3 data. In total the EPPS21 fit involves $N_{\rm param} = 24$ free parameters, see Table~\ref{Table:Params} ahead. Out of these 24 only 5 control the $A$ dependence of the parametrization and freeing more --- e.g. letting the $A$ dependence of the gluon antishadowing peak to vary independently of the valence quarks --- easily destabilizes the fit. Thus, there is more parametrization dependence e.g. in the gluon distributions of small nuclei in contrast to the case of heavy nuclei where the LHC data now provide strong constraints. To better control the $A$ dependence, e.g. pO runs at the LHC would be most welcome \cite{Paakkinen:2021jjp}. 

As in EPPS16, the deuteron is still taken to be free from nuclear effects, $R_i^A(x,Q^2) = 1$. In principle, as done e.g. in Ref.~\cite{Walt:2019slu}, one could include NMC data \cite{Arneodo:1996qe} on $F_2^{\rm D}/F_2^{\rm p}$ to constrain the deuteron nuclear effects simultaneously with the other nuclear data. The nuclear effects in deuteron are expected to be below $2\%$ \cite{Martin:2012da}. However, these deuteron data are already included in the CT18 fit \cite{Hou:2019efy} of the free proton PDFs (our baseline) neglecting the deuteron nuclear corrections \cite{Accardi:2021ysh}. Using CT18 for deuteron (with no additional corrections) thus effectively accounts also for the deuteron nuclear effects. As a result, including these NMC data in our analysis here would thus be inconsistent, leading also to some double counting. The way the deuteron is now handled is admittedly a bit unsatisfactory and once more underscores the fact that the era of fitting the free-proton and nuclear PDFs separately starts to come to its end. 

\subsection{Negativity features}

The parametrization of $R_i^{{\rm p}/A}(x,Q^2)$ is not restricted to be strictly positive definite at the parametrization scale -- whether or not the $\overline{\rm MS}$ PDFs should be non-negative is still an open question \cite{Candido:2020yat,Collins:2021vke}. In our analysis, particularly $R_g^{{\rm p}/A}$ at the parametrization scale gets easily negative at small $x$. However, since the central CT18 gluon distribution is valence-like -- in practice zero at small $x$ -- this does not easily lead to negative cross sections even at $Q^2 = Q^2_0$. The negativity problem of the gluon also goes away immediately above the parametrization scale. Despite the modified $A$ dependence in Eq.~(\ref{eq:Adep2}), some error sets involve a negative strange-quark nuclear PDF at small $x$. Also this feature disappears rapidly with the scale evolution and negative cross sections can presumably be avoided by adopting a high-enough factorization scale. Finally, some bound-proton modifications $R^{{\rm p}/A}_i$ for up and down quarks can display negative values at small $x$. However, this is irrelevant from the practical viewpoint as when combined to the full nucleus level in Eq.~(\ref{eq:FullPDF}) such negativity features disappear due to the anticorrelation between up- and down-quark nuclear modifications in bound protons (see Fig.~10 of Ref.~\cite{Eskola:2016oht}). The negativity issues in our analysis are similar as met in other recent global analyses by other groups, see Fig.~\ref{fig:absolutePDFsothers} ahead. 

\section{Observables}
\label{Observables}

\begin{figure*}[htb!]
\centering
\includegraphics[width=0.8\linewidth]{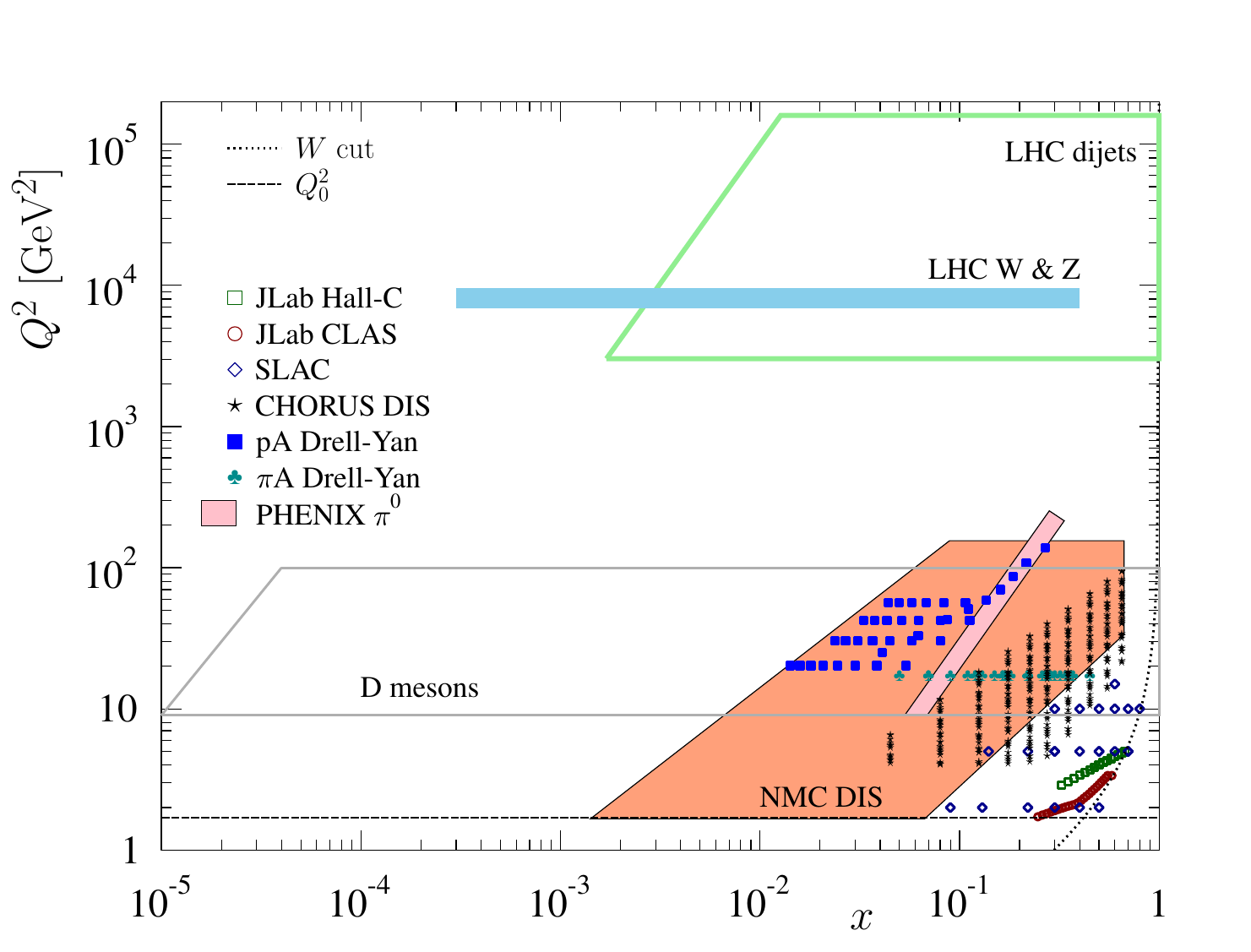}
\caption{The data included in the EPPS21 laid schematically on the $(x,Q^2)$ plane.}
\label{fig:xQ}
\end{figure*}

The observables we include in the analysis are chosen such that theoretical uncertainties from higher-order perturbative corrections, non-perturbative fragmentation functions (applied in the case of inclusive pion and D-meson production) and from the choice of the proton baseline PDFs are kept as small as possible. In practice, this means e.g. taking ratios of cross sections between different collision systems. Also, observables where final-state nuclear effects could be sizable --- e.g. quarkonia hadroproduction --- are not included. In Fig.~\ref{fig:xQ} we sketch the $x$ and $Q^2$ regions probed by the data that are included in the present analysis. 

\subsection{Lepton-nucleus DIS}
\label{leptonnucleusDIS}

The foundations of the global fits of nuclear PDFs are built on the muon-nucleus and electron-nucleus neutral-current DIS data. There is a significant legacy of data available from SLAC \cite{Gomez:1993ri}, NMC \cite{Amaudruz:1995tq,Arneodo:1995cs,Arneodo:1996rv,Amaudruz:1995tq,Arneodo:1996ru}, and EMC \cite{Ashman:1992kv} collaborations which we include in the analysis. As in EPPS16, we undo the isoscalar corrections that the experiments have in some cases imposed on the data, see Sect.~3.1 of the EPPS16 paper \cite{Eskola:2016oht}. As a new ingredient, we now include also very recent electron-nucleus data from the JLab CLAS \cite{Schmookler:2019nvf} and Hall-C \cite{Seely:2009gt} experiments. The CLAS data were already analyzed in Ref.~\cite{Paukkunen:2020rnb} in the context of PDF reweighting and a good compatibility with EPPS16 was found. 

The neutral-current DIS data mentioned here come either as ratios of $F_2$ structure functions at fixed virtuality $Q^2$ and Bjorken $x$,
\begin{align}
R^A_{{\rm F_2}}(x,Q^2) = \frac{F_2^A(x,Q^2)}{F_2^{\rm D}(x,Q^2)}  \,,
\end{align}
or as ratios of cross sections,
\begin{align}
R^A_{\sigma}(x,Q^2) = \frac{d^2\sigma^{\ell^-A}}{dxdQ^2} \bigg / \frac{d^2\sigma^{\ell^-{\rm D}}}{dxdQ^2} \,. 
\end{align}

We also include the CHORUS (anti)neutrino-nucleus charged-current DIS data \cite{Onengut:2005kv} which help in constraining the strange-quark content of the nucleons as well as the up-quark vs. down-quark flavour separation. These neutrino data are available as absolute cross sections. To reduce the experimental and theoretical uncertainties we do as first proposed in Ref.~\cite{Paukkunen:2013grz} and then elaborated in the EPPS16 analysis, and normalize the data by the integrated cross section at the corresponding beam energy. For exact treatment, see Sect.~3.2 of the EPPS16 paper \cite{Eskola:2016oht}.

In the EPPS16 analysis we did not pay too much attention to the treatment of normalization errors of the neutral-current DIS data i.e. the normalization uncertainties were simply added in quadrature, point-by-point. We are now more careful in this respect and, when available, treat the normalization errors more consistently in the definition of $\chi^2_{\rm global}$, see Eq.~(\ref{eq:chi2wnorm}) ahead.

All the DIS cross sections are computed using the SACOT-$\chi$ (simplified Aivazis-Collins-Olness-Tung) versions of the general-mass variable flavour number scheme (GM-VFNS) \cite{Kramer:2000hn,Collins:1998rz,Thorne:2008xf}. This is the same scheme that is used in the CT18 analysis \cite{Hou:2019efy}. In the case of neutrino DIS, we include also the dominant electroweak \cite{Arbuzov:2004zr} effects as well as target-mass  corrections following Ref.~\cite{Accardi:2008ne}. For the neutral-current DIS there were no target-mass corrections in the EPPS16 analysis. However, they begin to play a role in the case of JLab kinematics and we thus now account for the dominant target-mass corrections also in neutral-current DIS. This is done as in Ref.~\cite{Paukkunen:2020rnb} where the effects were studied in more detail -- see also Ref.~\cite{Segarra:2020gtj}. Only those DIS data points with the hadronic invariant mass $W \geq 1.8\,{\rm GeV}$ are included in the analysis.

\subsection{Proton-nucleus Drell-Yan}
\label{pnucleusDY}

The low-mass Drell-Yan production in proton-nucleus collisions helps in disentangling between the nuclear effects in valence and sea quarks. Here, we include the E772 \cite{Alde:1990im} and E866 \cite{Vasilev:1999fa} data sets in the form of nuclear ratios,
\begin{align}
\frac{d^2\sigma^{{\rm p}A}}{dx_2} \bigg/ \frac{d^2\sigma^{\rm pD}}{dx_2} \,, \ \ \
\frac{d^2\sigma^{{\rm p}A}}{dMdx_1} \bigg/ \frac{d^2\sigma^{\rm pBe}}{dMdx_1} \,,
\end{align}
where $M$ is the invariant mass of the produced lepton pair and $x_{1,2} = (M/\sqrt{s})e^{\pm y}$, where $y$ is the rapidity of the lepton pair. The differential cross sections are calculated ``on fly'' with no precomputed grids.

\subsection{Dijet production}
\label{Dijetproduction}

In the EPPS16 analysis, we used the first CMS $5 \, {\rm TeV}$ single-differential dijet pPb data \cite{Chatrchyan:2014hqa} in the form of a forward-to-backward ratio. Now, a double-differential analysis \cite{Sirunyan:2018qel} of the same data sample has become available and this is what we use in the present analysis. We have already scrutinized these data in Ref.~\cite{Eskola:2019dui} where they were found to put dramatically strong constraints on the nuclear modification of the gluon PDFs in the shadowing and antishadowing regions. The observable we fit is a double ratio,
\begin{align}
& R^{\rm norm.}_{\rm pPb} \big(\eta_{\rm dijet},p_{\rm T}^{\rm ave}\big) = \label{eq:selfnormdijet} \\ 
& \frac{1}{d\sigma^{\rm pPb}/dp_{\rm T}^{\rm ave}} \frac{d^2\sigma^{\rm pPb}}{d\eta_{\rm dijet}dp_{\rm T}^{\rm ave}}
\bigg /
\frac{1}{d\sigma^{\rm pp}/dp_{\rm T}^{\rm ave}} \frac{d^2\sigma^{\rm pp}}{d\eta_{\rm dijet}dp_{\rm T}^{\rm ave}} \nonumber \,,
\end{align}
where $\eta_{\rm dijet}$ and $p_{\rm T}^{\rm ave}$ are the average pseudorapidity and average transverse momentum of the two jets that make up the dijet, 
\begin{align}
\eta_{\rm dijet} & = \frac{1}{2} \Big(\eta^{\rm leading } + \eta^{\rm subleading }\Big) \,, \\
p_{\rm T}^{\rm ave} & = \frac{1}{2} \Big(p_{\rm T}^{\rm leading } + p_{\rm T}^{\rm subleading }\Big) \,.
\end{align}
By self-normalizing the spectra separately in pp and pPb collisions, a major part of the experimental systematic uncertainties cancel and the measurement is therefore very precise. Without the self-normalization, the systematic uncertainties in typical jet measurement can reach tens of percents. In Ref.~\cite{Eskola:2019dui} the ratio of Eq.~(\ref{eq:selfnormdijet}) was also found to be very insensitive to the choice of the baseline proton PDFs as well as to the factorization/renormalization scale variations around the central choice $\mu = p_{\rm T}^{\rm ave}$. The NLO look-up tables (see Sect.~\ref{Lookuptables}) are constructed by using the public NLOjet++ \cite{Nagy:2003tz} code. For more details on the implementation of the dijet cross sections, see Ref.~\cite{Eskola:2019dui}.

\subsection{W$^\pm$ and Z production}
\label{Electroweakbosons}

In the EPPS16 fit, we already included the $5\,{\rm TeV}$ W$^\pm$ and Z production data from CMS and ATLAS \cite{Khachatryan:2015hha,Khachatryan:2015pzs,Aad:2015gta}. These data were included as rapidity-binned forward-to-backward ratios,
\begin{align}
R_{\rm FB}^{Z, W^\pm}(y,\sqrt{s}) = \frac{d\sigma^{Z, W^\pm}_{\rm pPb}(y,\sqrt{s})}{dy} \bigg / \frac{d\sigma^{Z, W^\pm}_{\rm pp}(-y, \sqrt{s})}{dy} \,,
\end{align}
where, depending on the case, the rapidity variable $y$ refers either to the rapidity of the produced Z boson, or to the rapidity of the charged lepton resulting from the W$^\pm$ decay. These data are included in the EPPS21 analysis in the same way. Also the CMS $8.16\,{\rm TeV}$ W$^\pm$ data from pPb collisions are now available \cite{Sirunyan:2019dox} and they are significantly more precise than the $5\,{\rm TeV}$ data. To make the most out these new data, we form a nuclear modification ratio by using the $8\,{\rm TeV}$ W$^\pm$ data from the p-p measurement \cite{Khachatryan:2016pev}, 
\begin{align}
R_{\rm pPb}^{W^\pm}(y) = \frac{d\sigma^{W^\pm}_{\rm pPb}(\sqrt{s}=8.16\,{\rm TeV})/dy}{d\sigma^{W^\pm}_{\rm pp}(\sqrt{s}=8\,{\rm TeV})/dy} \,. \label{eq:mixedR}
\end{align}
In addition to the slightly different c.m. energies, the pp and pPb measurements do not share an exactly common rapidity binning and for some rapidity bins the ratio in Eq.~(\ref{eq:mixedR}) is taken between two nearby rapidity bins. The differences between the c.m. energies and the rapidity bins are, however, small enough that the uncertainties from the baseline proton PDFs still cancel rather well. This is demonstrated explicitly in Ref.~\cite{Eskola:2022rlm}, where also the rapidity binning of the mixed-energy ratio and the composition of the covariance matrix are explained. We acknowledge that there are also a few LHCb \cite{Aaij:2014pvu} and ALICE \cite{Alice:2016wka,Acharya:2020puh} data points available on cross sections for producing electroweak bosons in pPb collisions. These data are not included in the current analysis as a longer lever arm in rapidity would be required to place significant constraints through nuclear modifications in the shape of the rapidity distributions. The NLO look-up tables (see Sect.~\ref{Lookuptables}) for W$^\pm$ and Z production are constructed by using the public MCFM \cite{Boughezal:2016wmq} program, taking the renormalization/factorization scale equal to the pole mass of the produced heavy boson.

\subsection{D-meson production}
\label{Dmesonproduction}

\begin{figure}[htb!]
\centering
\includegraphics[width=1.0\linewidth]{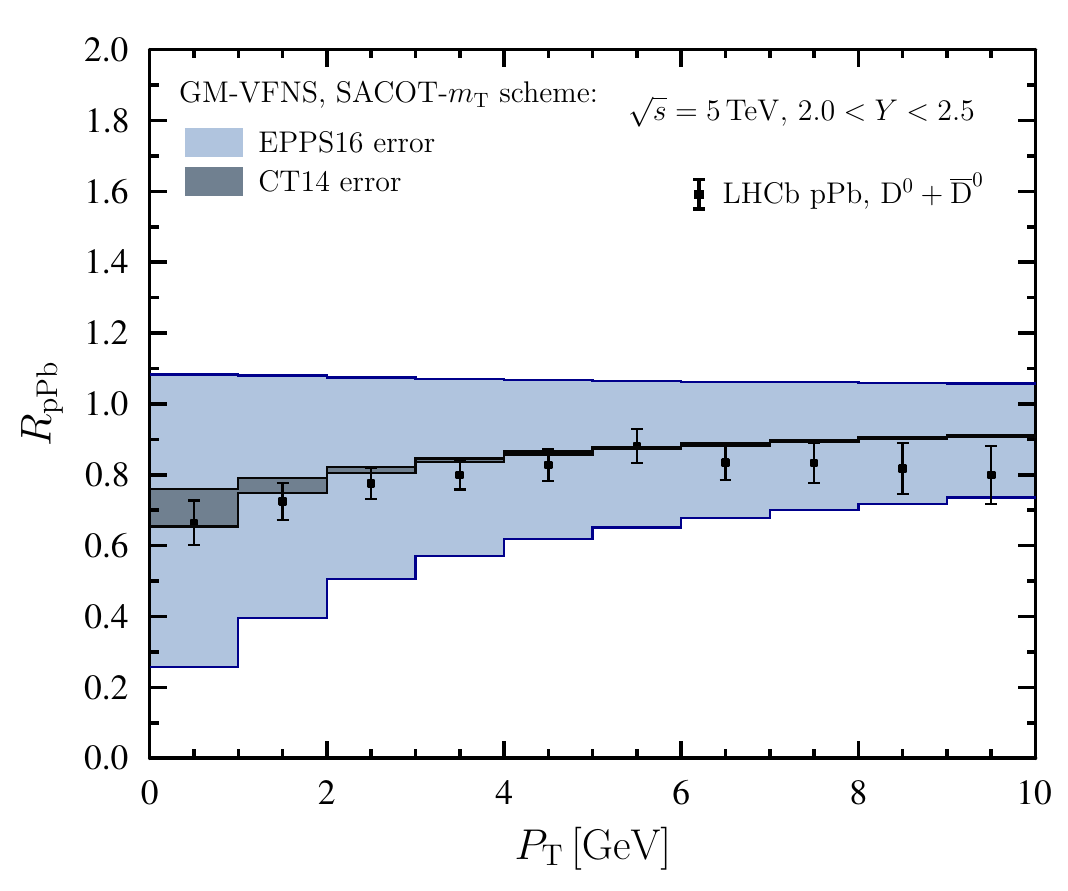}
\includegraphics[width=1.0\linewidth]{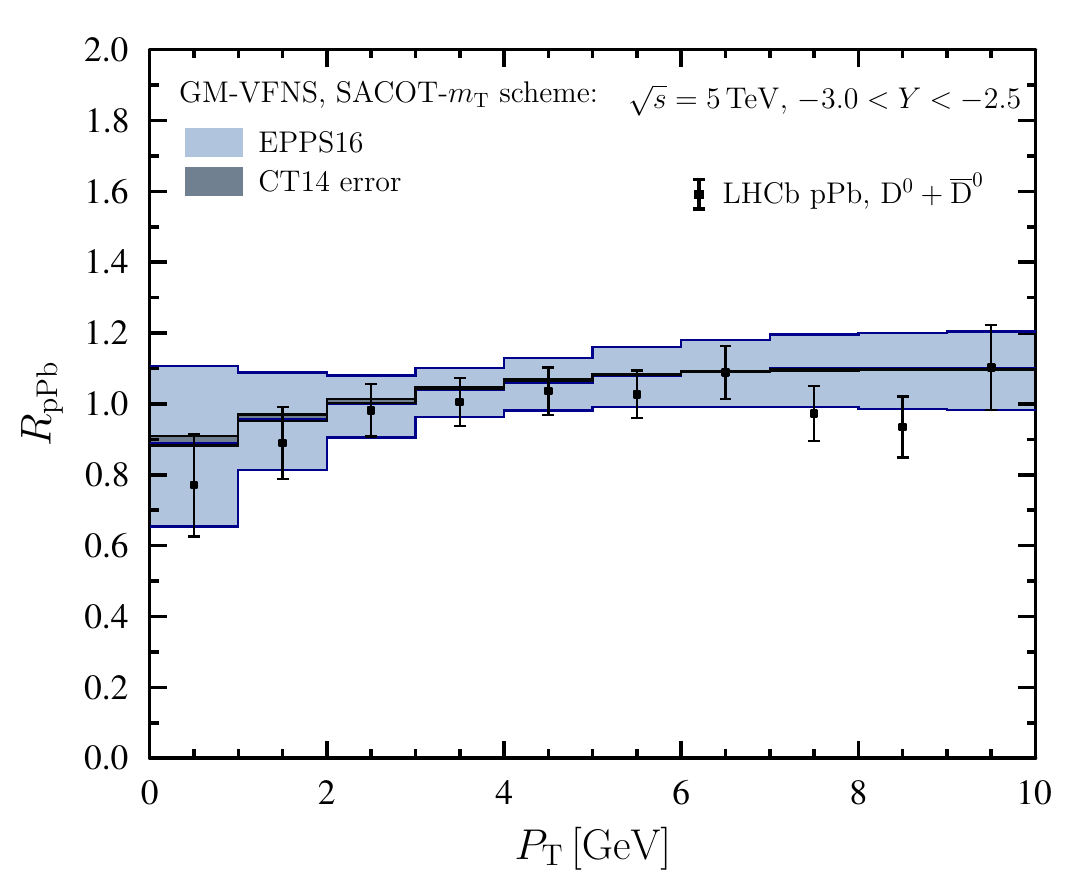}
\caption{The EPPS16 and CT14NLO uncertainties for D-meson production in forward (upper panel) and backward (lower panel) direction in p-Pb collisions. The data are from the LHCb collaboration \cite{Aaij:2017gcy}.
}
\label{fig:Dmesonperror}
\end{figure} 

Inclusive D-meson production in proton-nucleus collisions is known to carry a significant sensitivity on the gluon PDFs \cite{Cacciari:2015fta,Gauld:2015yia,Gauld:2016kpd,Zenaiev:2015rfa,Gauld:2015lxa,Kusina:2017gkz,Kramer:2017gct,Kusina:2020dki,Eskola:2019bgf}. Our EPPS21 analysis includes the double differential nuclear modification factors 
\begin{align}
R^{\rm D}_{\rm pPb}(p_{\rm T}, y) = \frac{d^2\sigma_{\rm pPb}}{dy dp_{\rm T}} \bigg / \frac{d^2\sigma_{\rm pp}}{dy dp_{\rm T}} \,,
\end{align}
measured by the LHCb collaboration at $\sqrt{s}= 5 \, {\rm TeV}$ \cite{Aaij:2017gcy}. The calculations are performed in the SACOT-$m_{\rm T}$ GM-VFNS scheme \cite{Helenius:2018uul} which accounts for the finite masses of the charm quark and the produced D meson. We use the KKKS08 fragmentation functions \cite{Kneesch:2007ey}. At very low values of $p_{\rm T}$, theoretical uncertainties from the QCD scale choices and treatment of the finite-mass effects become more important \cite{Eskola:2019bgf}. To reduce the impact of these uncertainties, a cut $p_{\rm T} > 3 \, {\rm GeV}$ is applied in the present analysis. Also the uncertainties from the baseline proton PDFs turn out to be negligible in the fitted region. This is demonstrated in Fig.\ref{fig:Dmesonperror} where the CT14NLO \cite{Dulat:2015mca} and EPPS16 uncertainties are shown for two rapidity intervals. For more details about the NLO setup, see Ref.~\cite{Eskola:2019bgf}.

Compared to the analysis in Ref.~\cite{Eskola:2019bgf}, we take here the luminosity uncertainties as correlated and not point-by-point also for this observable. Due to the separate (beam-direction reversed) data taking of the pPb forward and backward configurations, the normalization shifts can be different in the different configurations, as is allowed in our fit. The normalization uncertainties are computed as the quadratic sum of the given pPb and p-p uncertainties, separately for each beam direction.

\begin{figure*}[htb!]
\centering
\includegraphics[width=1.0\linewidth]{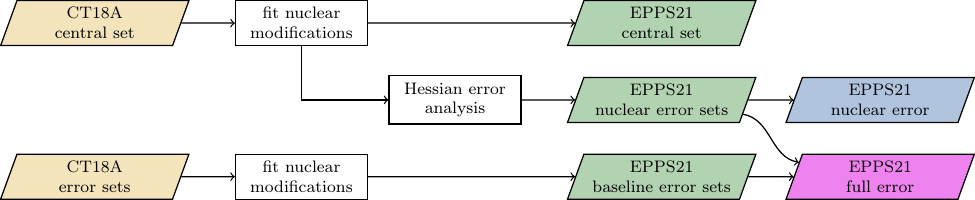}\vspace{0.3cm}
\caption{An outline of the analysis procedure of EPPS21.}
\label{fig:Flow}
\end{figure*} 

\subsection{Inclusive pion production}
\label{Inclusivepionproduction}

Historically, the inclusive pion production at RHIC was the first direct gluon constraint in the fits on nuclear PDFs \cite{deFlorian:2003qf,Eskola:2008ca}. Until that point only weak gluon constraints were obtained from the $Q^2$ slopes of $R_{{\rm F}_2}$ at small $x$ \cite{Eskola:2002us}. In the present analysis we still include the PHENIX 2007 $\pi^0$ data \cite{Adler:2006wg} taken in dAu collisions. However, due to including the D-meson and dijet data, the role of these pion data is no longer as prominent as it used to be e.g. in the EPS09 analysis \cite{Eskola:2009uj}. They are still included to highlight the compatibility. Recently, the nCTEQ collaboration has comprehensively studied the inclusion of also other light-meson data from RHIC and LHC \cite{Duwentaster:2021ioo} including the uncertainties coming from the fragmentation functions. The very recent PHENIX data \cite{PHENIX:2021dod}, however, seems to indicate a much stronger $p_{\rm T}$ dependence of the nuclear modifications in comparison e.g. to the EPPS16 predictions, which puts a new question mark on whether these newer data can still be consistently included in the global fits. Exploring the full implications of these new data is left for future work. The look-up tables (see Sect.~\ref{Lookuptables}) for inclusive pion production are constructed by using the INCNLO \cite{Aurenche:1999nz} code interfaced with the KKP fragmentation functions \cite{Kniehl:2000fe}, taking the factorization, renormalization and fragmentation scales equal to the $p_{\rm T}$ of the produced pion. As with the D-mesons, we use a $p_{\rm T} > 3 \, {\rm GeV}$ cut when fitting these data.

 
\subsection{Pion-nucleus Drell-Yan}
\label{PionnucleusDrellYan}

The pion-nucleus Drell-Yan process has potential to constrain the flavour decomposition of the valence quarks \cite{Paakkinen:2016wxk,Paakkinen:2017eat}. Unfortunately, the precision of the data is not high enough to provide significant discrimination power on top of the DIS data. We nevertheless include the same pion-induced Drell-Yan data sets that were there in the EPPS16 fit: The E615 data \cite{Heinrich:1989cp} is for the ratios,
\begin{align}
\frac{d\sigma^{\pi^+{\rm W}}}{dx_2} \bigg/ \frac{d\sigma^{\pi^-{\rm W}}}{dx_2} \,, \ \ \
\end{align}
the NA3 data \cite{Badier:1981ci} for ratios,
\begin{align}
\frac{d\sigma^{\pi^-{\rm Pt}}}{dx_2} \bigg/ \frac{d\sigma^{\pi^-{\rm H}}}{dx_2} \,, \ \ \
\end{align}
and the NA10 data \cite{Bordalo:1987cs} for ratios,
\begin{align}
\frac{d\sigma^{\pi^-{\rm W}}}{dx_2} \bigg/ \frac{d\sigma^{\pi^-{\rm D}}}{dx_2} \,. \ \ \
\end{align}
The look-up tables (see Sect.~\ref{Lookuptables}) are constructed with MCFM \cite{Boughezal:2016wmq} using the GRV pion PDFs \cite{Gluck:1991ey}. The isospin correction applied by the NA10 collaboration is accounted for as explained in Sect.~3 of Ref.~\cite{Paakkinen:2016wxk}, and the data normalization is treated separately for the high and low energy data. 


\section{Analysis procedure}
\label{Analysisprocedure}

As our preceding fits, the core of EPPS21 is based on $\chi^2$ minimization followed by the Hessian uncertainty analysis. As a new ingredient we now also chart the uncertainties that originate from the uncertainties of the baseline free proton PDF. The flow of the analysis is illustrated in Fig.~\ref{fig:Flow} whose content is explained below in more detail.

\subsection{The global $\chi^2$ function}
\label{TheChi2function}

The analysis is based on minimizing the global $\chi_{\rm global}^2$ figure-of-merit function. For each data set the individual $\chi^2$ is of the form, 
\begin{align}
\chi^2 = \sum_{ij} \big(T_i\{a_k \} - D_i\big) C_{ij}^{-1} \big(T_j\{a_k \} - D_j\big) \,,
\end{align}
where $T_i\{a_k \}$ denote the theory values, $D_i$ are the corresponding data values, and $C_{ij}^{-1}$ is the inverse of the experimental covariance matrix. The global $\chi_{\rm global}$ is then a sum of individual $\chi^2$s. In many cases, the overall normalization uncertainty is given separately and in this case,
\begin{align}
\chi^2 & = \sum_{ij} \big(T_i\{a_k \} - fD_i\big) \tilde C_{ij}^{-1} \big(T_j\{a_k \} - fD_j\big) \label{eq:chi2wnorm} \\
       & + \left(\frac{1-f}{\delta^{\rm norm.}}\right)^2 \nonumber  \,,
\end{align}
with
\begin{align}
\tilde C_{ij} = C_{ij} - \big(\delta^{\rm norm.}\big)^2 D_iD_j \,,
\end{align}
and the $\chi^2$ is minimized also with respect to $f$. When comparing the fit results with the theoretical values, we have multiplied the data values with the optimal normalization factor $f$, where appropriate. Effectively, we are thus neglecting here the possible D'Agostini bias \cite{DAgostini:1993arp}. The list of parameters that define our central fit is given in Table~\ref{Table:Params} and in Table~\ref{Table:Data} we list the individual data sets together with their central values of $\chi^2$.

\begin{table}[!htb]
\caption[]{\small Values of parameters that define the central EPPS21 nuclear PDFs at $Q_0^2=1.69\,{\rm GeV}^2$. The 24 parameters that were kept free in the fit are indicated in bold.}
\label{Table:Params}
\begin{tabular}{c|llllll}
 Parameter              &  ${u_{\rm V}}$    & ${d_{\rm V}}$     &  ${\overline{u}}$ \\
\hline
   $y_0(A_{\rm ref})$	& {sum rule}        & {sum rule}        &  \textbf{0.870} \\
   $\gamma_{y_0}$       & {sum rule}        & {sum rule}        &  \textbf{0.401} \\ 
   $a_2$                & 0, fixed          & 0, fixed          & 0, fixed    \\
   $x_a$                &  \textbf{0.0577}  & as $u_{\rm V}$    &  \textbf{0.110} \\
   $x_e$                &  \textbf{0.700}   & as $u_{\rm V}$    & as $u_{\rm V}$  \\
   $y_a(A_{\rm ref})$   &  \textbf{1.07}    &  \textbf{1.04}    &  \textbf{0.992}  \\
   $\gamma_{y_a}$       &  \textbf{0.221}   & as $u_{\rm V}$    & 0, as $u_{\rm V}$   \\
   $y_e(A_{\rm ref})$   &  \textbf{0.877}   &  \textbf{0.968}   &  \textbf{0.956} \\
   $\gamma_{y_e}$       &  \textbf{0.176}   & as $u_{\rm V}$    & as $u_{\rm V}$  \\
   $c_0$                &  1.8, fixed       & 1.8, fixed        & 1.8, fixed    \\
   $\beta$              & \textbf{2.20}     & as $u_{\rm V}$    & 1.3, fixed   \\
   $f_3$                & \textbf{0.291}    & as $u_{\rm V}$    & as $u_{\rm V}$    \\
   $f_6$                & \textbf{0.495}    & as $u_{\rm V}$    & as $u_{\rm V}$    \\
\hline \\
 Parameter              &  ${\overline{d}}$ &  ${s}$            & $g$      \\
\hline
   $y_0(A_{\rm ref})$   & \textbf{0.921}    & \textbf{0.403}    & {sum rule}      \\
   $\gamma_{y_0}$       & as $\overline{u}$ & as $\overline{u}$ & {sum rule}      \\ 
   $a_2$                & 0, fixed          & 0, fixed          & \textbf{3.66}    \\
   $x_a$                & as $\overline{u}$ & as $\overline{u}$ & \textbf{0.0975} \\
   $x_e$                & as $u_{\rm V}$    & as $u_{\rm V}$    & as $u_{\rm V}$  \\
   $y_a(A_{\rm ref})$   & \textbf{0.971}    &  \textbf{1.09}    &  \textbf{1.10}  \\
   $\gamma_{y_a}$       & $u_{\rm V}$       & $u_{\rm V}$       & as $u_{\rm V}$  \\
   $y_e(A_{\rm ref})$   & as $\overline{u}$ & as $\overline{u}$ & \textbf{0.852}  \\
   $\gamma_{y_e}$       & as $u_{\rm V}$    & as $u_{\rm V}$    & as $u_{\rm V}$  \\
   $c_0$                & 1.8, fixed        & 1.8, fixed        & 1.8, fixed    \\   
   $\beta$              & 1.3, fixed        & 1.3, fixed        & 1.3, fixed   \\
   $f_3$                & as $u_{\rm V}$    & as $u_{\rm V}$    & as $u_{\rm V}$    \\
   $f_6$                & as $u_{\rm V}$    & as $u_{\rm V}$    & as $u_{\rm V}$    \\
\end{tabular}
\end{table}

\begin{table*}
\begin{center}
{
\caption{The data used in the EPPS21 analysis. The new data with respect to the EPPS16 analysis are marked with a star.}
\label{Table:Data}
\begin{tabular}{lclrrcl}
 Experiment & Observable &  Collisions & Data points & $\chi^2$ & Normalization & Ref. \\
\hline
\hline
 JLab Hall C${^\bigstar}$         & DIS	& $e^-$He(3), $e^-$D            & 15 & 4.47    & 1.027 & \cite{Seely:2009gt} \\
 \\
 JLab Hall C${^\bigstar}$         & DIS	& $e^-$He(4), $e^-$D            & 15 & 4.33    & 0.985 & \cite{Seely:2009gt} \\
 SLAC E139                        & DIS	& $e^-$He(4), $e^-$D            & 16 & 7.75    & 0.997 & \cite{Gomez:1993ri} \\
 CERN NMC 95, re.                 & DIS	& $\mu^-$He(4), $\mu^-$D        & 16 & 17.90   & 1.000 & \cite{Amaudruz:1995tq} \\
 \\
 CERN NMC 95, $Q^2$ dep.          & DIS	& $\mu^-$Li(6), $\mu^-$D        & 153 & 159.74  & 1.002 & \cite{Arneodo:1995cs} \\
  \\
 JLab Hall C${^\bigstar}$         & DIS	& $e^-$Be(9), $e^-$D            & 15 & 4.72     & 0.971 & \cite{Seely:2009gt} \\
 SLAC E139                        & DIS	& $e^-$Be(9), $e^-$D            & 15 & 15.19    & 0.990 & \cite{Gomez:1993ri}  \\
 CERN NMC 96                      & DIS	& $\mu^-$Be(9), $\mu^-$C        & 15 & 4.84     & 1.000 & \cite{Arneodo:1996rv} \\
 \\
 JLab Hall C${^\bigstar}$        & DIS	& $e^-$C(12), $e^-$D            & 15  & 2.58    & 0.981 & \cite{Seely:2009gt} \\
 SLAC E139	                     & DIS	& $e^-$C(12), $e^-$D            &  6  & 4.89    & 0.998 & \cite{Gomez:1993ri}  \\
 CERN NMC 95, $Q^2$ dep.         & DIS	& $\mu^-$C(12), $\mu^-$D        & 165 & 131.25  & 0.997 & \cite{Arneodo:1995cs}  \\ 
 CERN NMC 95, re.                & DIS	& $\mu^-$C(12), $\mu^-$D        & 16  & 16.99   & 0.998 & \cite{Amaudruz:1995tq} \\
 CERN NMC 95, re.                & DIS	& $\mu^-$C(12), $\mu^-$Li(6)    & 20  & 16.27   & 0.997 & \cite{Amaudruz:1995tq} \\
 JLab CLAS${^\bigstar}$          & DIS	& $e^-$C(12), $\mu^-$D(6)       & 25  & 19.41   & 0.996 &\cite{Schmookler:2019nvf} \\
 FNAL E772                       & DY	& pC(12), pD                    &  9  & 8.20    & - & \cite{Alde:1990im}    \\
\\
 SLAC E139                       & DIS	& $e^-$Al(27), $e^-$D           & 15  &  10.58  & 0.994 & \cite{Gomez:1993ri}   \\
 CERN NMC 96    	             & DIS 	& $\mu^-$Al(27), $\mu^-$C(12)   & 15  & 7.02    &  1.000 & \cite{Arneodo:1996rv} \\
 JLab CLAS${^\bigstar}$          & DIS	& $e^-$Al(27), $e^-$D           & 25  & 20.68   & 1.004 & \cite{Schmookler:2019nvf} \\
\\
SLAC E139	                     & DIS	& $e^-$Ca(40), $e^-$D           &  6 & 3.91     & 0.989  & \cite{Gomez:1993ri}   \\
CERN NMC 95, re. 	             & DIS	& $\mu^-$Ca(40), $\mu^-$D       & 15 & 30.45    & 1.004 & \cite{Amaudruz:1995tq} \\
CERN NMC 95, re. 	             & DIS	& $\mu^-$Ca(40), $\mu^-$Li(6)   & 20 & 17.08    & 0.998 & \cite{Amaudruz:1995tq} \\
CERN NMC 96    	                 & DIS	& $\mu^-$Ca(40), $\mu^-$C(12)   & 15 & 8.35     & 1.001 & \cite{Arneodo:1996rv} \\
FNAL E772 	                     &  DY	& pCa(40), pD                   &  9 & 2.59     & - & \cite{Alde:1990im}  \\
\\
SLAC E139	                     & DIS	& $e^-$Fe(56), $e^-$D           & 20 & 23.86    & 1.002 & \cite{Gomez:1993ri}   \\
CERN NMC 96    	                 & DIS	& $\mu^-$Fe(56), $\mu^-$C(12)   & 15 & 11.11    & 1.001 & \cite{Arneodo:1996rv} \\
JLab CLAS${^\bigstar}$           & DIS	& $e^-$Fe(56), $e^-$D           & 25 & 26.74    & 1.005 & \cite{Schmookler:2019nvf} \\
FNAL E772 	                     & DY	& $e^-$Fe(56), $e^-$D           &  9 & 2.03     & - & \cite{Alde:1990im}    \\
FNAL E866  	                     & DY   & pFe(56), pBe(9)               & 28 & 21.04    & - & \cite{Vasilev:1999fa} \\
\\ 
CERN EMC                         & DIS  & $\mu^-$Cu(64), $\mu^-$D       & 19 & 15.13    & - & \cite{Ashman:1992kv}   \\
\\
SLAC E139	                     & DIS	& $e^-$Ag(108), $e^-$D          &  6 & 8.13     & 0.990 & \cite{Gomez:1993ri}   \\
\\
CERN NMC 96  	                 & DIS  & $\mu^-$Sn(117), $\mu^-$C(12)  & 15 & 10.90    & 0.999 & \cite{Arneodo:1996rv} \\
CERN NMC 96, $Q^2$ dep.          & DIS	& $\mu^-$Sn(117), $\mu^-$C(12)  & 144 & 84.44   & 0.999 & \cite{Arneodo:1996ru} \\
\\
FNAL E772 	                     & DY	& pW(184), pD     	            &  9  & 5.93    & - & \cite{Alde:1990im}    \\
FNAL E866 	                     & DY	& pW(184), pBe(9)               & 28  & 25.82   & - & \cite{Vasilev:1999fa} \\
CERN NA10                        & DY	&  $\pi^-$W(184), $\pi^-$D      & 10  & 10.87   & 1.040(h.e), 1.116(l.e)  & \cite{Bordalo:1987cs}\\
FNAL E615                        & DY	&  $\pi^+{\rm W}(184)$, $\pi^-{\rm W}$(184)  &  11  &  13.26 & - & \cite{Heinrich:1989cp} \\
\\
CERN NA3                         & DY	&  $\pi^-$Pt(195), $\pi^-$H    & 7  &  4.70     & - & \cite{Badier:1981ci} \\
\\
SLAC E139	                     & DIS	& $e^-$Au(197), $e^-$D          & 16 & 19.70    & 0.999 & \cite{Gomez:1993ri}   \\
RHIC PHENIX                      & $\pi^0$ 	& dAu(197), pp              & 17 & 6.68     & 1.008 & \cite{Adler:2006wg}\\

\\
CERN NMC 96                      & DIS	        & $\mu^-$Pb(207), $\mu^-$C(12)           & 15  & 4.29       & 1.000 & \cite{Arneodo:1996rv} \\
JLab CLAS${^\bigstar}$           & DIS	& $e^-$Pb(208), $e^-$D                   & 25  & 15.39   & 0.994 & \cite{Schmookler:2019nvf} \\
CERN CHORUS                      & DIS	        & $\nu$Pb(208), $\overline{\nu}$Pb(208)  & 824 & 990.95   & - & \cite{Onengut:2005kv} \\
CERN CMS 5{\rm TeV}              & W$^\pm$	    & pPb(208)                           & 10  & 11.82      & - & \cite{Khachatryan:2015hha}\\
CERN CMS 8{\rm TeV}${^\bigstar}$ & W$^\pm$	    & pPb(208), pp               & 44  & 41.30      &  0.996 & \cite{Sirunyan:2019dox} \\
CERN CMS                         & Z	        & pPb(208)                               & 6   & 6.80       & - & \cite{Khachatryan:2015pzs}\\
CERN ATLAS                       & Z	        & pPb(208)                               & 7   & 8.91       & - & \cite{Aad:2015gta}\\
CERN CMS${^\bigstar}$            & dijet	    & pPb(208)                   & 83  & 123.81     & - & \cite{Chatrchyan:2014hqa}\\
CERN LHCb${^\bigstar}$           & D meson	    & pPb(208)           & 48  & 45.71       & 0.999(fwd.), 1.010(bwd.) & \cite{Aaij:2017gcy} \\
\\
 \hline		   
 Total 		     &        &     &  2077 & 2058.5 &  &                     \\
\end{tabular}
}
\end{center}
\end{table*}

\subsection{Uncertainty analysis}
\label{Uncertaintyanalysis}

\begin{figure*}[htb!]
\centering
\includegraphics[width=1.0\linewidth]{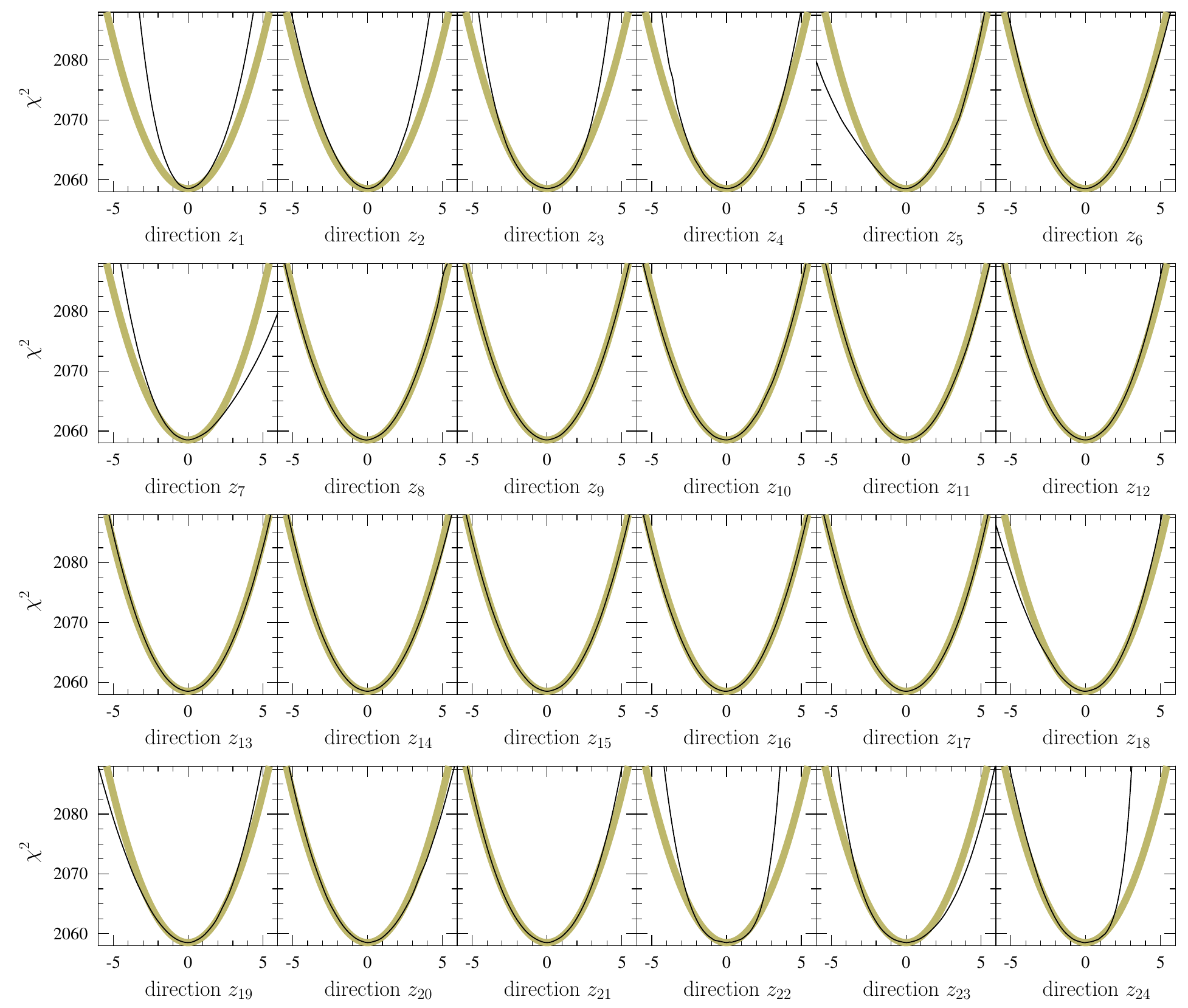}
\caption{The $\chi_{\rm global}^2$ as a function of eigenvector directions $z_i$ (black curves) compared to the ideal quadratic form $\chi^2_0 + z_i^2$ (thicker curves).}
\label{fig:QuadraticTest}
\end{figure*}

Our uncertainty analysis leans on the standard Hessian method \cite{Pumplin:2001ct}. The global $\chi^2$ is expanded about the fitted minimum as, 
\begin{align}
\chi^2_{\rm global}\{a_k\} & \approx \chi^2_0 + \sum_{ij} \delta a_i H_{ij} \delta a_j \label{eq:chi2orig} 
\end{align}
where $\delta a_j \equiv a_j-a_j^0$ are deviations from the best-fit values and $\chi^2_0$ is the fitted minimum $\chi^2$. The Hessian matrix has a complete set of positive-definite eigenvalues $\epsilon_k$ and orthonormal eigenvectors $v_j^{(k)}$, 
\begin{align}
H_{ij} v_j^{(k)} & =  \epsilon_k v_i^{(k)} \, , \\
\sum_i v_i^{(k)} v_i^{(\ell)} & =  \sum_i v_k^{(i)} v_\ell^{(i)} = \delta_{k\ell}.
\end{align}
These are used to introduce new coordinates, 
\begin{align}
 z_k     & \equiv \sum_j D_{kj}  \delta a_j, \label{eq:diag} \\
 D_{kj}  & \equiv  \sqrt{\epsilon_k} v_j^{(k)} \,. \label{eq:directions}
\end{align}
In the new basis, the global $\chi^2$ function simplifies to
\begin{equation}
 \chi^2(\vec z) \approx \chi^2_0  + \sum_i  z_i^2 \, . \label{eq:chi2diagonalized}
\end{equation}
In Fig.~\ref{fig:QuadraticTest} we plot the $\chi^2$ profiles along each eigenvector direction. In most of the cases the quadratic approximation seems to hold very well (in the plotted range) but in some cases its imperfections are also clearly visible.

Our evaluation of the Hessian matrix follows the iterative procedure discussed more detailedly in Sect.~4.1 of Ref.~\cite{Eskola:2016oht}. The best fit corresponds to the origin of the $z$ space, $z_i=0$, and the PDF error sets $S_i^{\pm}$ are defined as those PDFs that correspond to definite points in the $z$ space, 
\begin{align}
S_1^{\pm} & \equiv f^A \Big(\delta z_1^\pm, 0, 0, \ldots , 0\Big) \nonumber \\
S_2^{\pm} & \equiv f^A \Big(0,\delta z_2^\pm, 0, \ldots , 0\Big) \\
& \ \, \vdots \nonumber \\
S_N^{\pm} & \equiv f^A \Big(0, 0, \ldots 0, \, \delta z_N^\pm\Big) \nonumber \,.
\end{align}
Since the Hessian matrix is diagonal in the $z$ space, the error sets $S_k^\pm$ can be seen to define the uncertainties due to uncorrelated sources of errors. As a result the total uncertainty for a given PDF-dependent quantity $X(S)$ can be taken to be e.g. of the form, 
\begin{align}
\delta X & = \sqrt{\sum_k \left[\left(\frac{\partial X}{\partial z_k} \right) \delta_z^\pm \right]^2 } \label{eq:errorprop} \\
& \approx \frac{1}{2} \sqrt{\sum_k \Big[X(S_k^\pm)-X(S_0) \Big]^2} \nonumber \,.
\end{align}
Often, one defines the upward uncertainty $\big( \delta X \big)^+$ and the downward uncertainty $\big( \delta X \big)^-$ separately by the prescription \cite{Nadolsky:2001yg}, 
\begin{align}
& \big( \delta X \big)^\pm \label{eq:errorprop2} \\
& = \sqrt{\sum_k \bigg[ \substack{ \max \\ \min}  \left[X(S_k^+)-X(S_0), X(S_k^-)-X(S_0), 0\right]\bigg]^2} \nonumber 
\end{align}
i.e. the deviations above (below) the central value are added in quadrature. It is this latter definition of uncertainties that we adopt in the present paper. 

\begin{figure*}[htb!]
\centering
\includegraphics[width=1.0\linewidth]{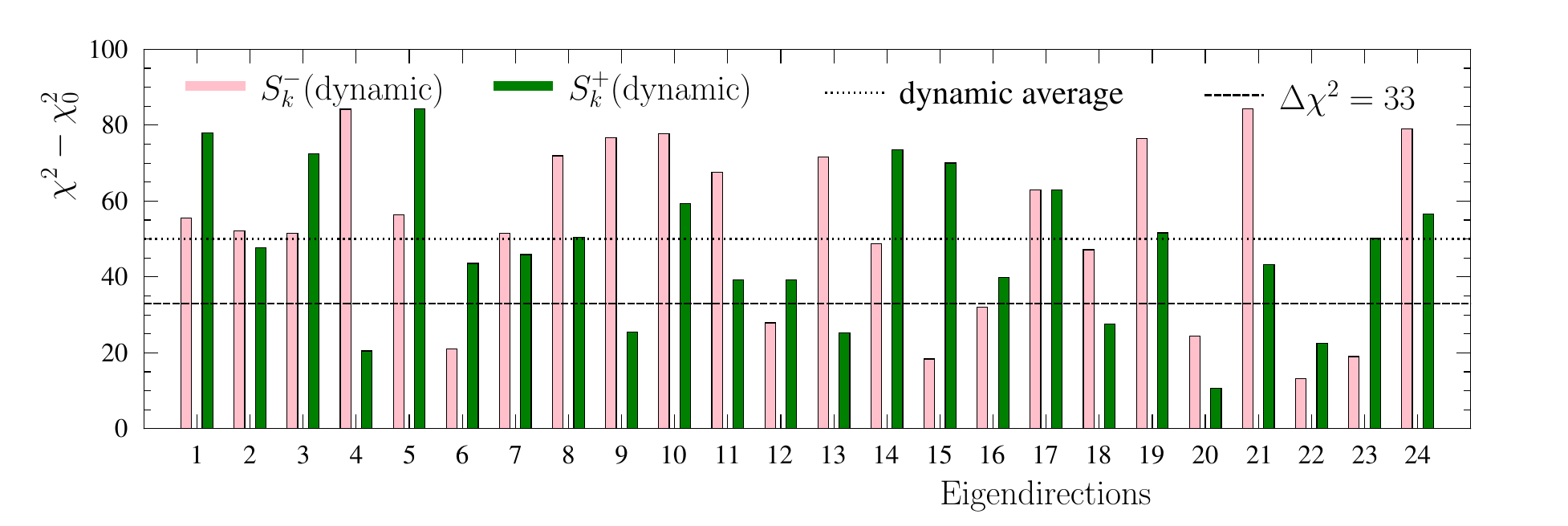}
\caption{The dynamically determined increases in the global $\chi^2$ corresponding to each eigenvector direction (see Ref.~\cite{Eskola:2016oht}, Sect.~4.1). The average of these, 50, and the EPPS21 tolerance $\Delta\chi^2 \approx 33$ are indicated as well.
}
\label{fig:dynamic}
\end{figure*} 

How to assign values for the parameters $\delta z_i^\pm$ is not unique. In the present analysis we use for transparency the prescription in which we assume the probability distribution for our best-fit parameters to be Gaussian. In other words, we assume that if we would have several instances of the entire EPPS21 data collection, the obtained parameters would form a Gaussian distribution centered around our current best-fit parameters. It follows that the probability to observe a $\chi^2$ less than $\chi^2_0 + \Delta \chi^2$ is given by
\begin{align}
P \left( \Delta\chi^2, N_{\rm param} \right) = \int_0^{\Delta\chi^2} dx \, p(x, N_{\rm param}) \,,
\end{align}
where $N_{\rm param}$ is the number of fit parameters and 
\begin{align}
p(x, N_{\rm param}) = \frac{\exp\left({-x/2}\right)}{2\Gamma(N_{\rm param}/2)} \left( \frac{x}{2} \right)^{N_{\rm param}/2-1} \,.
\end{align}
Requiring $P \left( \Delta\chi^2, N_{\rm param} \right) = 90\%$ results in $\Delta\chi^2 \approx 33$ for $N_{\rm param} = 24$. This defines the tolerance of EPPS21 and the values for the parameters $\delta z_i^\pm$ are then determined such that they increase the $\chi^2$ by this amount. The found values are listed in Table~\ref{Table:kaikkideltat}. We have also tested other ways to define the uncertainties: we have e.g. repeated the ``dynamical-tolerance analysis'' as done in EPPS16 (Ref.~\cite{Eskola:2016oht}, Sect.~4.1). The resulting individual, dynamically set values of $\chi^2(S_k^\pm) - \chi_0^2$ are shown in Fig.~(\ref{fig:dynamic}). In the big picture the resulting uncertainties are very similar as here: the resulting average dynamical tolerance is around 50, which means that the uncertainty bands differ roughly by a factor of $\sqrt{50/33} \approx 1.2$. In other words, to obtain the uncertainties defined in the similar manner as in EPPS16, the errorbands would need to be inflated approximately by this factor. However, e.g. the implicit parametrization bias is likely to be more important than this difference. 

\begin{table}[!htb]
\caption[]{\small The parameters $\delta z_i^\pm$ that define the EPPS21 error sets.}
\label{Table:kaikkideltat}
\begin{center}
\begin{tabular}{ll|ll}
$\delta z_i^-$       &  Value & $\delta z_i^+$ &  Value \\
\hline
$\delta z_1^-$      &  -3.40  & $\delta z_1^+$      &  4.59 \\
$\delta z_2^-$      &  -5.40  & $\delta z_2^+$      &  4.36   \\
$\delta z_3^-$      &  -4.76  & $\delta z_3^+$      &  4.37  \\
$\delta z_4^-$      &  -4.52  & $\delta z_4^+$      &  5.19 \\
$\delta z_5^-$      &  -7.41  & $\delta z_5^+$      &  5.51 \\
$\delta z_6^-$      &  -5.49  & $\delta z_6^+$      &  6.03  \\
$\delta z_7^-$      &  -4.73  & $\delta z_7^+$      &  7.48  \\
$\delta z_8^-$      &  -5.85  & $\delta z_8^+$      &  5.68  \\
$\delta z_9^-$      &  -5.73  & $\delta z_9^+$      &  5.81 \\
$\delta z_{10}^-$   &  -5.91  & $\delta z_{10}^+$   &  5.60 \\
$\delta z_{11}^-$   &  -5.70  & $\delta z_{11}^+$   &  5.93   \\
$\delta z_{12}^-$   &  -5.84  & $\delta z_{12}^+$   &  5.46  \\
$\delta z_{13}^-$   &  -5.58  & $\delta z_{13}^+$   &  5.85  \\
$\delta z_{14}^-$   &  -5.63  & $\delta z_{14}^+$   &  5.96  \\
$\delta z_{15}^-$   &  -5.71  & $\delta z_{15}^+$   &  5.84   \\
$\delta z_{16}^-$   &  -5.88  & $\delta z_{16}^+$   &  5.75     \\
$\delta z_{17}^-$   &  -5.83  & $\delta z_{17}^+$   &  5.77  \\
$\delta z_{18}^-$   &  -6.59  & $\delta z_{18}^+$   &  5.37  \\
$\delta z_{19}^-$   &  -6.37  & $\delta z_{19}^+$   &  5.19   \\
$\delta z_{20}^-$   &  -5.55  & $\delta z_{20}^+$   &  6.16  \\
$\delta z_{21}^-$   &  -5.60  & $\delta z_{21}^+$   &  5.26 \\
$\delta z_{22}^-$   &  -4.34  & $\delta z_{22}^+$   &  3.68 \\
$\delta z_{23}^-$   &  -4.72  & $\delta z_{23}^+$   &  6.25 \\
$\delta z_{24}^-$   &  -5.41  & $\delta z_{24}^+$   &  3.17
\end{tabular}
\end{center}
\end{table}

\subsection{Baseline proton uncertainties}
\label{Baselineprotonuncertainties}

Even though most of the observables in our analysis are ratios whose dependence on the baseline proton PDF is only weak, the cancellation is not perfect and some dependence on the chosen baseline PDF always persists. Also, the neutrino DIS cross sections that we include are normalized only to the integrated cross sections and one would expect a pronounced sensitivity on the baseline PDFs. In the present analysis, we will quantify the baseline dependence by using the CT18ANLO Hessian error sets. We do this by simply propagating the CT18ANLO uncertainty into nuclear quantities by using Eq.~(\ref{eq:errorprop2}). The quantity $X$ there could be, for example, a nuclear modification factor $R_i^{{\rm p}/A}$ or an absolute nuclear PDF itself, $f_i^{A}$, which both depend, in some complicated way through the $\chi^2$ minimization, on the free proton PDFs. Thus, to chart the sensitivity of nuclear PDFs on the baseline proton PDFs, we repeat the $\chi^2$ minimization by taking each free proton PDF error set $S^\pm_i$ separately as our baseline in Eq.~(\ref{eq:defnPDF}). By looping over all CT18ANLO error sets we can consequently propagate its uncertainties on any nuclear quantity that appears in our analysis. The full uncertainty is then formed by adding in quadrature the ``nuclear errors'' -- uncertainties that stem from the parametrization of nuclear modifications -- and the ``proton errors.'' The procedure here is on par with the method used in the nNNPDF fits \cite{AbdulKhalek:2020yuc}. However, here we will be able to also separate the uncertainties that stem from the nuclear parameters and those that originate from the free proton PDFs. In practice, the EPPS21 PDFs thus feature $24({\rm nuclear})+29({\rm proton})=53$ error directions (i.e. 106 error sets) and the full uncertainty is obtained by extending the sum in Eq.~(\ref{eq:errorprop2}) over all of these. In other words, the total uncertainty is built from two components, 
\begin{align}
& \big( \delta X \big)^\pm \label{eq:errorprop3} \\
& = \Bigg\{
\underbrace{
\sum_{k=1}^{24} \bigg[ \substack{ \max \\ \min}  \left[X(S_k^+)-X(S_0), X(S_k^-)-X(S_0), 0\right]\bigg]^2}_{\rm nuclear \ error} \nonumber \\
& +
\underbrace{
\sum_{k=25}^{53} \bigg[ \substack{ \max \\ \min}  \left[X(S_k^+)-X(S_0), X(S_k^-)-X(S_0), 0\right]\bigg]^2}_{\rm proton \ error} \Bigg\}^{\frac{1}{2}} \,.
\nonumber 
\end{align}
For example, if the quantity $X$ is a cross section $\sigma$ in pPb collisions, the required cross sections are obtained schematically as 
\begin{align}
S_0 & : \sigma(S^\pm_0) = f^{\rm p}_0  \otimes \hat \sigma \otimes f^{\rm Pb}_0 \nonumber \\[4pt]
S^\pm_{i=1,24} & : \sigma(S^\pm_{i=1,24}) = f^{\rm p}_0      \otimes \hat \sigma \otimes f^{\rm Pb}_{i,\pm} \\[4pt]
S^\pm_{i=25,53} & : \sigma(S^\pm_{i=25,53}) = f^{\rm p}_{i-24, \pm} \otimes \hat \sigma \otimes f^{\rm Pb}_{i, \pm} \nonumber \,,
\end{align}
where $f^{\rm p}_0$ and $f^{\rm Pb}_0$ are the central free-proton and lead PDFs, and $f^{\rm p}_{i, \pm}$ and $f^{\rm Pb}_{i, \pm}$ are the corresponding error sets.

\subsection{Look-up tables}
\label{Lookuptables}

To speed up the numerical analysis, the calculations of the LHC, RHIC, and pion-nucleus Drell-Yan processes are carried out by precomputing look-up tables. Following here the method used in the EPPS16 analysis (Sect.~3.3 of Ref.~\cite{Eskola:2016oht}), we precompute partial cross sections,
\begin{align}
\sigma_{i,k} = \sum_{j,n} f_j^{\rm p} \otimes \hat\sigma_{j,n} \otimes f^{A}_{n}[i,k] \,,
\end{align}
where $\hat\sigma_{j,n}$ are the partonic coefficient functions and
\begin{align}
f^{A}[i,k] & = {f^{A}} \bigg(R^{{\rm p}/A}_j=
\left\{
\begin{array}{cl}
1, & {\rm if} \ j=i \\
0, & {\rm otherwise}
\end{array}
\right.
,x, Q^2 \bigg) \nonumber \\
& \times \theta\big(x_{k-1} < x < x_k\big) \,,
\end{align}
where the points $x_0, x_1,\ldots x_N=1$ define a suitable grid in the $x$ variable. The true cross sections are then obtained as simple sums,
\begin{align}
\sigma = \sum_{i,k} \sigma_{i,k} R^{{\rm p}/A}_i \bigg( \frac{x_{k-1}+x_{k}}{2} , Q^2\bigg) \,. 
\end{align}
The look-up tables are computed by using the central CT18A proton PDFs. For a complete consistency we should recompute the look-up tables also with each CT18A error set to be used when fitting the nuclear modifications with that particular CT18A error set. However, the most important observables we use in conjunction with the look-up tables turn out not very sensitive to the choice of the baseline proton PDF and we thus always use the tables computed with the central CT18A proton PDFs. The baseline dependence cancels out particularly well in the case of dijet and D-meson observables which are the most constraining data sets included via the look-up-table method. For W$^\pm$ bosons the proton-PDF cancellation is less exact in the used ratios, but we have verified that the residual baseline uncertainties are small enough compared to the experimental uncertainties that they do not bias the nuclear-modification fitting~\cite{Eskola:2022rlm}, justifying the use of fixed look-up tables also in this case.


\section{The EPPS21 PDFs}
\label{TheEPPS21PDFs}

\begin{figure*}[htb!]
\centering
\includegraphics[width=0.210\linewidth]{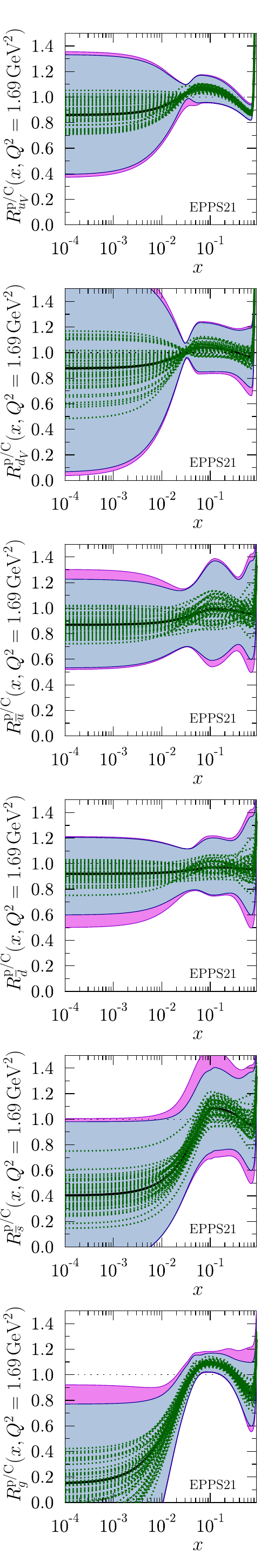}
\includegraphics[width=0.210\linewidth]{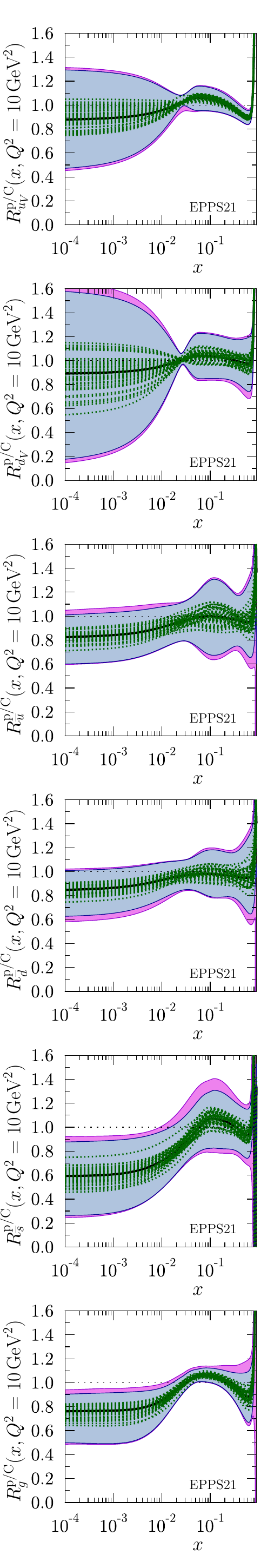}
\includegraphics[width=0.210\linewidth]{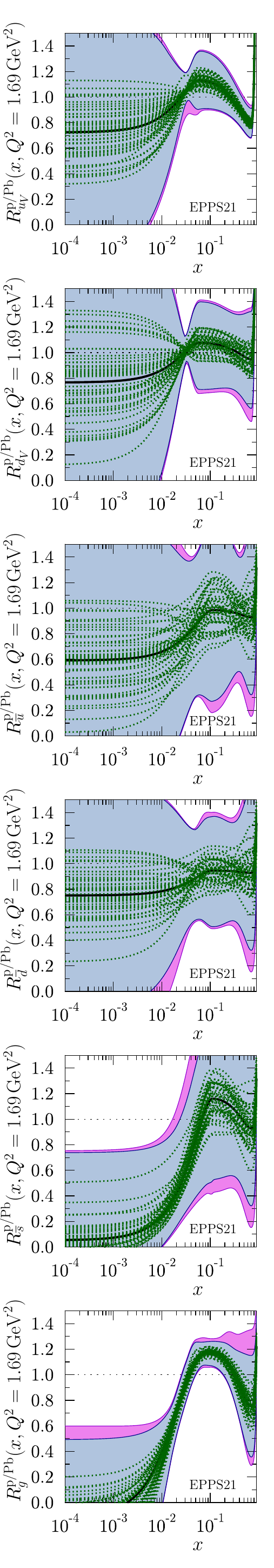}
\includegraphics[width=0.210\linewidth]{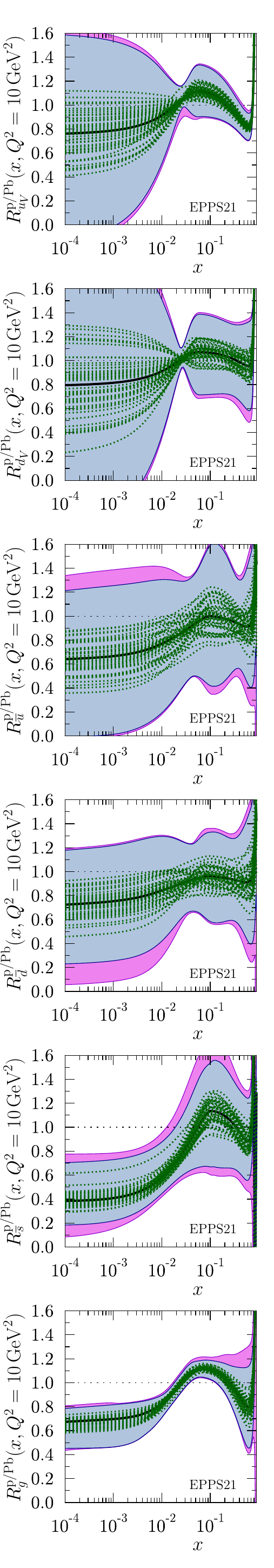}
\caption{
The EPPS21 nuclear modifications of bound protons in carbon (two leftmost columns) in lead (two rightmost columns) at the initial scale $Q^2=1.69\,{\rm GeV}^2$ and at $Q^2=10\,{\rm GeV}^2$. The central results are shown by thick black curves, and the nuclear error sets by green dotted curves. The blue bands correspond to the nuclear uncertainties and the purple ones to the full uncertainty (nuclear and baseline errors added in quadrature).}
\label{fig:allsets_lowQ2}
\end{figure*} 

\begin{figure*}[htb!]
\centering
\includegraphics[width=0.210\linewidth]{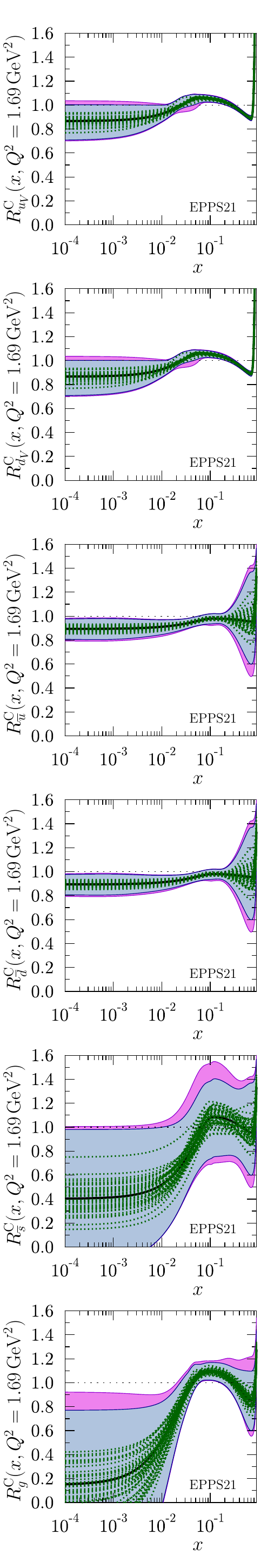}
\includegraphics[width=0.210\linewidth]{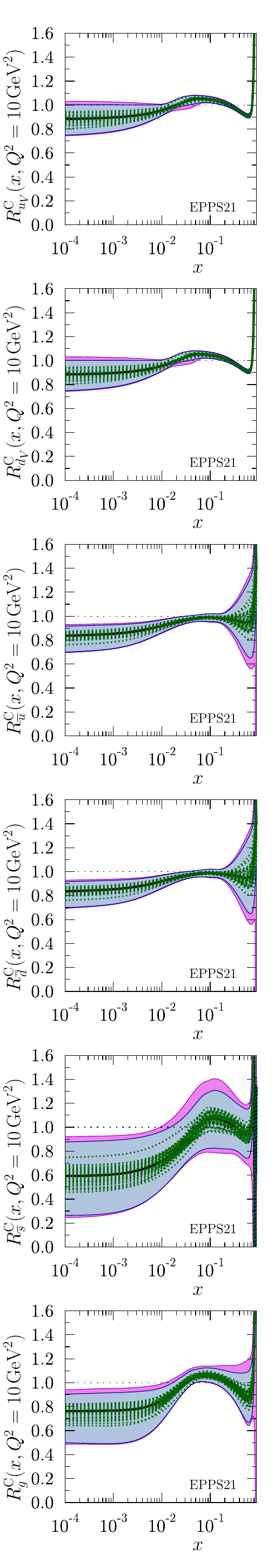}
\includegraphics[width=0.210\linewidth]{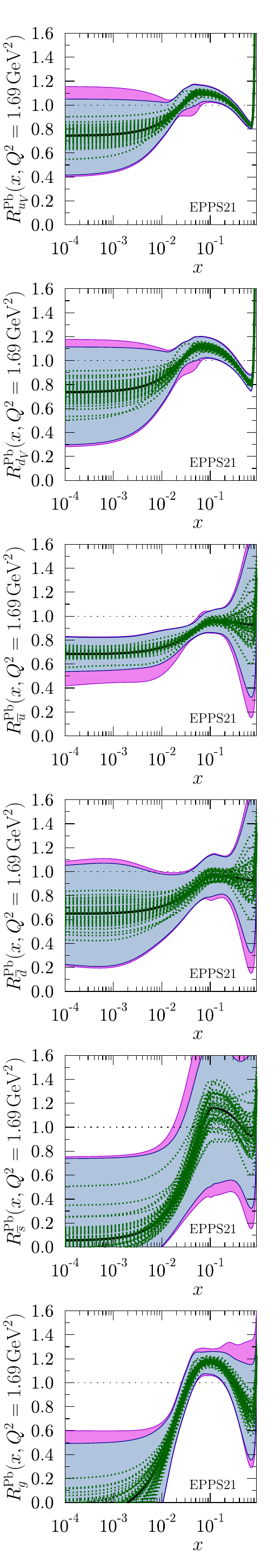}
\includegraphics[width=0.210\linewidth]{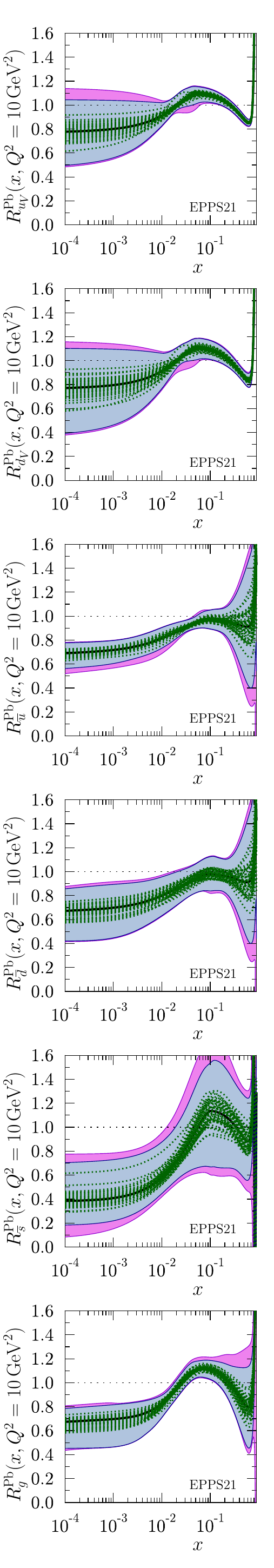}
\caption{
The EPPS21 nuclear modifications of average nucleons in carbon (two leftmost columns) in lead (two rightmost columns) at the initial scale $Q^2=1.69\,{\rm GeV}^2$ and at $Q^2=10\,{\rm GeV}^2$. The central results are shown by thick black curves, and the nuclear error sets by green dotted curves. The blue bands correspond to the nuclear uncertainties and the purple ones to the full uncertainty (nuclear and baseline errors added in quadrature).}
\label{fig:allfullsets_lowQ2}
\end{figure*} 

\begin{figure*}[htb!]
\centering
\includegraphics[width=0.329\linewidth]{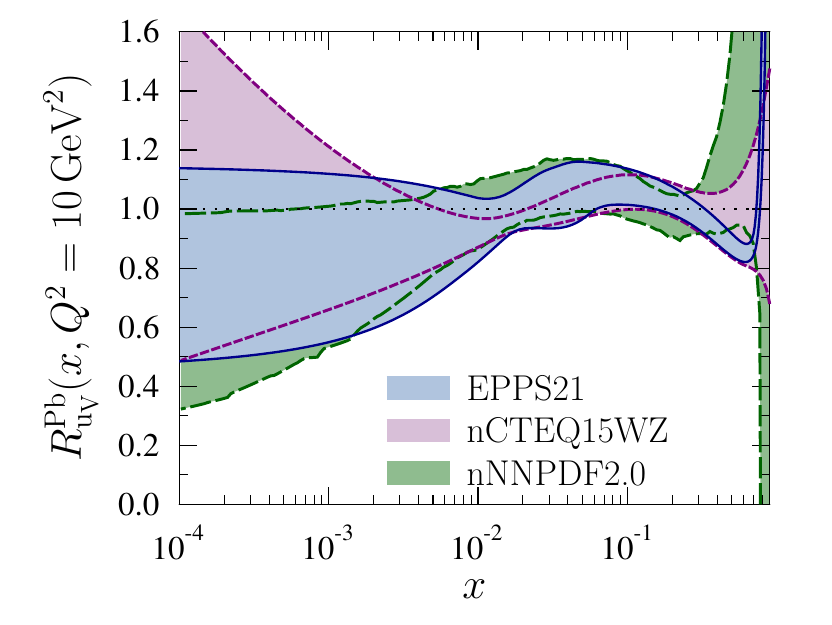}
\includegraphics[width=0.329\linewidth]{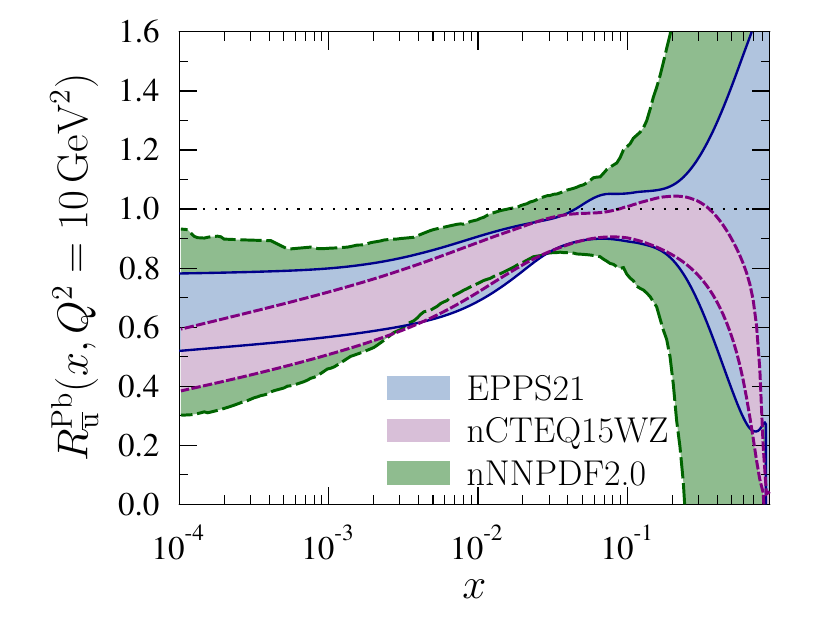}
\includegraphics[width=0.329\linewidth]{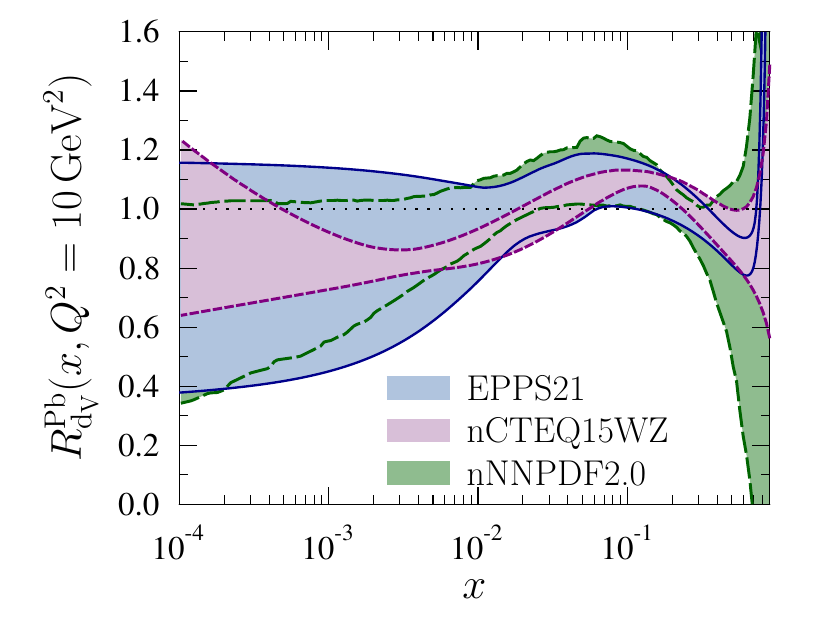}
\includegraphics[width=0.329\linewidth]{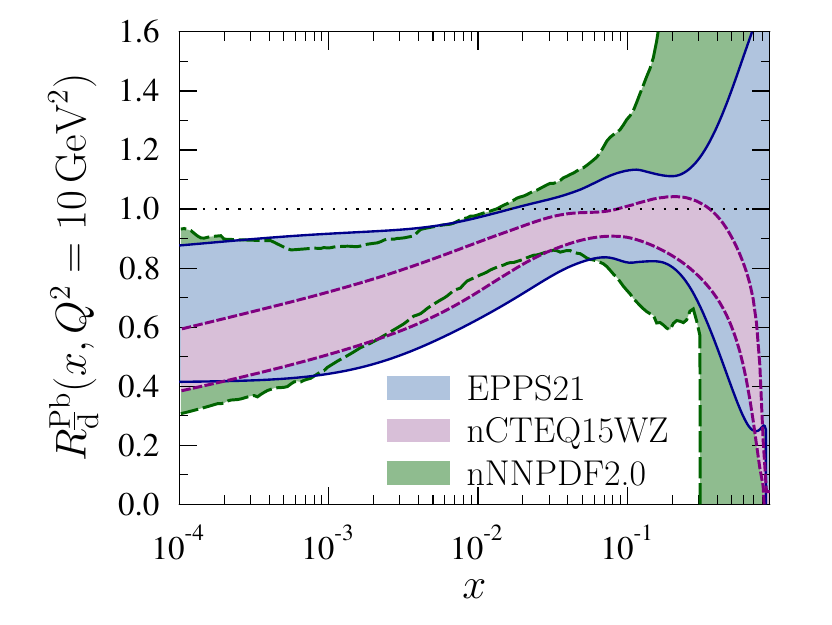}
\includegraphics[width=0.329\linewidth]{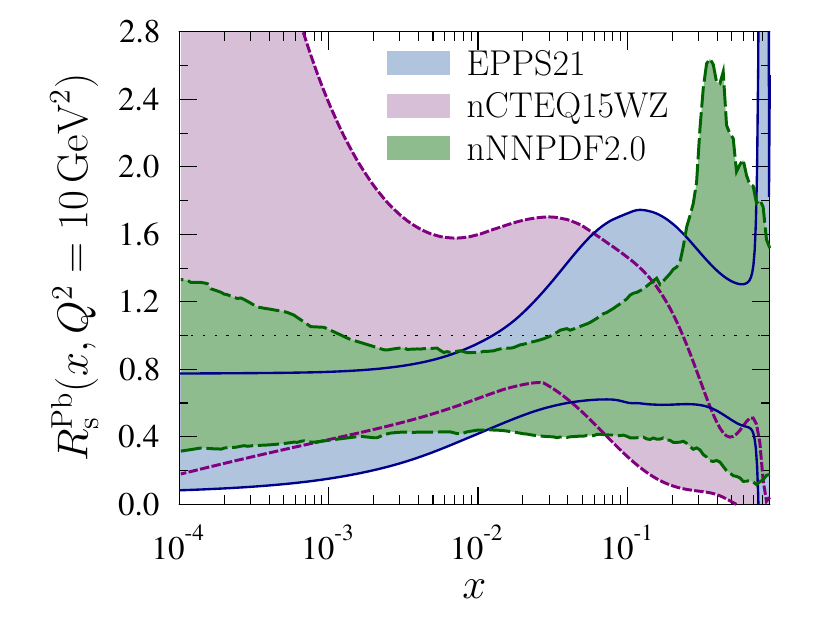}
\includegraphics[width=0.329\linewidth]{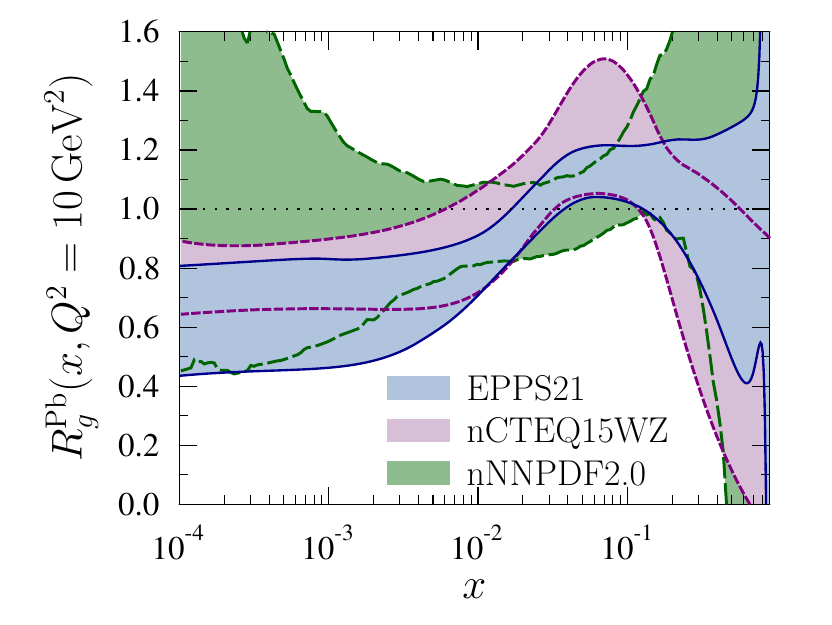}
\caption{
Comparison between the EPPS21 (blue), nCTEQ15WZ (purple) \cite{Kusina:2016fxy}, and nNNPDF2.0 (green) \cite{AbdulKhalek:2020yuc} average-nucleon nuclear modifications at $Q^2=10\,{\rm GeV}^2$. The EPPS21 and nNNPDF uncertainties include the free-proton uncertainties but the nCTEQ15WZ error bands only include the nuclear uncertainty. 
}
\label{fig:allfullsets_others}
\end{figure*} 

\begin{figure*}[htb!]
\centering
\includegraphics[width=0.329\linewidth]{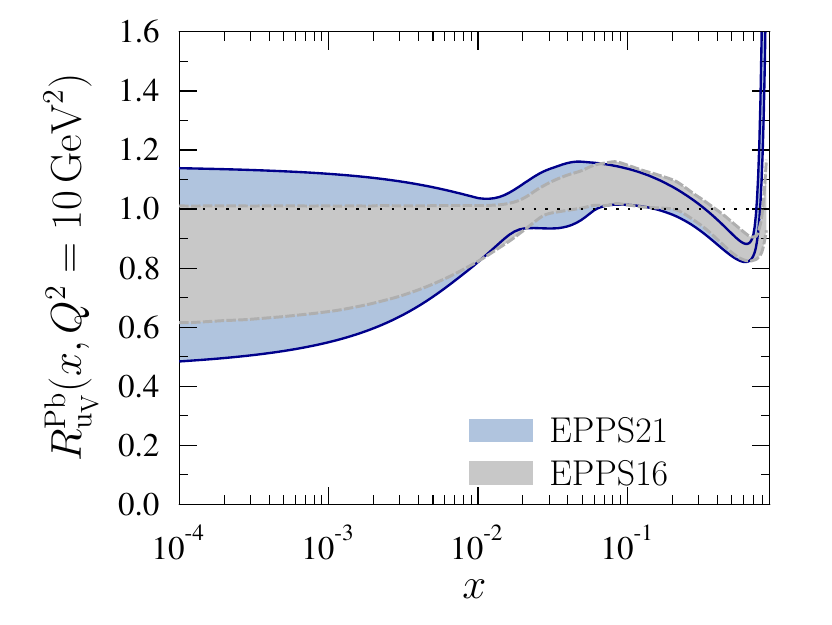}
\includegraphics[width=0.329\linewidth]{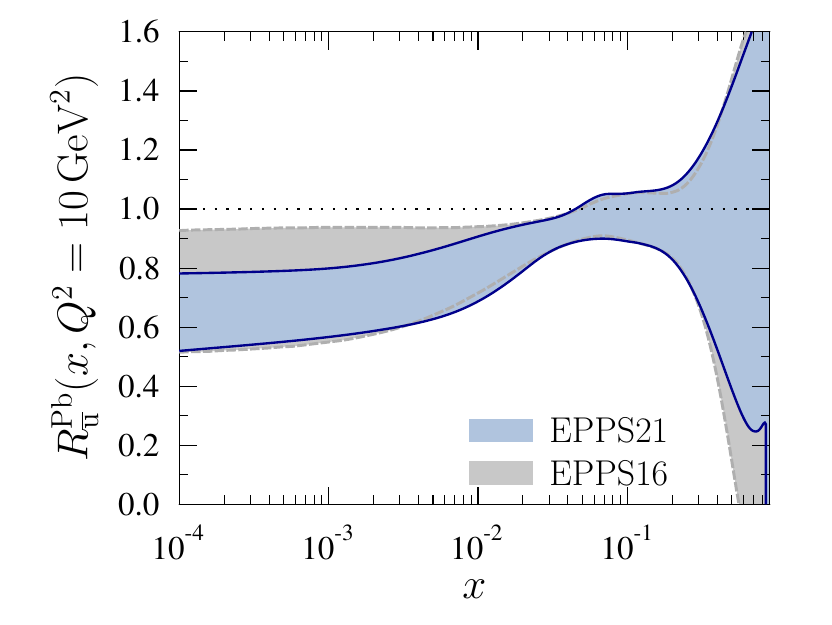}
\includegraphics[width=0.329\linewidth]{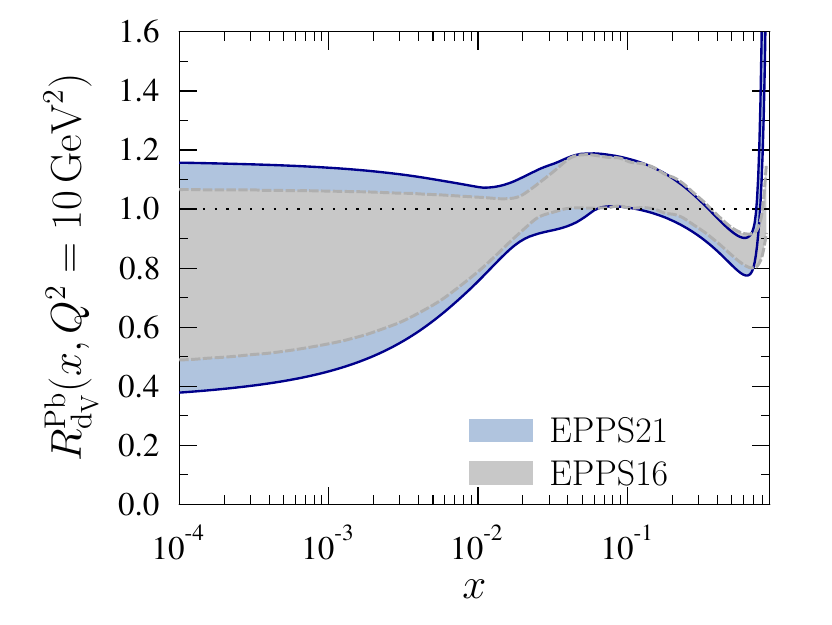}
\includegraphics[width=0.329\linewidth]{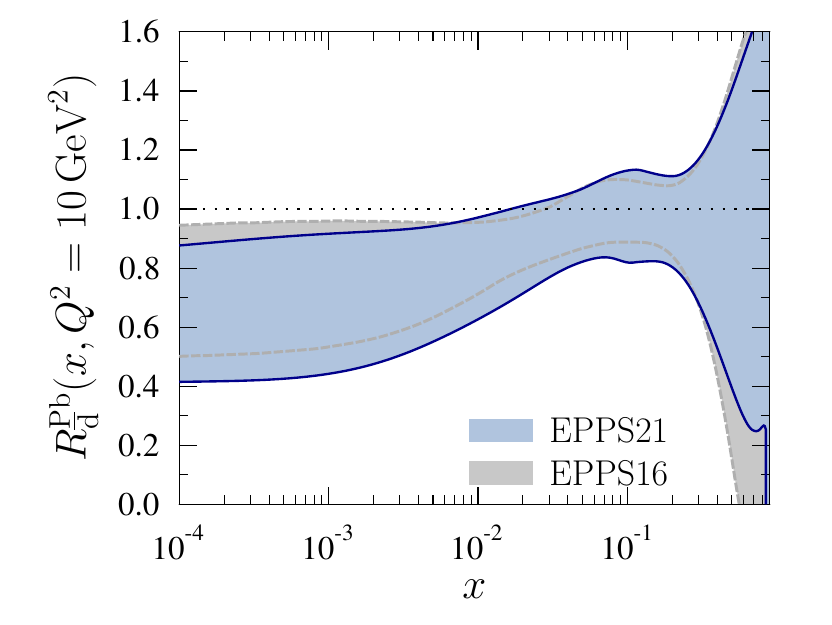}
\includegraphics[width=0.329\linewidth]{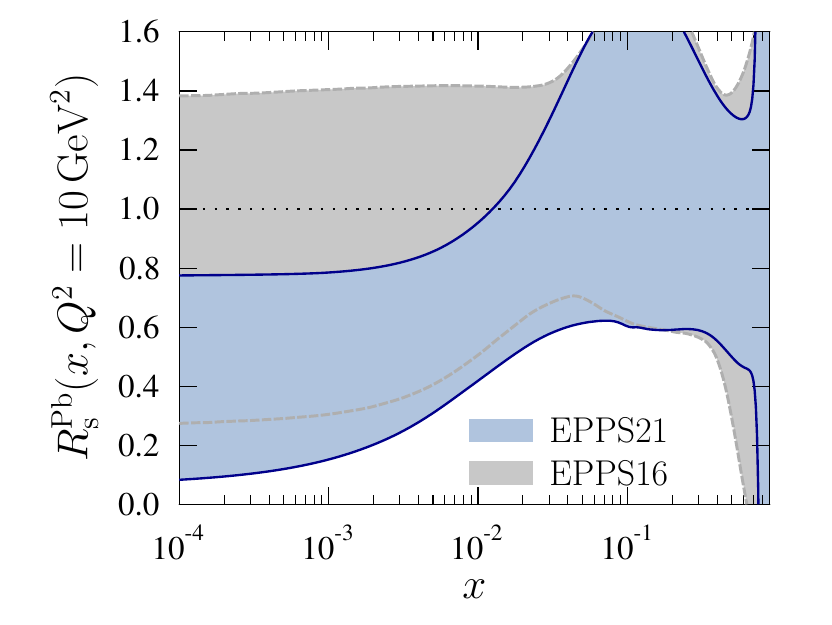}
\includegraphics[width=0.329\linewidth]{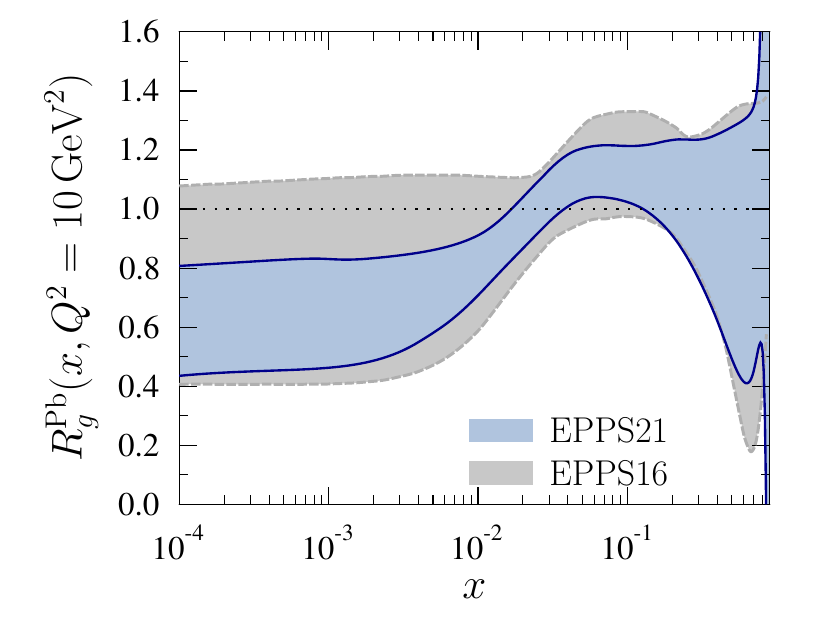}
\caption{
Comparison between the EPPS21 (blue) and the EPPS16 (gray) \cite{Eskola:2016oht} average-nucleon nuclear modifications at $Q^2=10\,{\rm GeV}^2$. The EPPS21 uncertainties include the free-proton uncertainties but the EPPS16 error bands only include the nuclear uncertainty. 
}
\label{fig:allfullsets_EPPS16}
\end{figure*} 

\begin{figure*}[htb!]
\centering
\includegraphics[width=\linewidth]{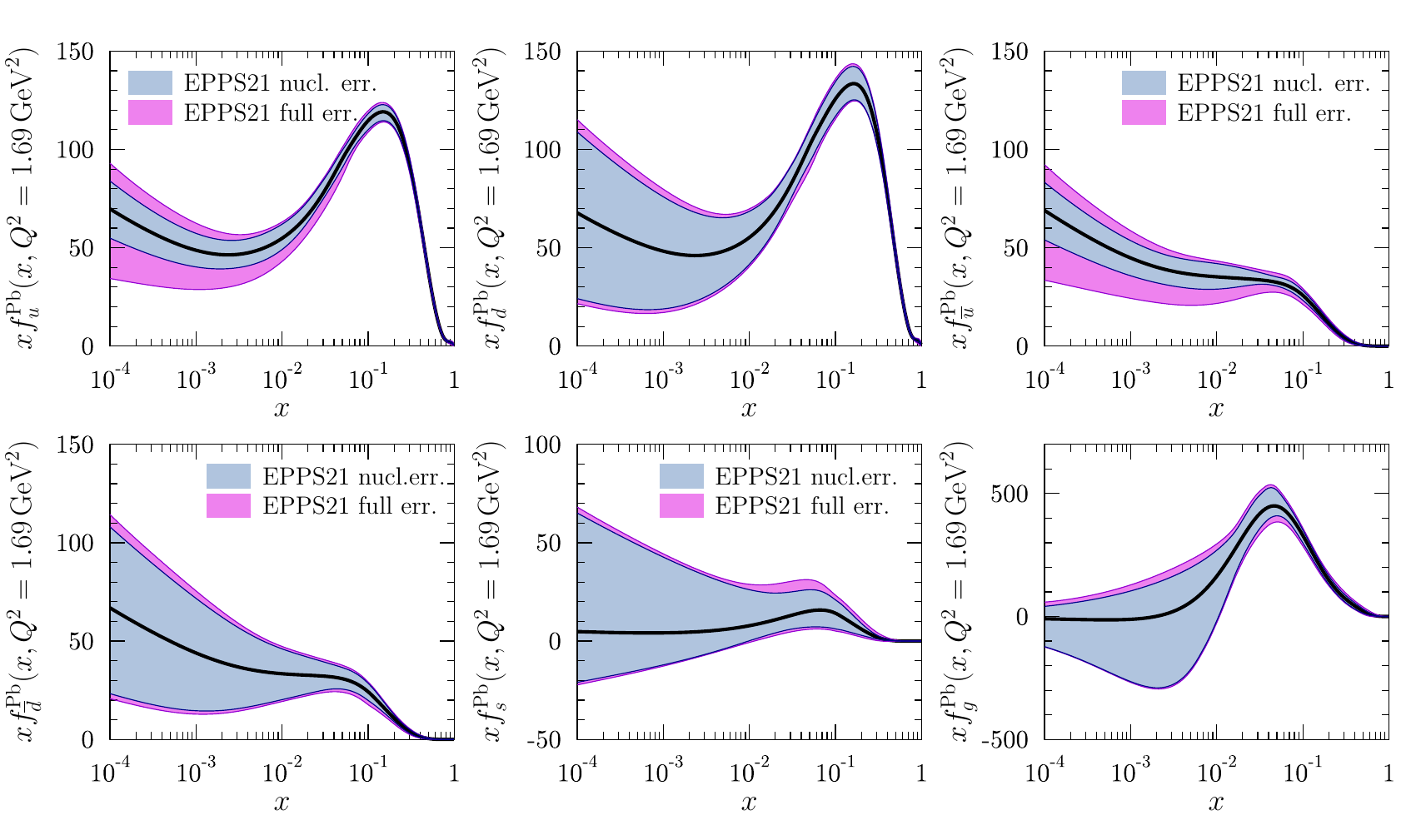}
\includegraphics[width=\linewidth]{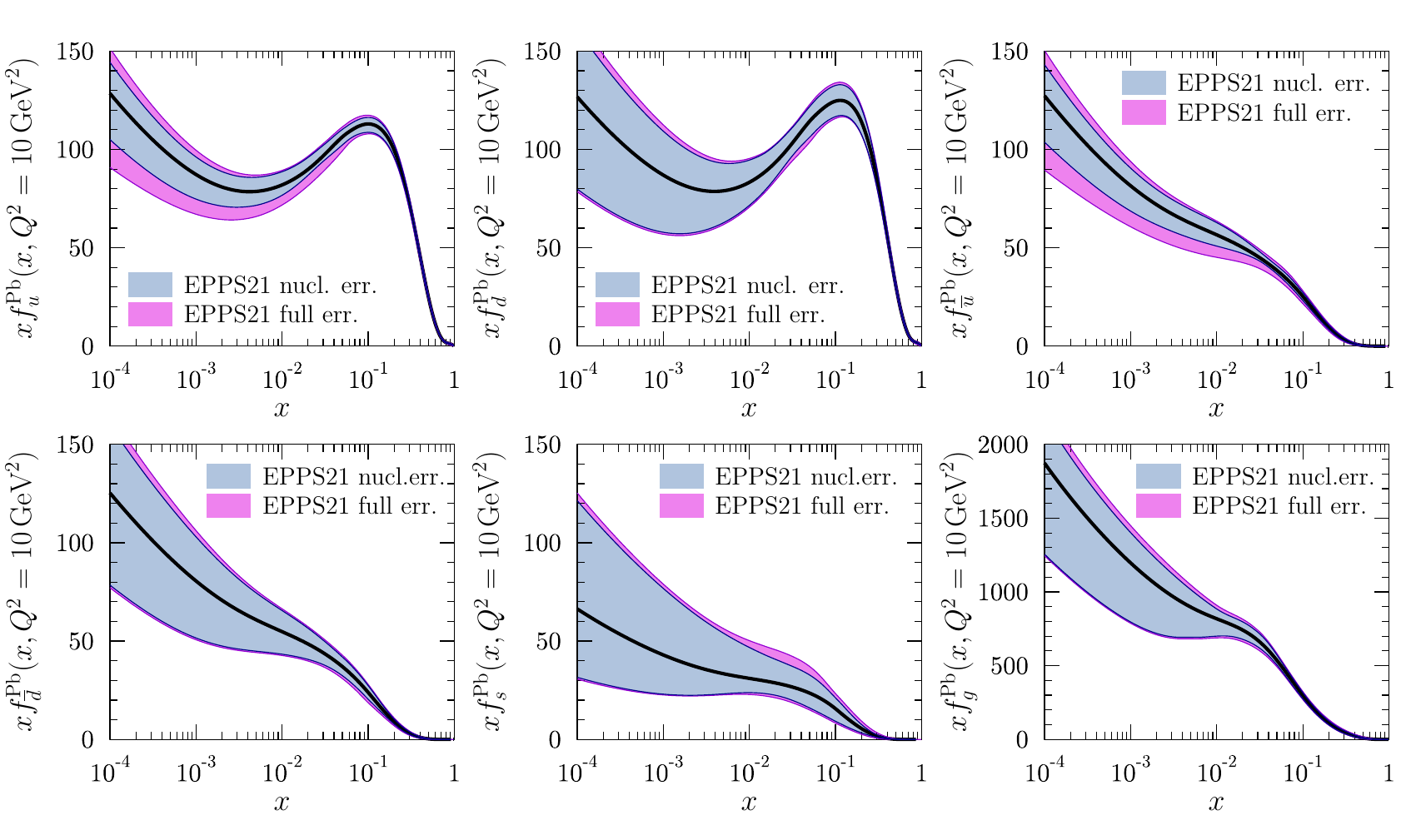}
\caption{
The 90\% confidence-level EPPS21 PDFs in lead nucleus at the parametrization scale $Q^2=1.69\,{\rm GeV}^2$ (upper panels) and at $Q^2=10\,{\rm GeV}^2$ (lower panels). The central results are shown by thick black curves, the blue bands correspond to the nuclear uncertainties and the purple ones to the full uncertainty (nuclear and baseline errors added in quadrature).}
\label{fig:absolutePDFs}
\end{figure*} 

\begin{figure*}[htb!]
\centering
\includegraphics[width=\linewidth]{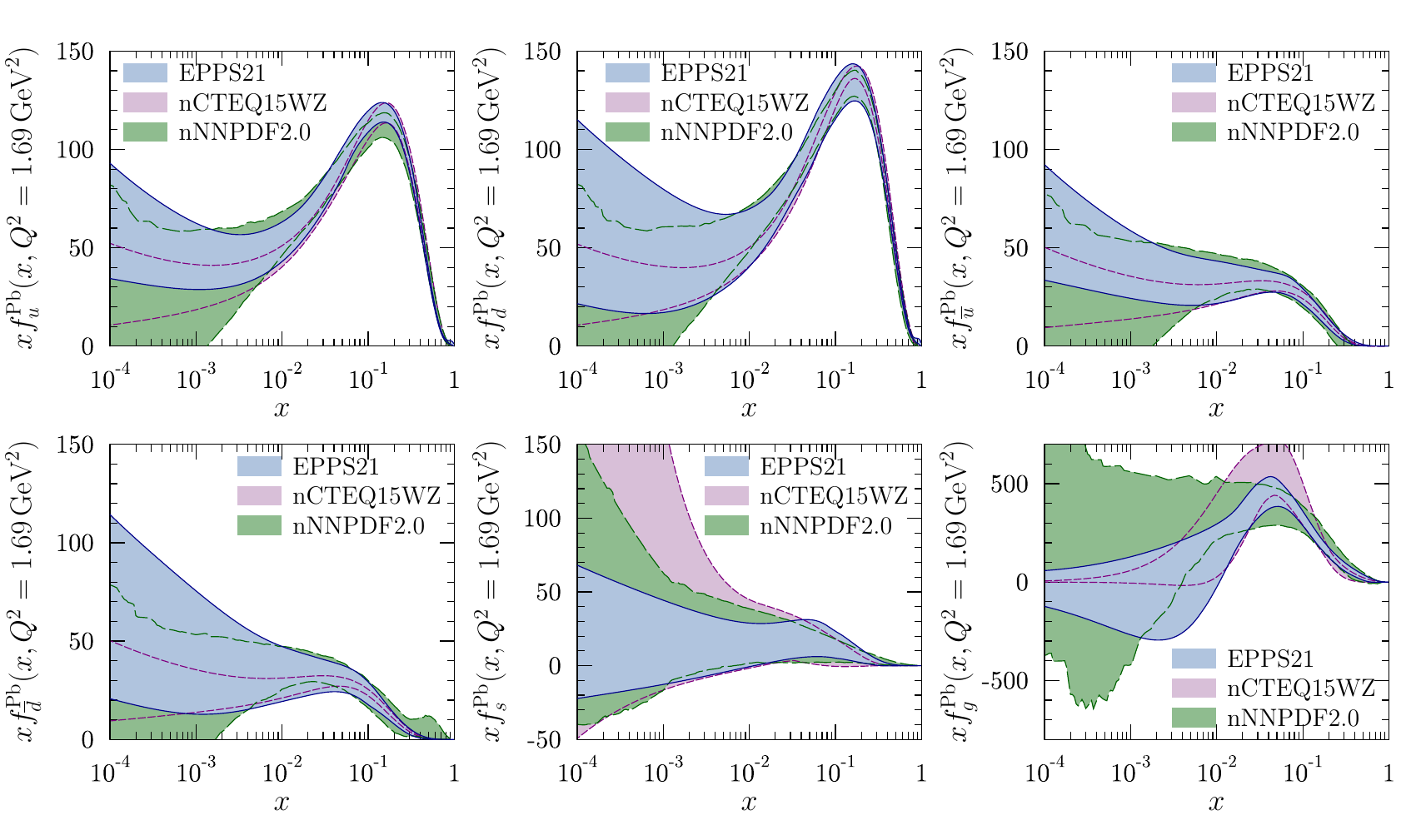}
\vspace{0.5cm}
\includegraphics[width=\linewidth]{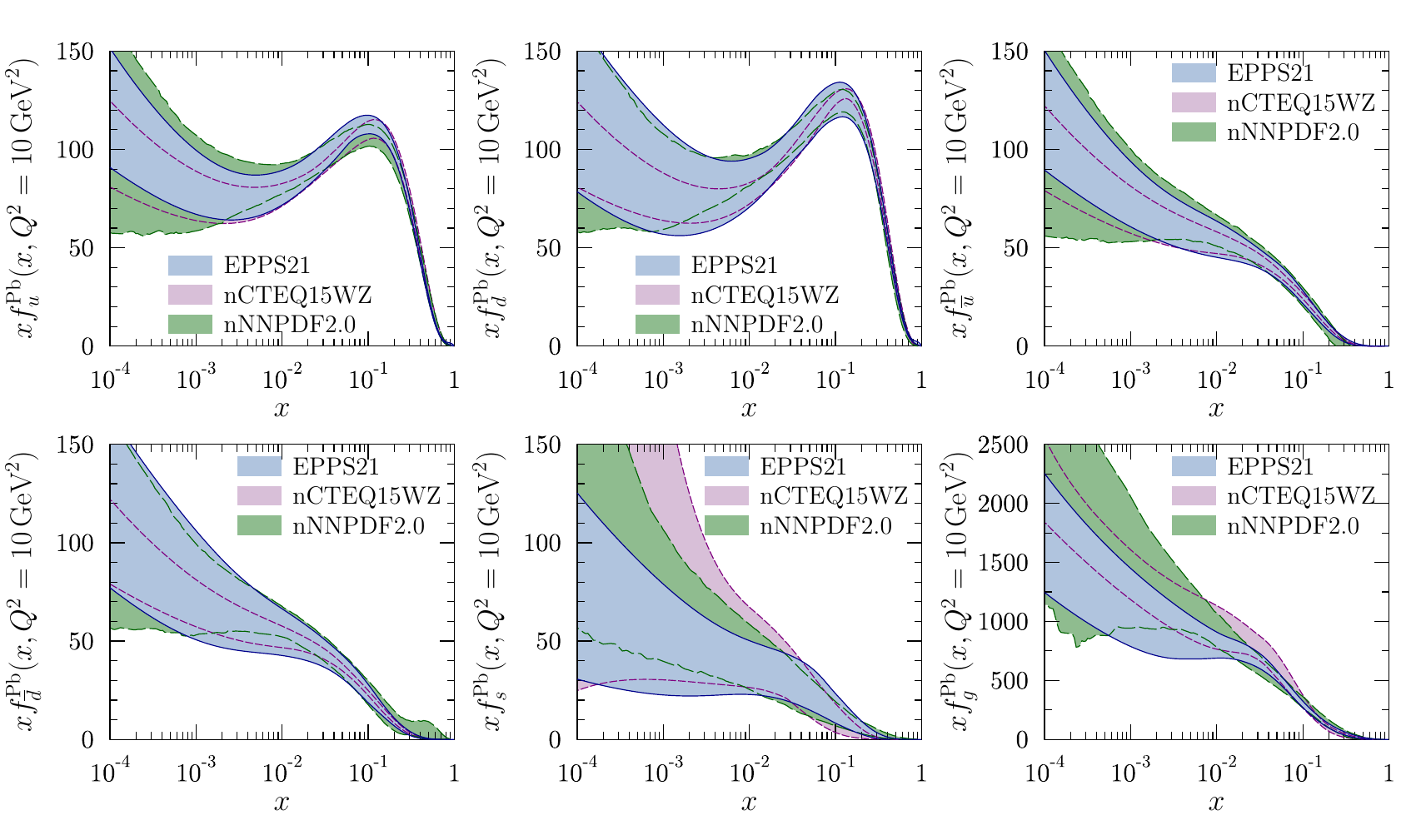}
\caption{
The 90\% confidence-level EPPS21 (blue), nCTEQ15WZ (purple) \cite{Kusina:2016fxy}, and nNNPDF2.0 (green) \cite{AbdulKhalek:2020yuc} PDFs in lead at $Q^2=1.69\,{\rm GeV}^2$ (upper panels) and at $Q^2=10\,{\rm GeV}^2$ (lower panels).}
\label{fig:absolutePDFsothers}
\end{figure*}

The EPPS21 bound-proton nuclear modifications are shown in Fig.~\ref{fig:allsets_lowQ2} for carbon and lead at the parametrization scale $Q^2=1.69 \, {\rm GeV}^2$ as well as at $Q^2=10 \, {\rm GeV}^2$. Here, as in most of the figures to follow, the blue bands denote the nuclear errors and in purple bands we indicate the full uncertainty in which the nuclear errors are added in quadrature with the uncertainties originating from the free proton PDFs. From the practical point of view, the bound-proton nuclear modifications $R_i^{\rm p/A}$ are not, however, the most relevant quantities as the cross sections only see the full PDFs involving appropriate linear combinations of protons and neutrons. In Fig.~\ref{fig:allfullsets_lowQ2} we therefore present the average-nucleon nuclear modifications $R_i^{A}$ from Eq.~(\ref{eq:FullR}) for carbon and lead at the parametrization scale $Q^2=1.69 \, {\rm GeV}^2$ as well as at $Q^2=10 \, {\rm GeV}^2$. Comparing Fig.~\ref{fig:allsets_lowQ2} and Fig.~\ref{fig:allfullsets_lowQ2} one clearly observes how the uncertainties in the nuclear modifications of average up- and down-quark PDFs are clearly smaller in comparison to the uncertainties in bound-proton nuclear modifications. This is to be expected as e.g. $R_{u_{\rm V}}^{\rm p/A}$ and $R_{d_{\rm V}}^{\rm p/A}$ are strongly anticorrelated as was demonstrated already in the context of EPPS16 analysis (Ref.~\cite{Eskola:2016oht}, Fig.~10). Since the average-nucleon modifications $R_{u_{\rm V}}^{\rm A}$ and $R_{d_{\rm V}}^{\rm A}$ are both linear combinations of $R_{u_{\rm V}}^{\rm p/A}$ and $R_{d_{\rm V}}^{\rm p/A}$, the uncertainties tend to diminish. Similar reasoning applies for the sea-quark nuclear modifications. From Fig.~\ref{fig:allfullsets_lowQ2} we can see that at small-$x$ the average up-sea modification for lead $R_{\overline u}^{\rm Pb}$ seems to be clearly better constrained than the average down-sea modification $R_{\overline d}^{\rm Pb}$. This is because of the factor of 4 difference between the electric charges of up and down quarks which makes the structure-function ratios four times more sensitive to $R_{\overline u}^{\rm Pb}$ than to $R_{\overline d}^{\rm Pb}$. For an isoscalar nucleus like carbon there is no such difference.

Towards smaller values of $x$ the DGLAP evolution efficiently reduces the uncertainties in particular for gluons, but also for the sea quarks. This is actually one of the reasons we do not try to build in too much additional flexibility for the parametrization at small $x$ -- such variations would anyway be wiped out very quickly towards higher values of $Q^2$ and thus be rather irrelevant for most of the applications. As we can see from Fig.~\ref{fig:allfullsets_lowQ2}, the strange-quark nuclear modifications are the ones that retain the largest overall uncertainties. They are indeed tricky to constrain: in principle the electroweak boson production at the LHC carries a significant contribution from the strange quarks \cite{Kusina:2020lyz}. However, these processes are sensitive to the PDFs at high interaction scales $Q^2 \sim 10000 \, {\rm GeV}^2$ where the behaviour of the strange-quark PDF depends very strongly on the initial gluon distribution. As a result, even the rather precise CMS $8 \, {\rm TeV}$ W$^\pm$ data \cite{Sirunyan:2019dox} are not able to restrict the strange-quark PDFs tightly. Unlike in our earlier analyses the central gluon nuclear modification does go negative for Pb and other large nuclei at the parametrization scale $Q^2=1.69 \, {\rm GeV}^2$. However, the negativity disappears extremely fast -- already at $Q^2=1.80 \, {\rm GeV}^2$ the central gluon is entirely positive. As a result, the negativity should not be a problem in practical applications. It can be expected that the negativity of the gluon PDF will be even a bigger issue at next-to-NLO fits as the evolution at small $x$ is faster than at NLO. It will eventually be interesting to see whether e.g. including the small $x$ resummation \cite{Bonvini:2016wki} in the DGLAP evolution kernels will reduce the tendency of the gluon to go negative similarly as in the case of free-proton fits \cite{Ball:2017otu}. Likewise, including non-linear, $1/Q^2$-type corrections in the DGLAP evolution is expected to slow down the evolution at small $x$ and possibly reduce the tendency of the gluon PDF to go negative \cite{Eskola:2002yc}.

Comparisons of the EPPS21 average-nucleon nuclear modifications -- i.e. those given by Eq.~(\ref{eq:FullR}) -- to the nCTEQ15WZ \cite{Kusina:2016fxy} and nNNPDF2.0 \cite{AbdulKhalek:2020yuc} nuclear PDFs are shown in Fig.~\ref{fig:allfullsets_others} at $Q^2=10\,{\rm GeV}^2$. The EPPS21 uncertainties correspond to the full uncertainty with the nuclear and free-proton uncertainties added in quadrature. Also the nNNPDF2.0 uncertainty includes the uncertainty that comes from the free proton. In the case of nCTEQ15WZ only the nuclear uncertainties are available. Within the plotted 90\% confidence-level errors all three are observed to agree with each other. However, the widths of the uncertainty bands differ quite significantly in places and they do so for varying reasons. For example, since there was no flavour separation between ${\overline u}$ and ${\overline d}$ in nCTEQ15WZ, the corresponding uncertainty for $R^{\rm Pb}_{\overline u}$ and $R^{\rm Pb}_{\overline d}$ tends to be smaller than in EPPS21 or nNNPDF2.0. The nNNPDF2.0 uncertainties for $R^{\rm Pb}_{\overline u}$ and $R^{\rm Pb}_{\overline d}$, in turn, are larger than those of EPPS21 presumably because nNNPDF2.0 does not include the fixed-target Drell-Yan data. The strange-quark uncertainty is the largest in nCTEQ15WZ which probably follows from the exclusion of all the neutrino DIS data in the nCTEQ15WZ fit. Both EPPS21 and nNNPDF2.0 do include neutrino DIS data. A bit surprisingly, the nCTEQ15WZ has the smallest small-$x$ gluon error even though the analysis does not involve any dijet or D-meson data. The reason for the smaller uncertainty is most likely in the more restrictive form of the adopted fit function. The largeness of the nNNPDF2.0 gluon uncertainty -- particularly at small $x$ -- in comparison to EPPS21 reflects the effect of including (EPPS21) or not including (nNNPDF2.0) the dijet or D-meson data. In the case of valence-quark nuclear modifications there seems to be no clear systematics in the widths of the error bands. In principle, EPPS21 has the least-restricted $W$ cut such that more large-$x$ data are included in the analysis than e.g. in the nCTEQ15WZ fit. However, the nCTEQ15WZ valence-quark errors are still smaller in places. Also here, the form of the parametrization is likely to play a role.

In Fig.~\ref{fig:allfullsets_EPPS16} we compare the average-nucleon nuclear modifications from EPPS21 and our previous analysis EPPS16. The largest differences are in the sectors of strange quarks and gluons. Thanks to the new D-meson and dijet data, the EPPS21 gluons are now much better constrained than what they are in EPPS16. The EPPS21 errors of the strange-quark nuclear modifications at small $x$ also appear significantly smaller than those of EPPS16. In part, the smaller strange-quark errors must follow from the better constrained gluons as the two are intertwined through the DGLAP evolution.

In Fig.~\ref{fig:absolutePDFs} we present the absolute EPPS21 PDFs for the full lead nucleus at $Q^2=1.69\,{\rm GeV}^2$ and at $Q^2=10\,{\rm GeV}^2$ together with the uncertainties broken down to the nuclear- and free-proton errors. The main conclusion to be made here is that, in most of the cases, the nuclear uncertainty is the dominant one also for the absolute PDFs. Only in $f_u^{\rm Pb}$ and $f_{\overline u}^{\rm Pb}$ the free-proton originating uncertainties can exceed the nuclear uncertainties in the plotted range. This is mainly so, as the uncertainty of $R_{\overline u}^{\rm Pb}$ is clearly smaller than that of $R_{\overline d}^{\rm Pb}$, as we saw in Fig.~\ref{fig:allfullsets_lowQ2}. Comparison to the two other contemporary nuclear PDFs, nCTEQ15WZ \cite{Kusina:2016fxy} and nNNPDF2.0 \cite{AbdulKhalek:2020yuc}, are shown in Fig.~\ref{fig:absolutePDFsothers}. Again, all three can be observed to agree within the presented 90\% confidence-level errors. The systematics between the sizes of the error bands follows the same pattern as was already seen in Fig.~\ref{fig:allfullsets_others}. 

\begin{figure*}[htb!]
\centering
\includegraphics[width=0.410\linewidth]{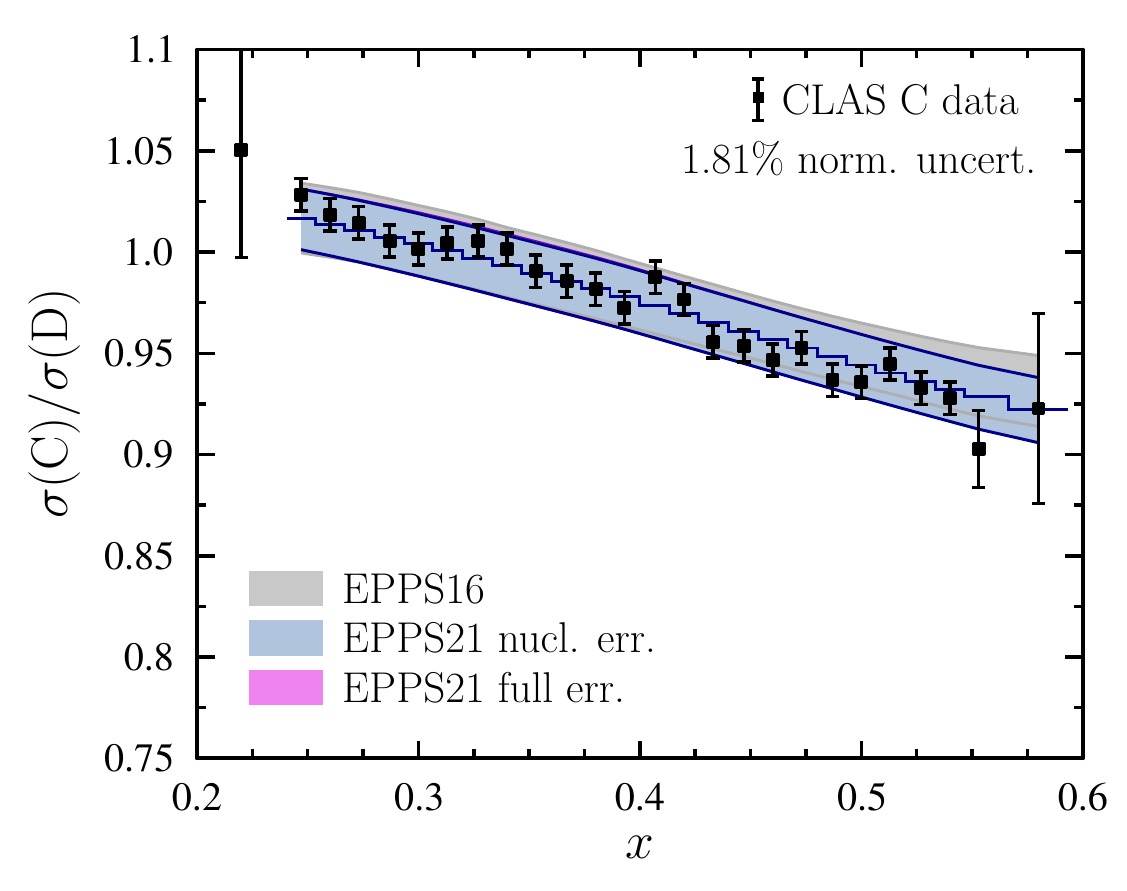}
\includegraphics[width=0.410\linewidth]{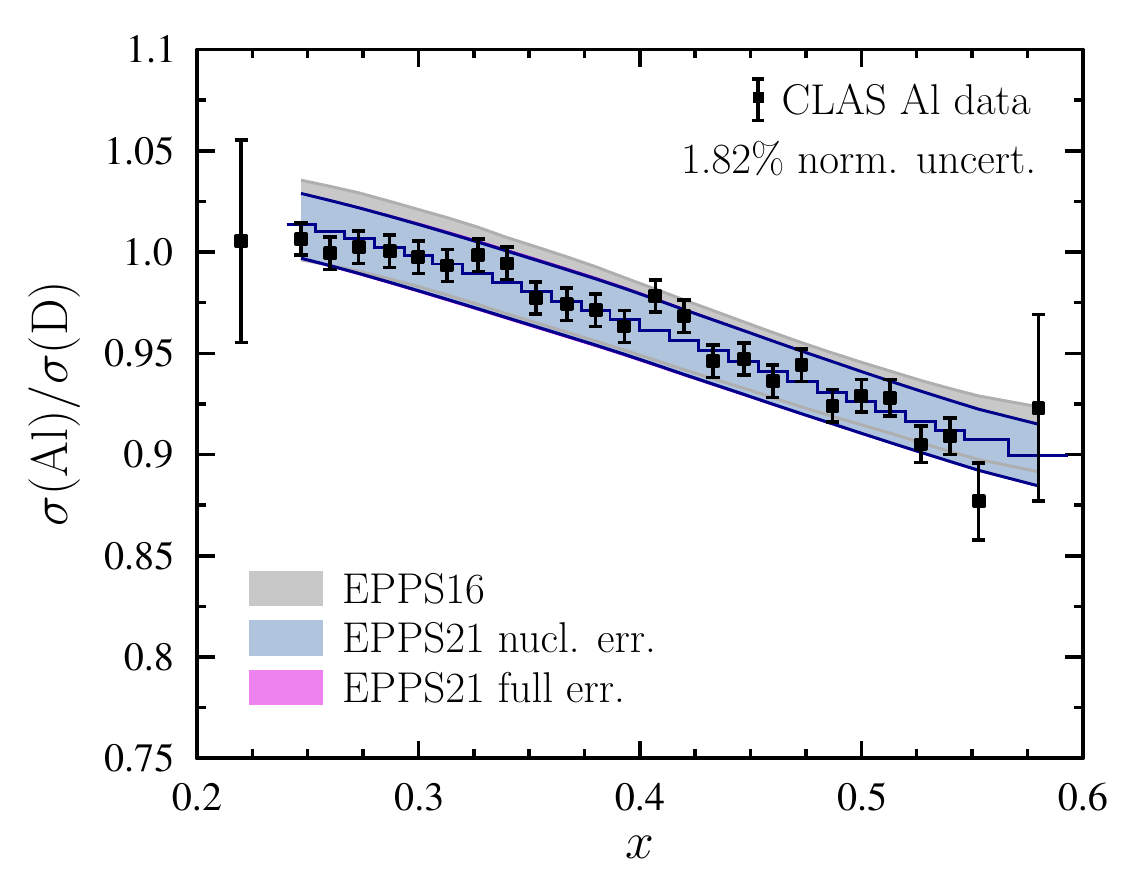}

\vspace{-0.35cm}
\includegraphics[width=0.410\linewidth]{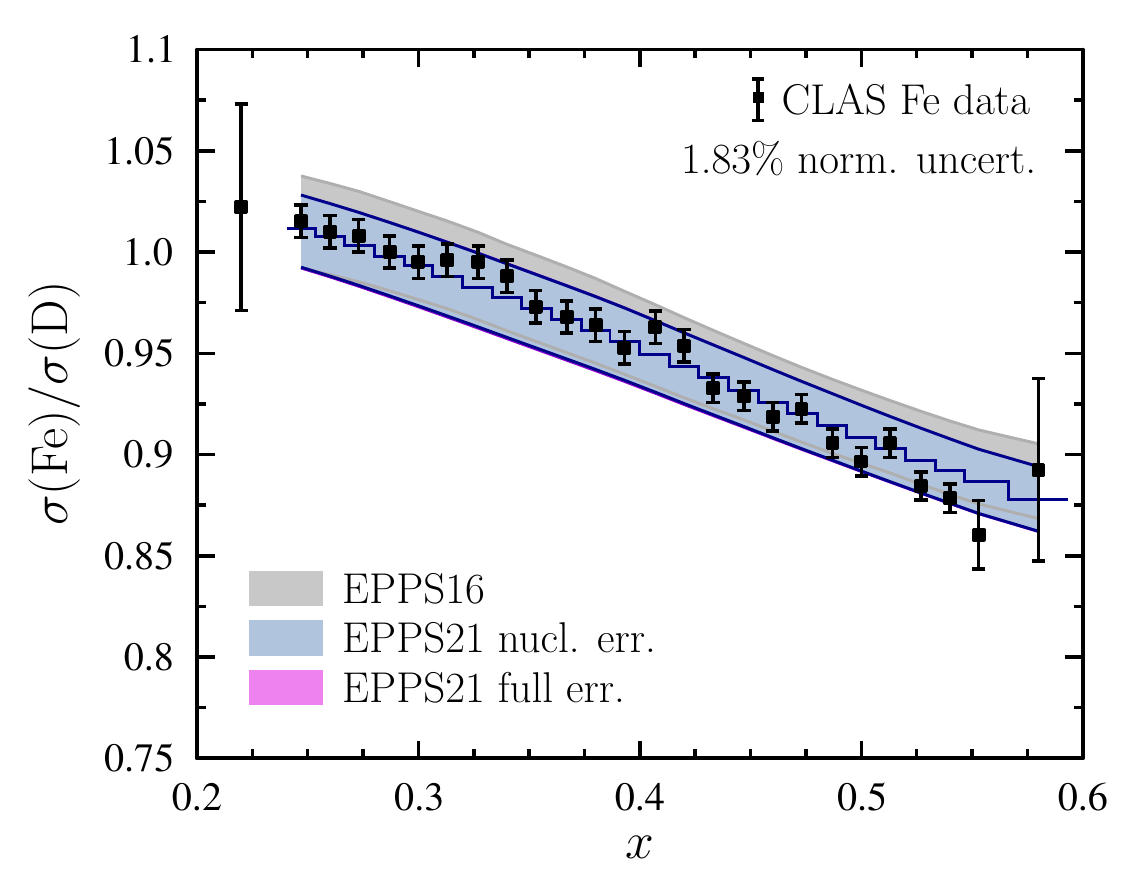}
\includegraphics[width=0.410\linewidth]{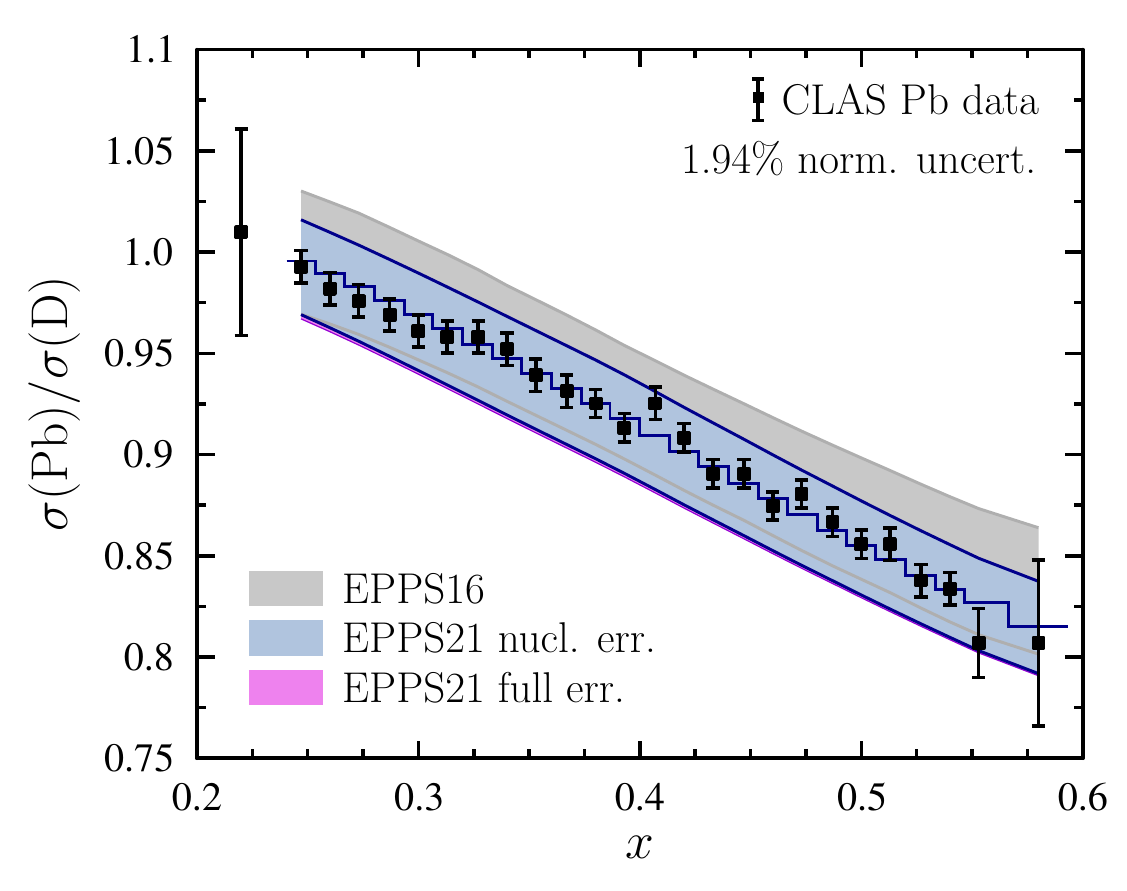}

\vspace{-0.35cm}
\includegraphics[width=0.410\linewidth]{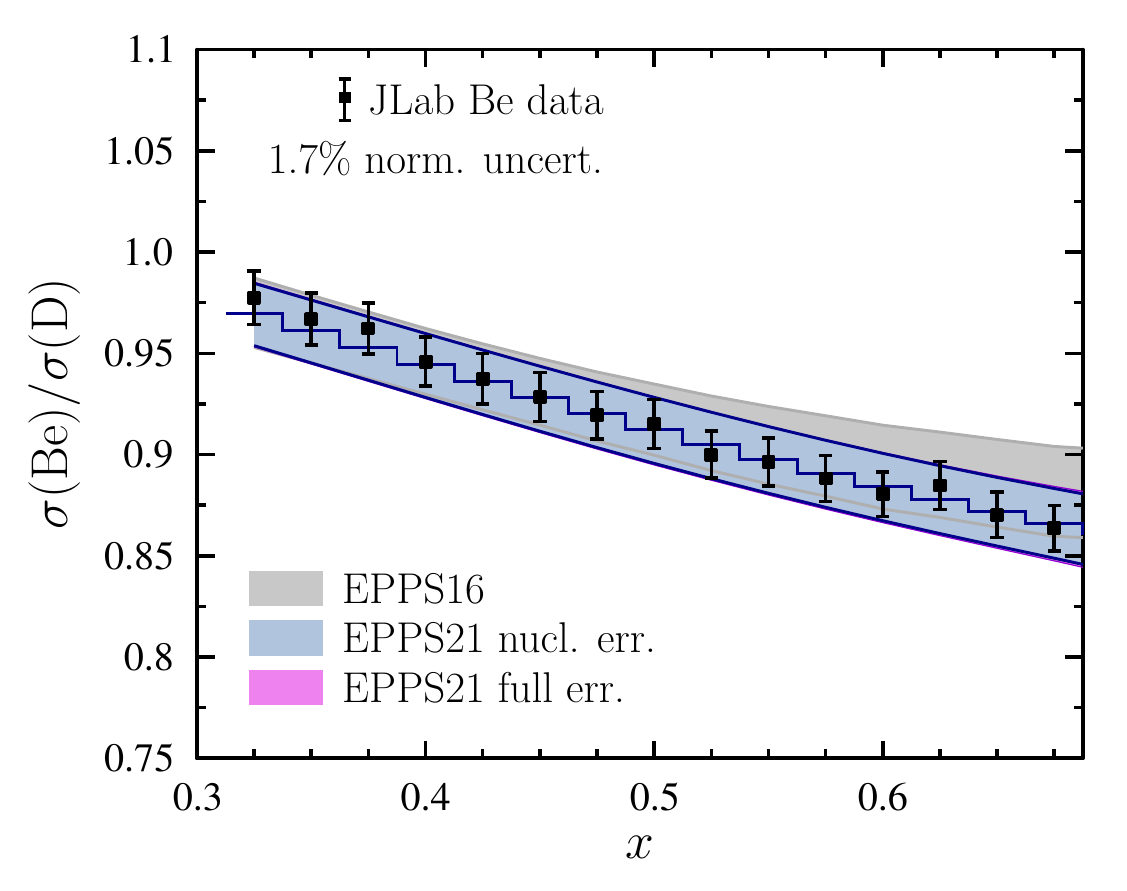}
\includegraphics[width=0.410\linewidth]{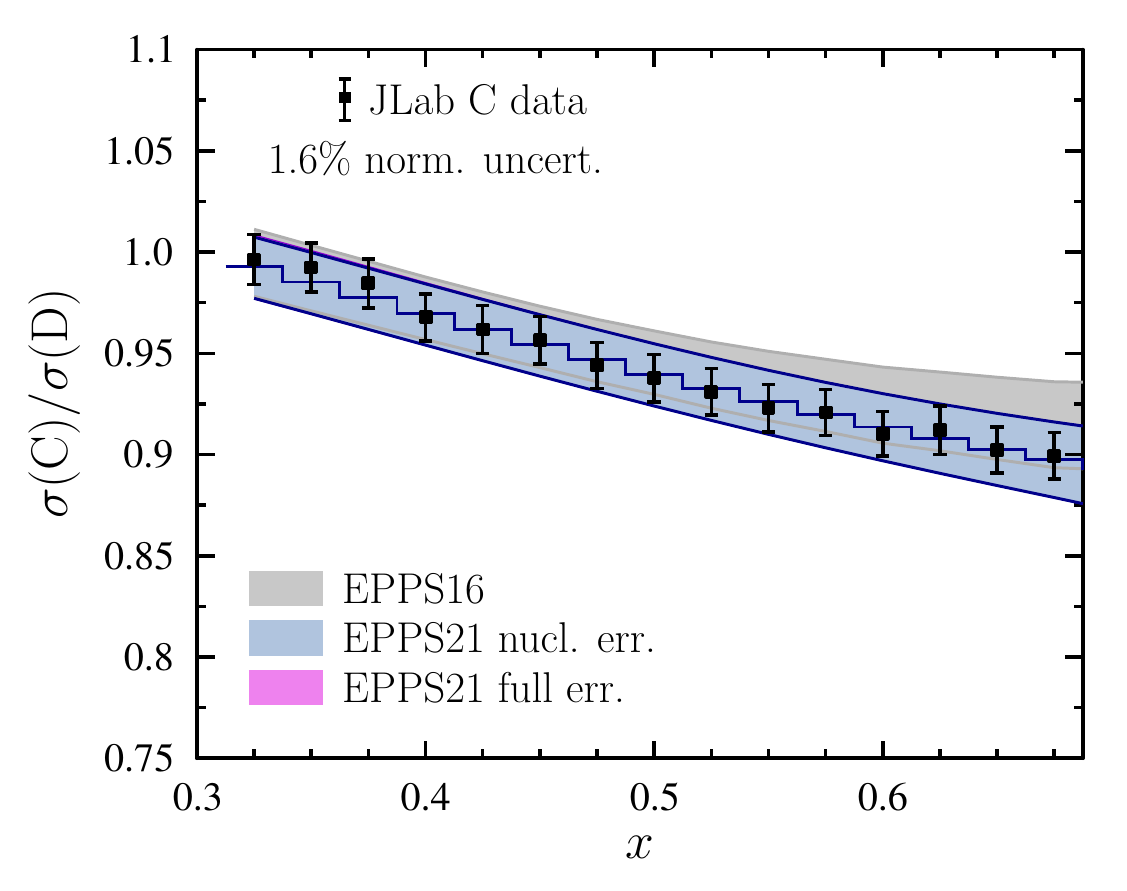}

\vspace{-0.35cm}
\includegraphics[width=0.410\linewidth]{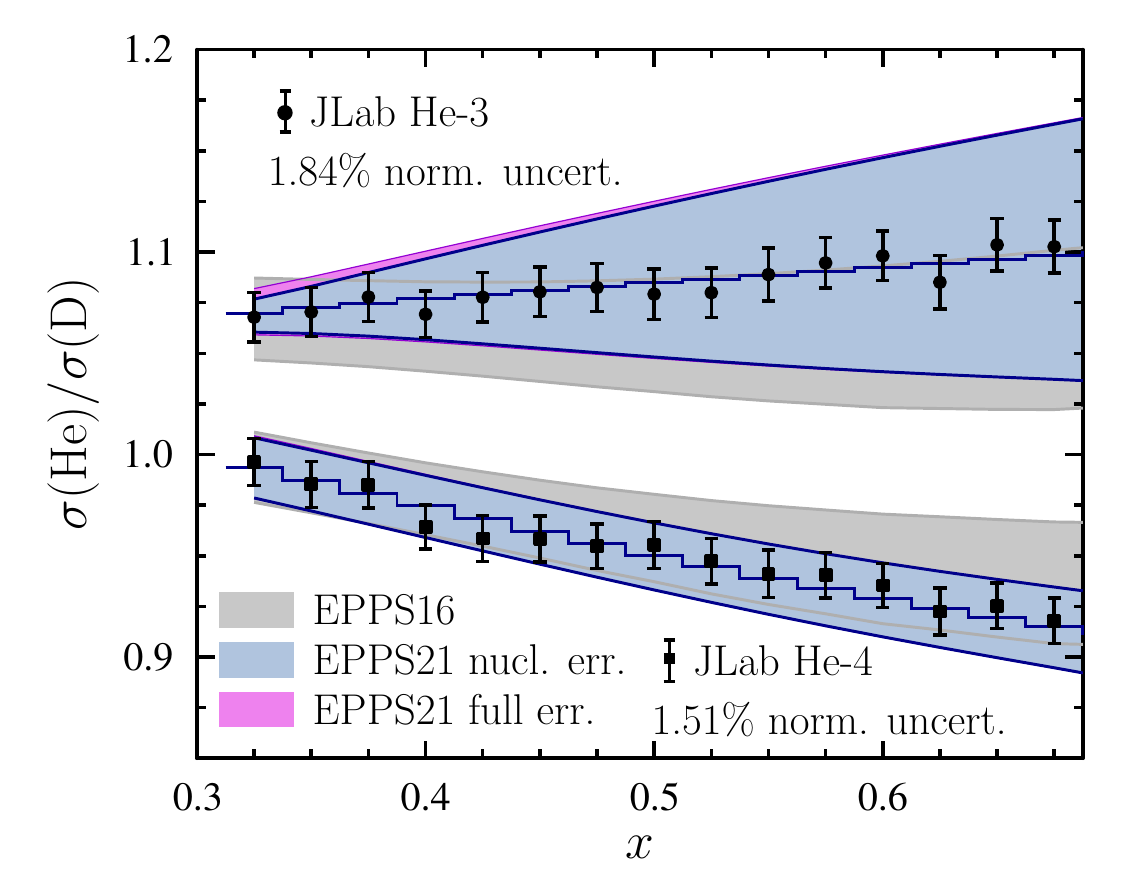}
\caption{The JLab CLAS \cite{Schmookler:2019nvf} and Hall-C data \cite{Seely:2009gt} compared with the EPPS21 and EPPS16 analyses. The solid blue lines show our central results, inner blue bands the nuclear uncertainties, and the purple bands the total uncertainty. The grey bands correspond to the EPPS16 results. The experimental data have been scaled with the normalization factors indicated in Table~\ref{Table:Data}.
}
\label{fig:CLASdata}
\end{figure*} 

\begin{figure*}[htb!]
\centering
\includegraphics[width=0.335\linewidth]{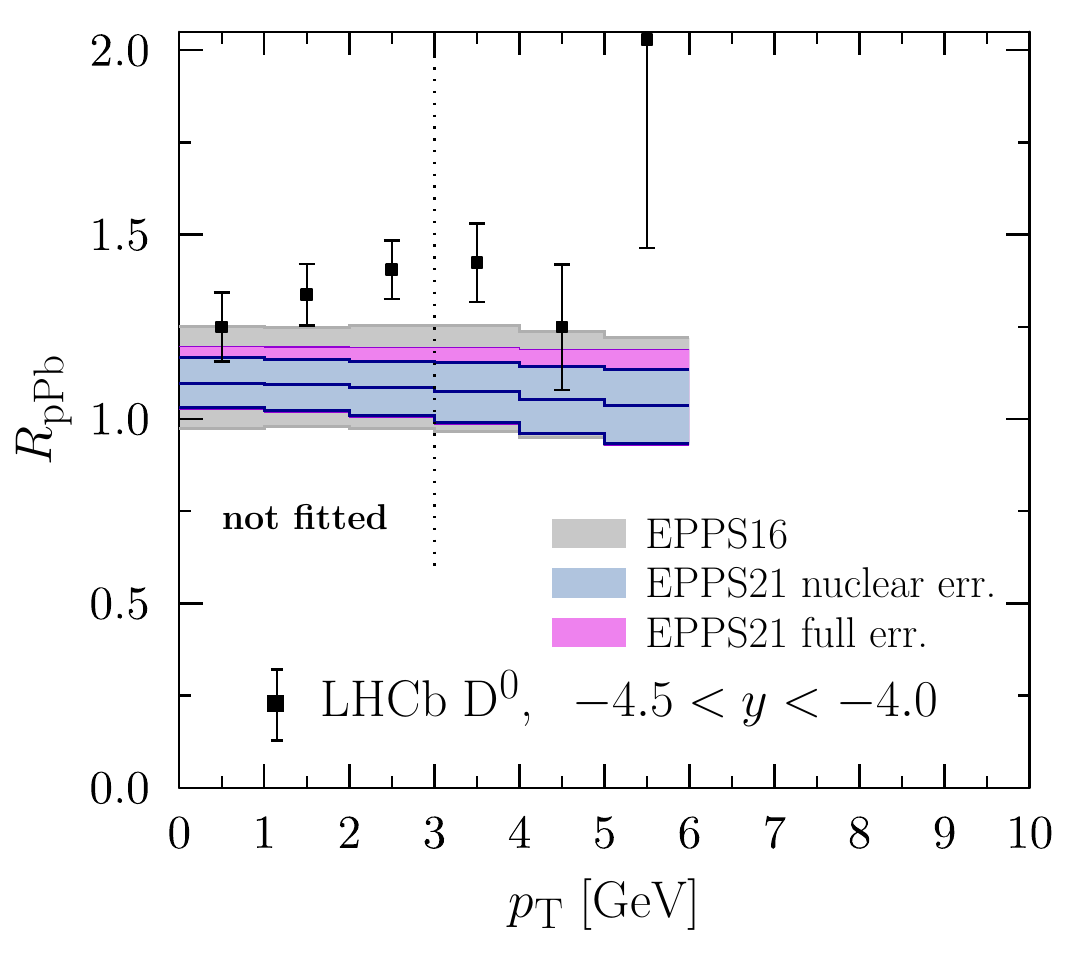}
\includegraphics[width=0.335\linewidth]{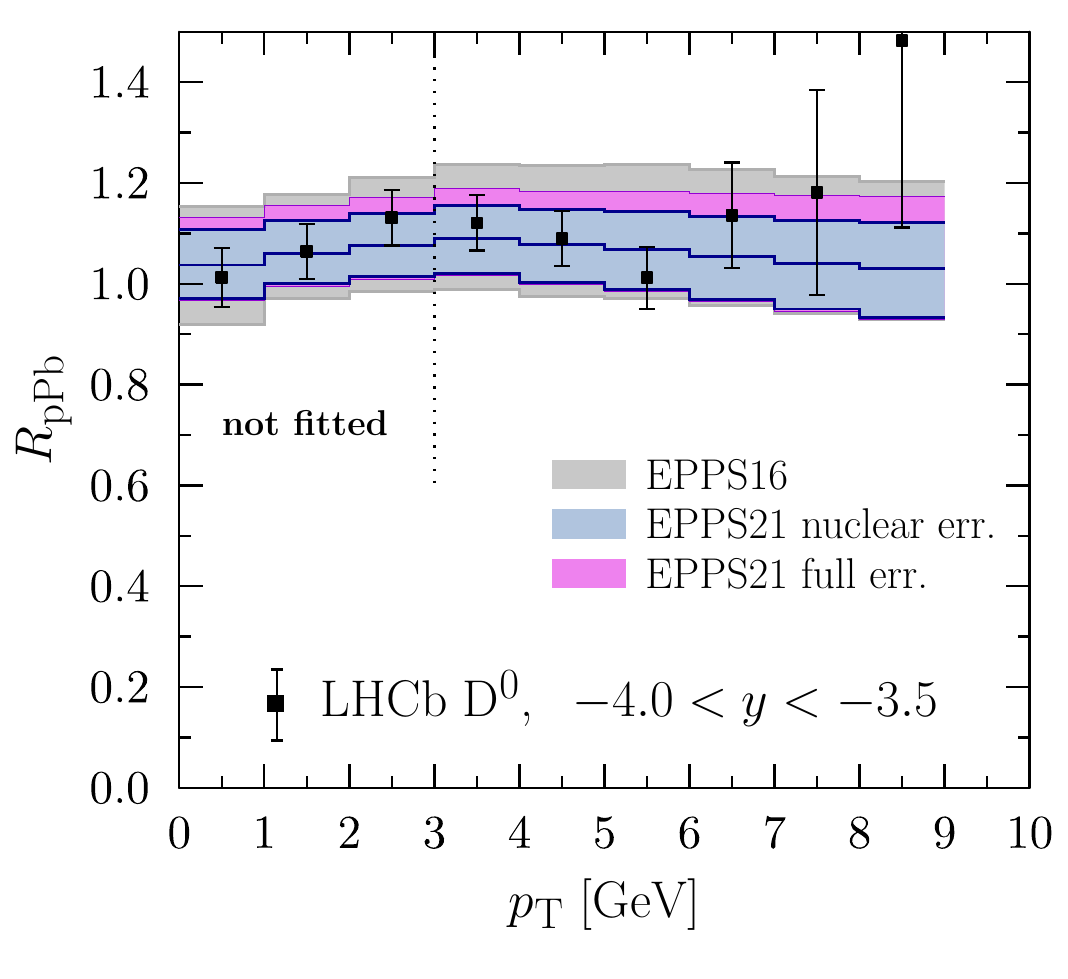}
\includegraphics[width=0.335\linewidth]{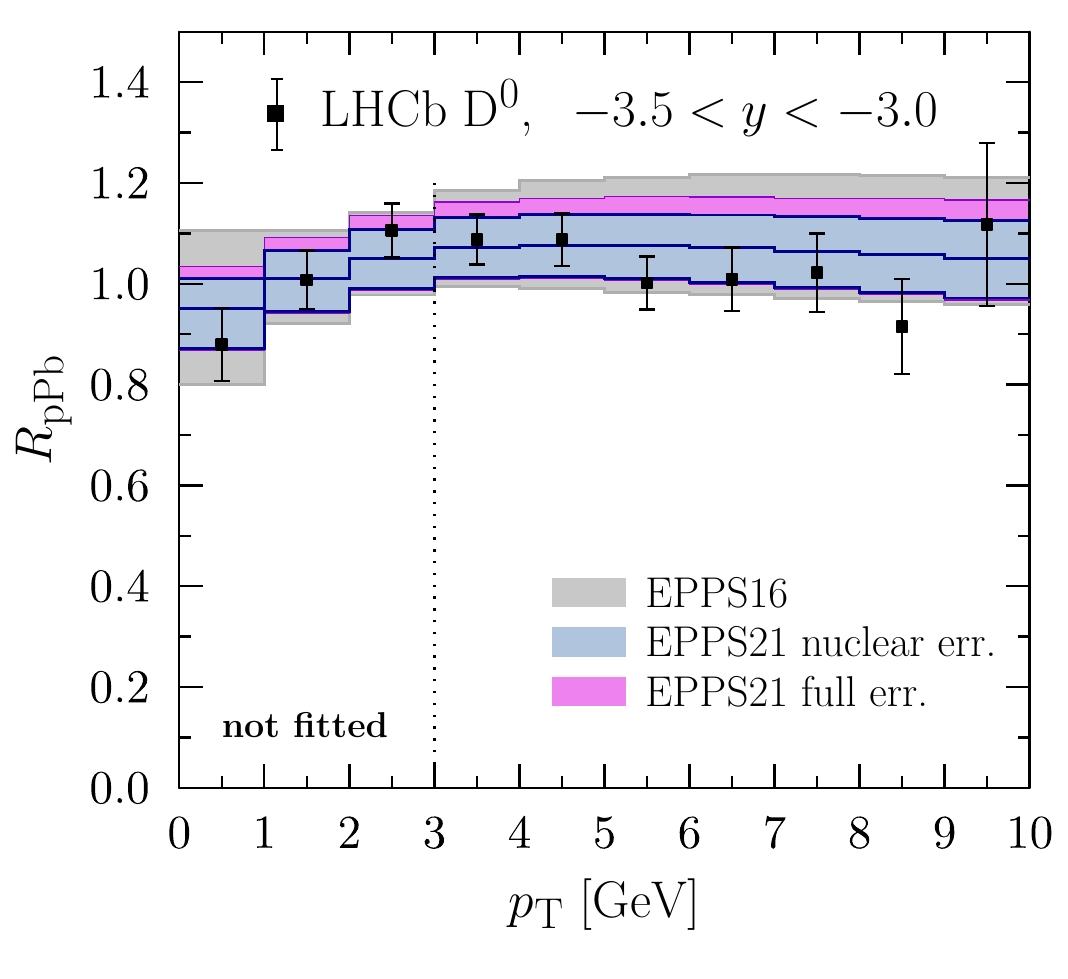}
\includegraphics[width=0.335\linewidth]{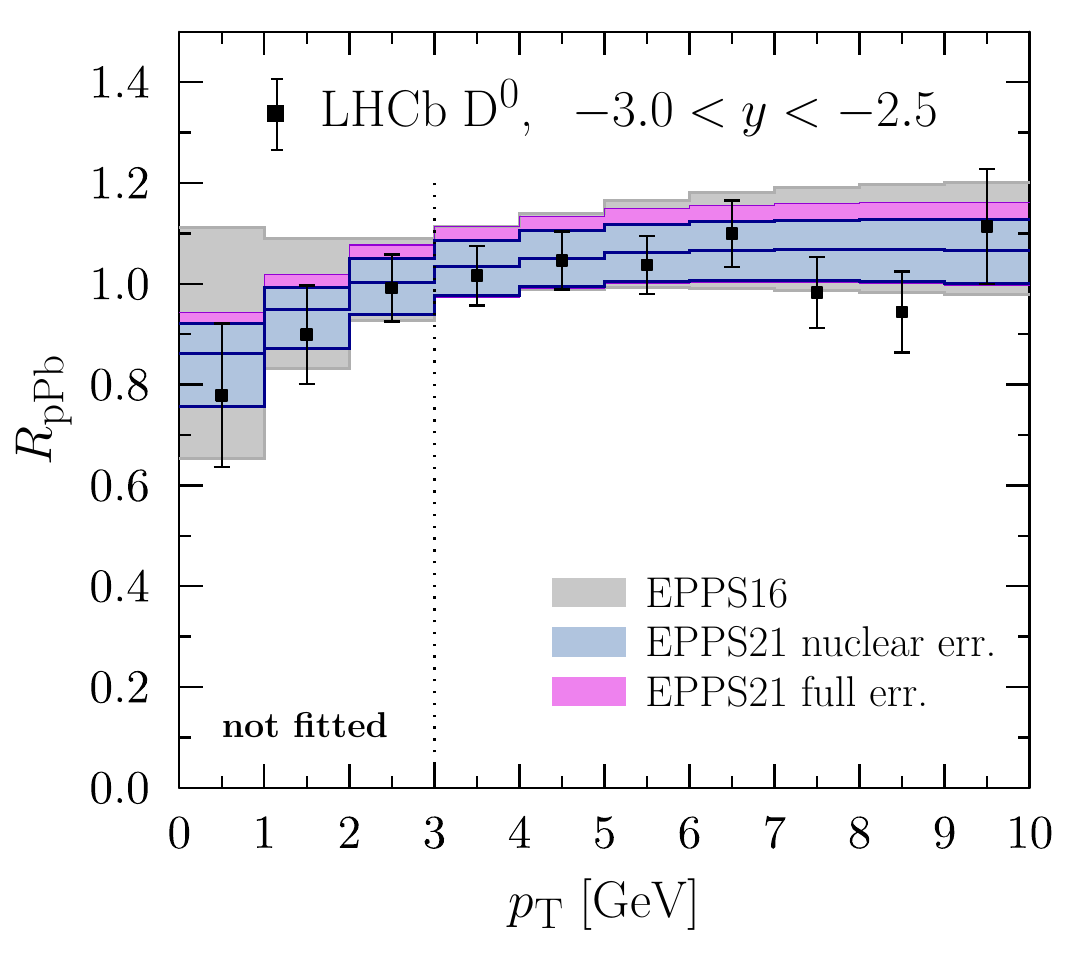}
\includegraphics[width=0.335\linewidth]{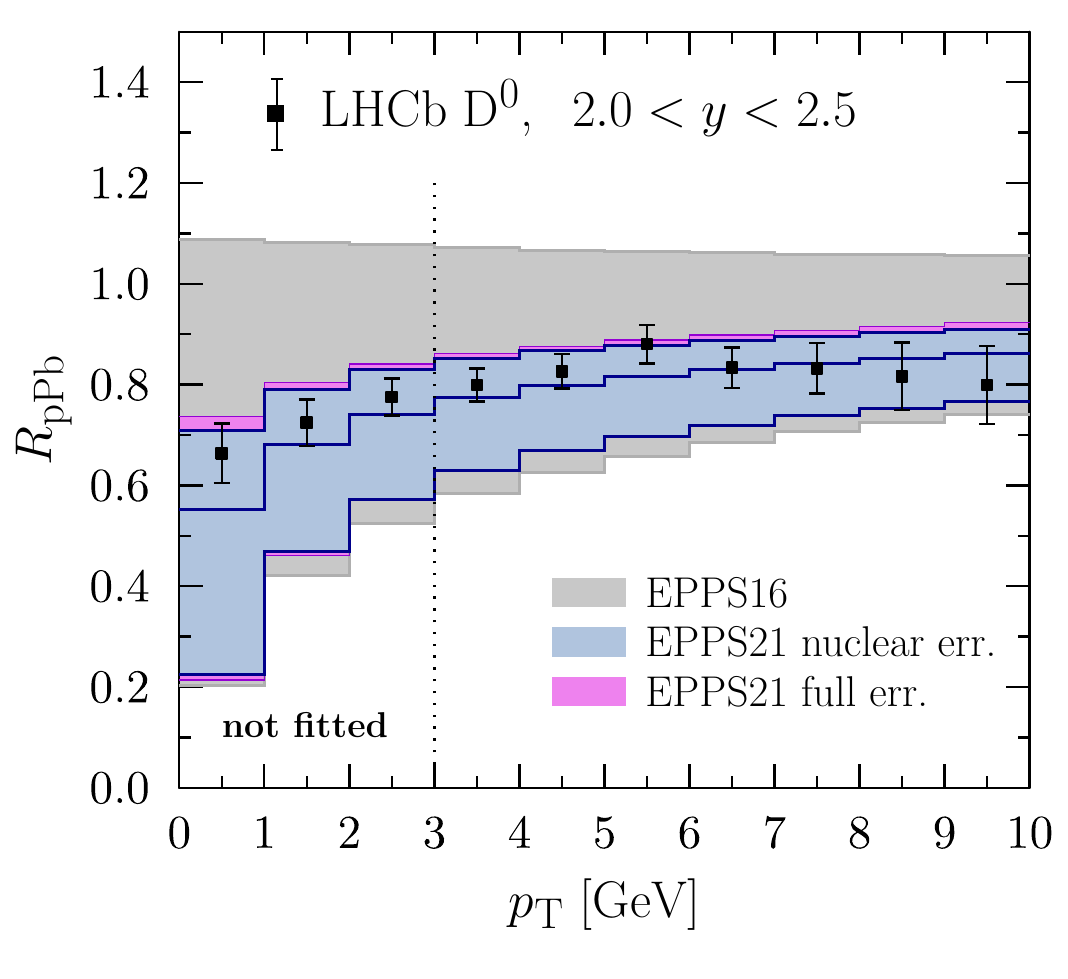}
\includegraphics[width=0.335\linewidth]{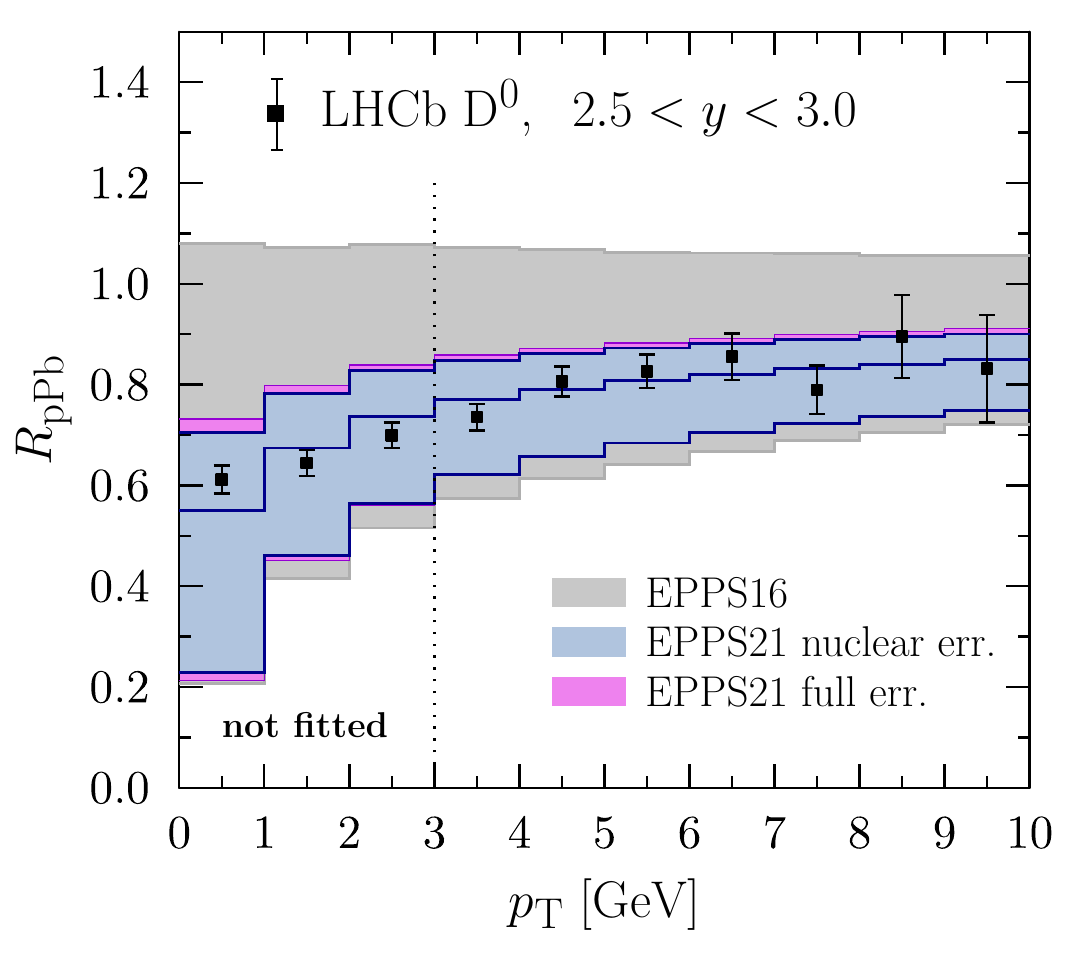}
\includegraphics[width=0.335\linewidth]{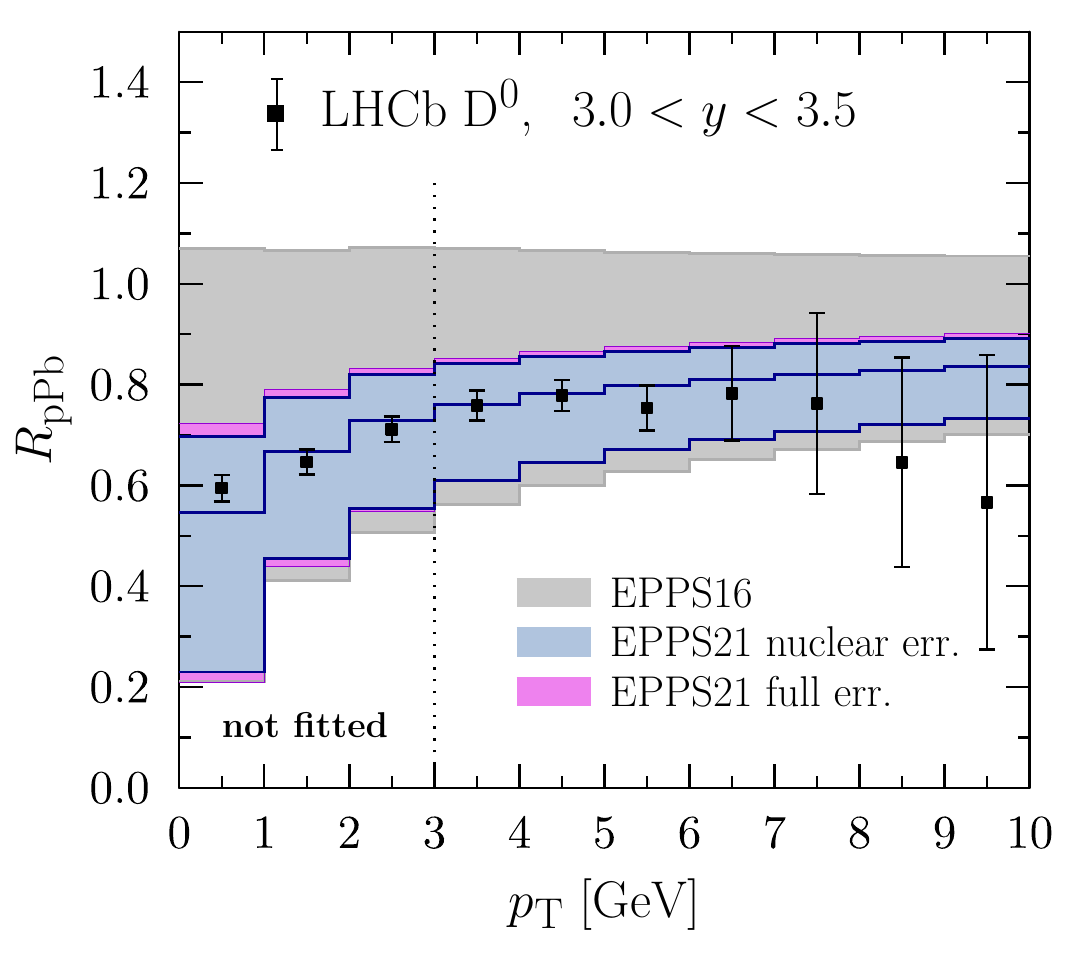}
\includegraphics[width=0.335\linewidth]{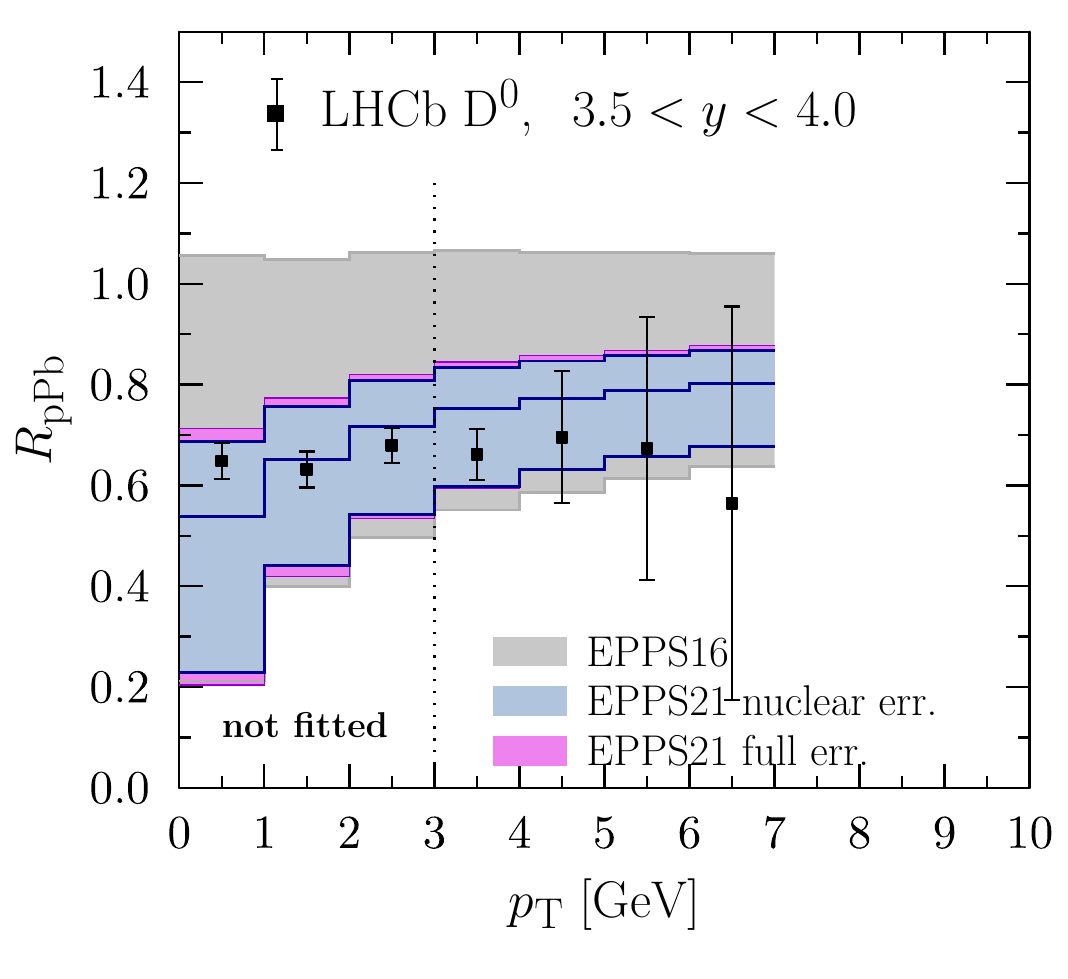}
\caption{The LHCb D-meson data \cite{Aaij:2017gcy} compared with the EPPS21 and EPPS16 analyses. The solid blue lines show our central results, inner blue bands the nuclear uncertainties, and the purple bands the total uncertainty. The grey bands correspond to the EPPS16 results. The points below $p_{\rm T} = 3\,{\rm GeV}$ were not included in the analysis. The experimental data have been scaled with the normalization factors indicated in Table~\ref{Table:Data}.}
\label{fig:LHCbdata}
\end{figure*} 

\begin{figure*}[htb!]
\centering
\includegraphics[width=0.45\linewidth]{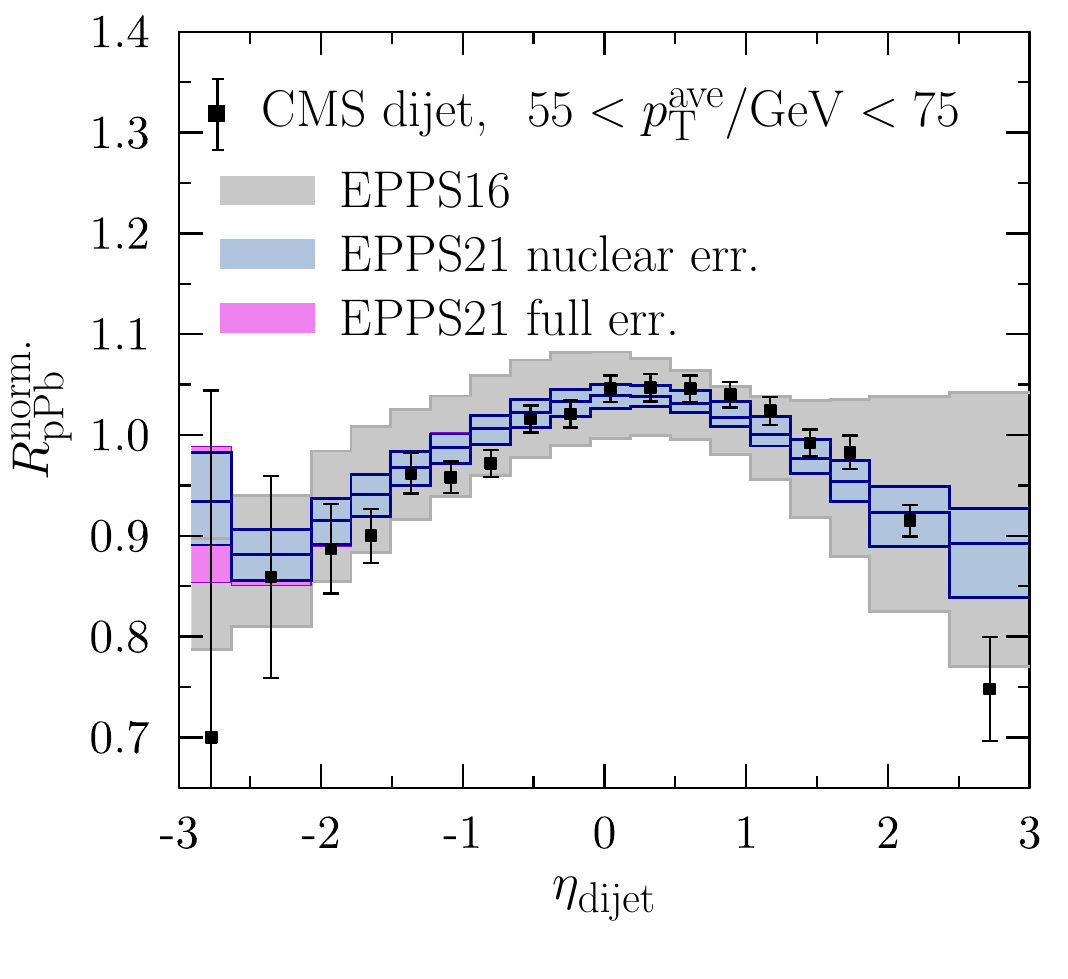}
\includegraphics[width=0.45\linewidth]{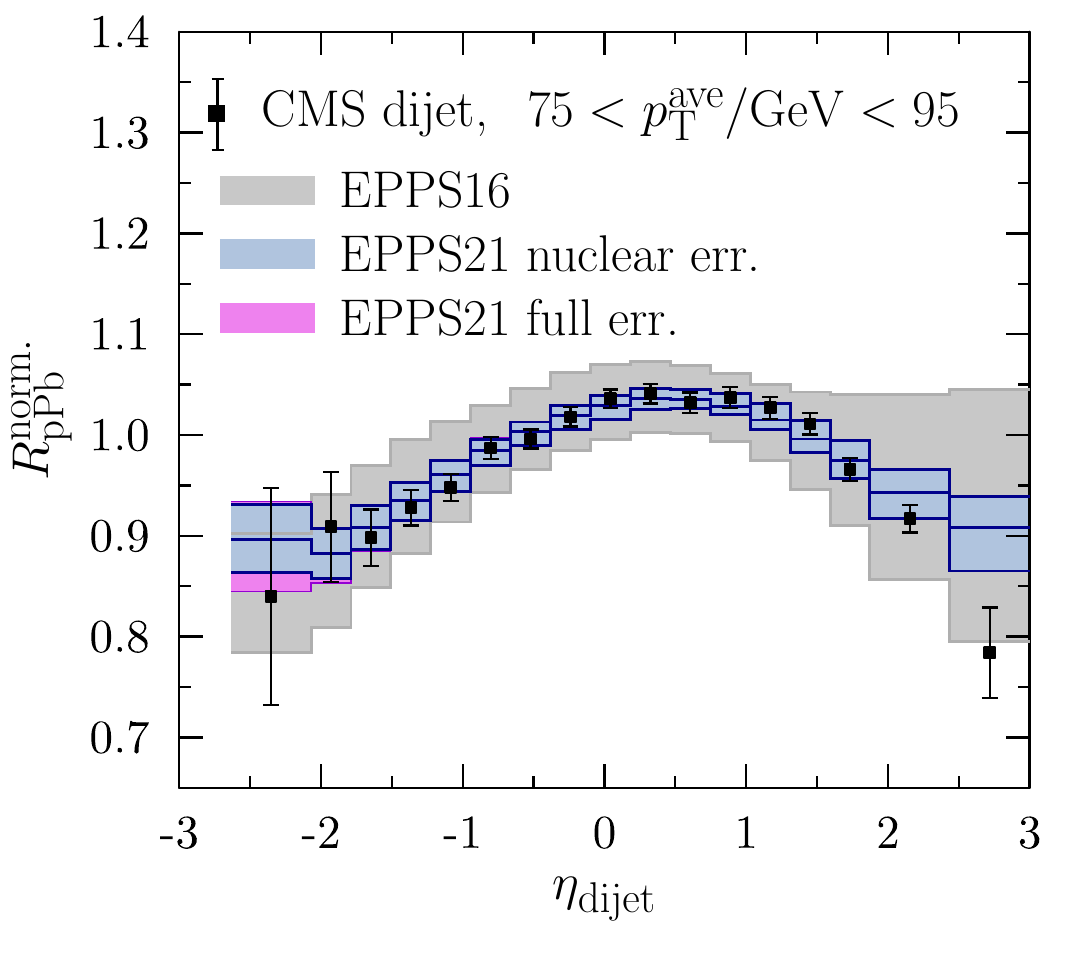}
\includegraphics[width=0.45\linewidth]{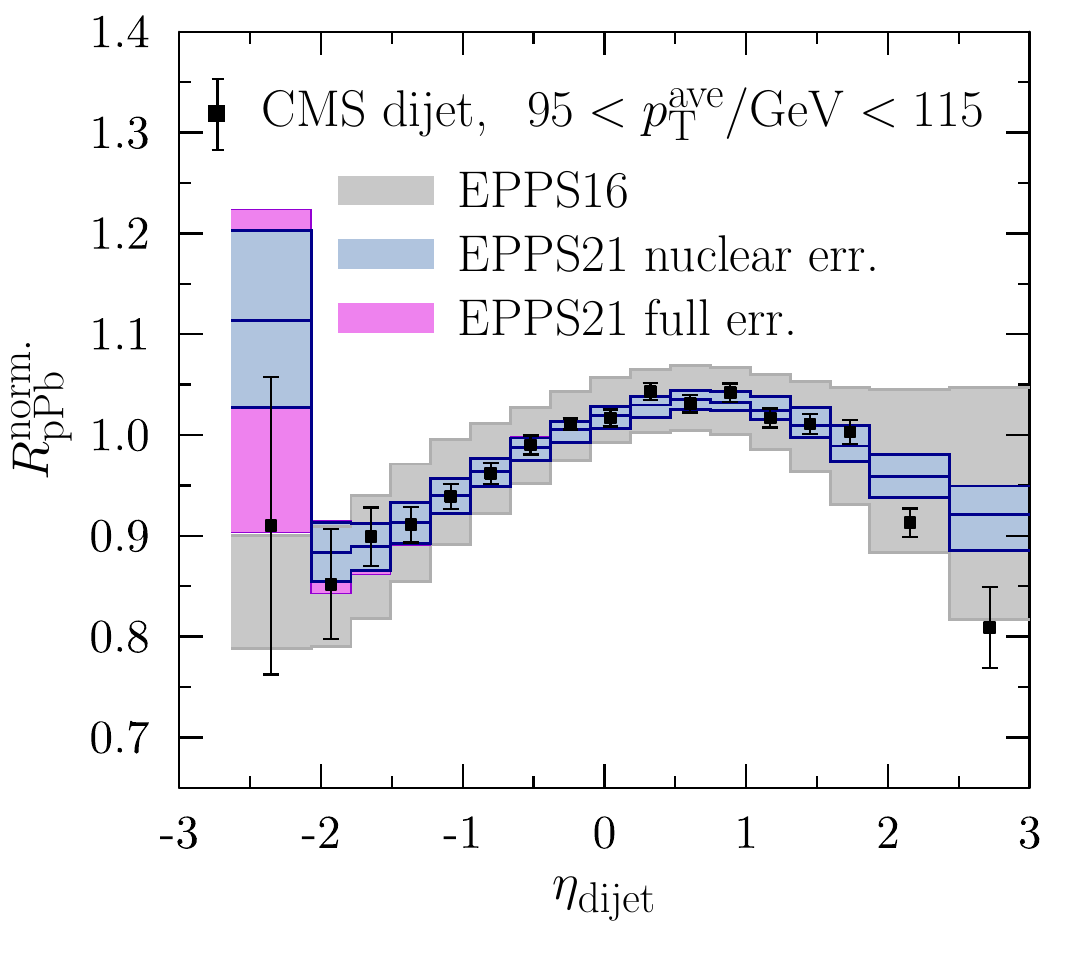}
\includegraphics[width=0.45\linewidth]{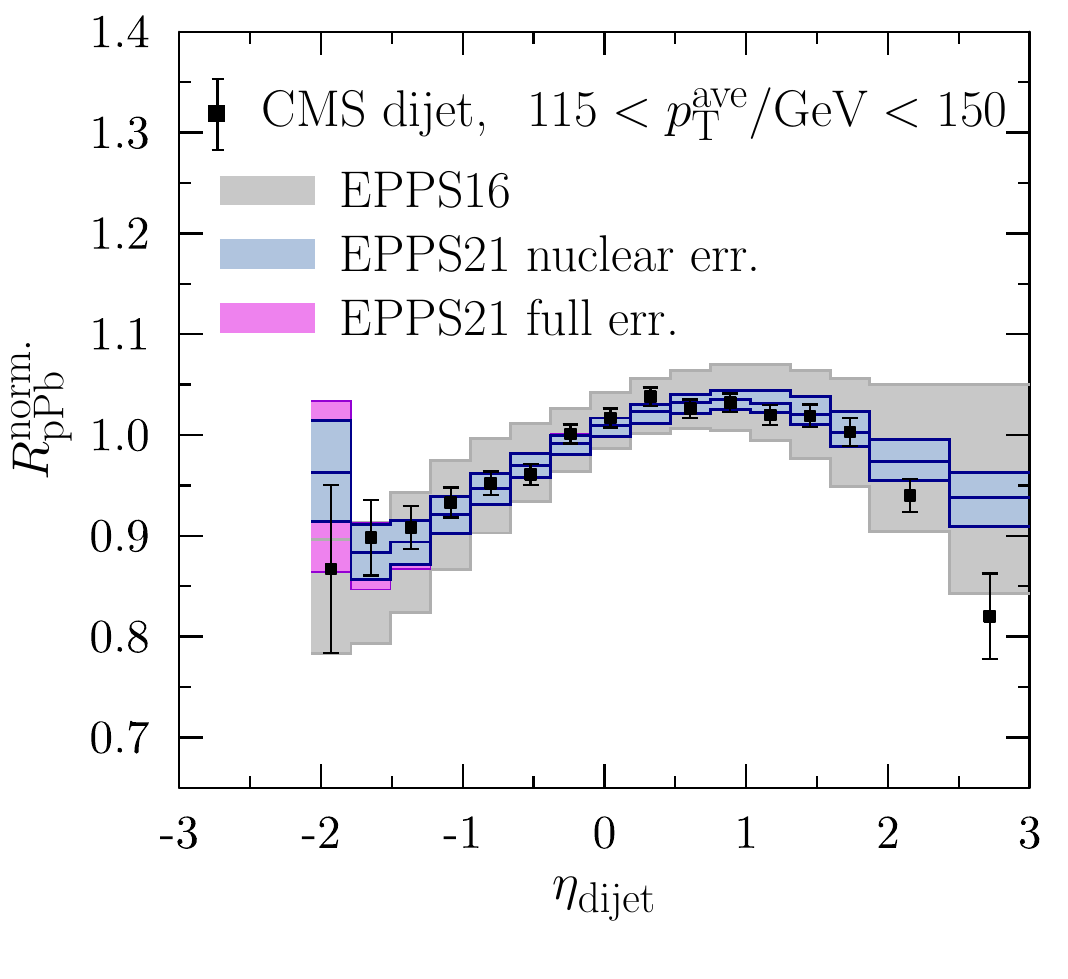}
\includegraphics[width=0.45\linewidth]{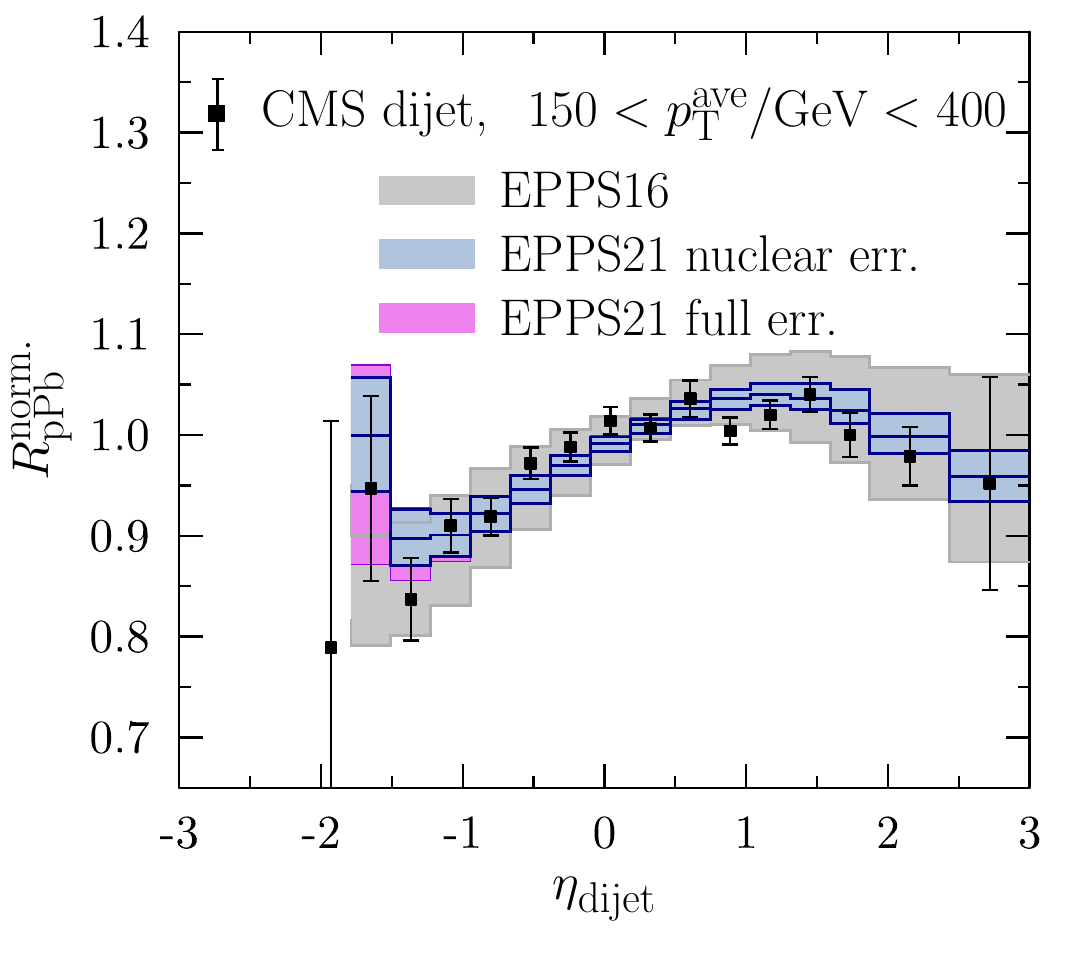}
\caption{The CMS dijet data \cite{Sirunyan:2018qel}  compared with the EPPS21 analysis. The solid blue lines show our central results, inner blue bands the nuclear uncertainties, and the purple bands the total uncertainty. The grey bands correspond to the EPPS16 results.}
\label{fig:CMSdijetdata}
\end{figure*}

\begin{figure*}[htbp!]
\centering
\includegraphics[width=0.45\linewidth]{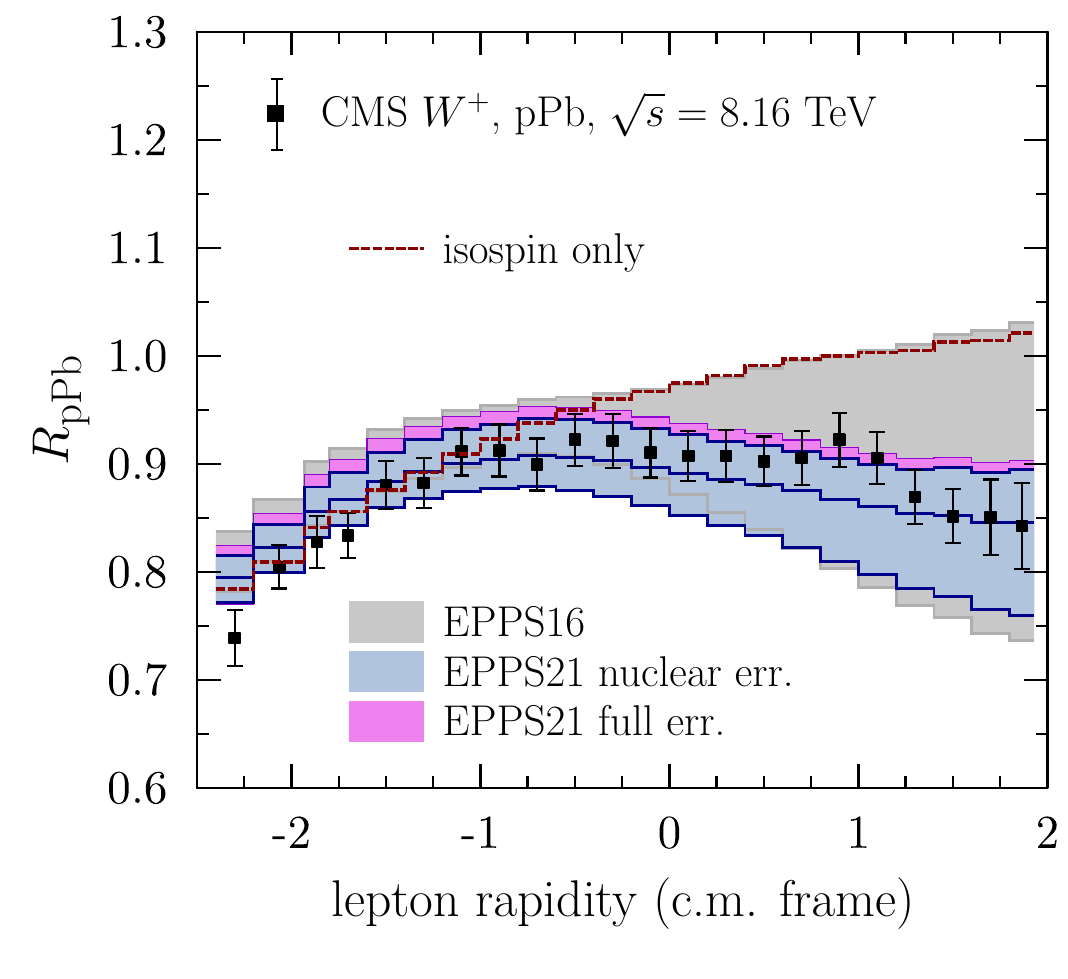}
\includegraphics[width=0.45\linewidth]{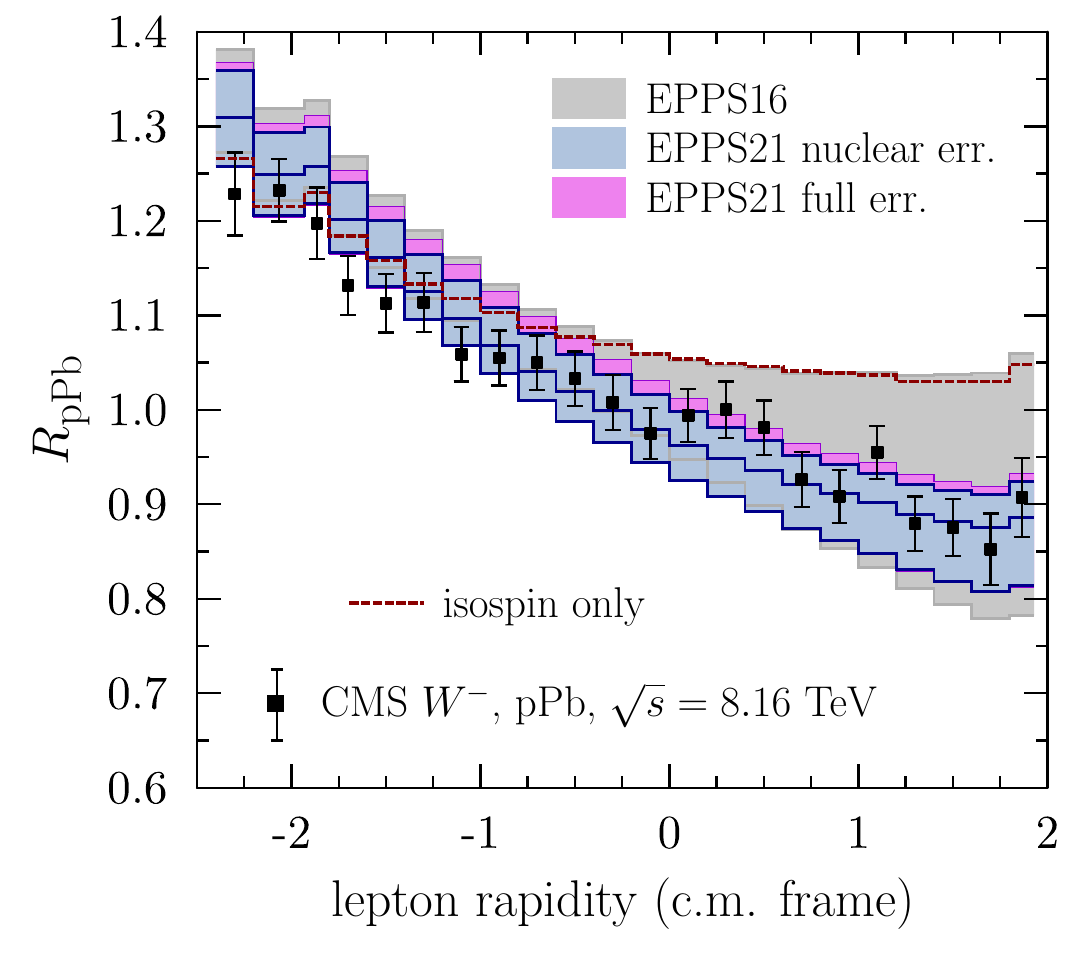}
\caption{
The CMS $8.16\,{\rm TeV}$ W$^\pm$ data \cite{Sirunyan:2019dox} for nuclear modification $R_{\rm pPb}$ compared with the EPPS21 and EPPS16 fits. The solid blue lines show our central results, inner blue bands the nuclear uncertainties, and the purple bands the total uncertainty. The grey bands correspond to the EPPS16 results. The red dashed curve corresponds to the isospin effects only (only isospin effects in PDFs). The experimental data have been scaled with the normalization factors indicated in Table~\ref{Table:Data}.
}
\label{fig:CMS8Wdatat}
\end{figure*}

\section{Comparison with the new data}
\label{Comparison_with_the_data}

\subsection{Comparison with the fitted data}
\label{sec:fitted}

In the following we demonstrate the compatibility of the EPPS21 nuclear PDFs with the experimental data. Here, we will discuss only the new data sets included in the EPPS21 analysis -- comparisons of the EPPS21 results with the data that were already included in the EPPS16 analysis are provided in \ref{sec:app}. In all cases the plotted data have been multiplied by the optimal normalization factors indicated in Table~\ref{Table:Data}.

The recent JLab CLAS \cite{Schmookler:2019nvf} and Hall-C light-nucleus \cite{Seely:2009gt} DIS measurements are shown in Fig.~\ref{fig:CLASdata}. These data are new in comparison to the EPPS16 fit, and we therefore also show the EPPS16 errorbands for comparison. In the case of CLAS, the new EPPS21 results tend to lie below the EPPS16 ones but the $x$ derivatives are almost equal. These CLAS data overlap with the older SLAC data \cite{Gomez:1993ri} in $x$ but probe the nuclear PDFs typically at lower $Q^2$ than the SLAC measurements. The good simultaneous fit to both hints that no sizable $A$-dependent power corrections that would scale like $Q^{-2n}$ emerge in the probed $Q^2$ region. The Hall-C data are also new in comparison to the EPPS16 analysis and the EPPS21 errorbands are now visibly smaller for all other than the He-3 data. The larger uncertainty for He-3 follows from the new free parameter $f_3$, using a fixed global tolerance, and that these 15 JLab data points constitute the only constraints for this new parameter. We note that only those Hall-C data points with $W \geq 1.8\,{\rm GeV}$ are included in the $\chi^2$ analysis. However, the exact choice of the $W$ cut is not critical here but we have checked that e.g. the first data points towards higher $x$ (lower $W$) are still reproduced so that the fit is stable against small variations in the $W$ cut.

The LHCb D-meson data shown in Fig.~\ref{fig:LHCbdata} provide stringent constraints for the nuclear gluons \cite{Eskola:2019bgf} for a wide range in $x$. The differences between the EPPS16 and the new EPPS21 uncertainty bands are large, the EPPS21 errors being now much smaller. Since the $p_{\rm T}$ slopes of the data are rather mild in the fitted region $p_{\rm T} > 3\,{\rm GeV}$, the normalization uncertainty can effectively compensate for variations in the gluon nuclear modifications and, as a result, the uncertainty bands appear larger than the data errors at low $p_{\rm T}$. We note that even in the non-fitted region $p_{\rm T} < 3\,{\rm GeV}$ the data are still well reproduced by the EPPS21 PDFs and in this sense we see no signs of novel small-$x$ dynamics beyond the linear DGLAP. The data at negative rapidities rise above unity which is consistent with having a gluon antishadowing. At the very backward-most bin, the data are clearly above the predictions -- we have no explanation for this behaviour and it cannot be explained by nuclear effects in PDFs. 

In Fig.~\ref{fig:CMSdijetdata} we compare the CMS double-differential dijet measurements \cite{Sirunyan:2018qel} with the EPPS21 and EPPS16 calculations. Similarly to the case of D mesons, the differences between the EPPS16 and EPPS21 uncertainty bands are significant, and the new data have squeezed the EPPS21 uncertainties considerably from EPPS16. This translates into much better constrained nuclear modifications of the gluons \cite{Eskola:2019dui}. Towards negative values of $\eta_{\rm dijet}$ one becomes increasingly sensitive to the valence quarks in the Pb nucleus which, in part, causes the downward trend towards negative $\eta_{\rm dijet}$. In the bin with the largest transverse momentum $p_{\rm T}^{\rm ave}$ the point near $\eta_{\rm dijet} \approx -2$ is very near to the edge of the phase space i.e. the produced dijet carries nearly all the available energy. Effects like hadronization (one has to allow the nucleon remnants to have energy to hadronize), threshold logarithms, and target-mass corrections become arguably large in this particular bin and it is left out from the fit. Towards positive values of $\eta_{\rm dijet}$ a clear downward trend is again observed which in the calculations originates from the gluon shadowing. However, the very forward data points would prefer a clearly stronger gluon shadowing than what the fit allows for. We have tested that gluon PDFs that can reproduce the most forward datapoints fail in reproducing the data at other values of $\eta_{\rm dijet}$. We have also varied the small-$x$ part of our gluon parametrization to see whether the difficulty would be associated with a too stringent form of the fit function. However, varying the parametrization does not lead to any better result than what is seen in Fig.~\ref{fig:CMSdijetdata}. This can also be understood as follows: Estimating the probed value of nuclear $x$ by $x^{\rm Pb} \approx 2p_{\rm T}^{\rm ave}/\sqrt{s}e^{-\eta_{\rm dijet}+0.465}$, the values for the $\eta_{\rm dijet} \approx 2.75$ bin with $p_{\rm T}^{\rm ave} = 95\,{\rm GeV}$ and the $\eta_{\rm dijet} \approx 2.25$ bin with $p_{\rm T}^{\rm ave} = 55\,{\rm GeV}$ are approximately the same. Both thus probe the nuclear modification $R^{\rm Pb}_g(x,Q^2)$ at the same values of $x$. As the scale dependence of $R^{\rm Pb}_g(x,Q^2)$ is very weak at such large interaction scales it appears impossible to reproduce these to data points with a same set of PDFs. What causes the most forward data points to be so strongly suppressed remains to be understood. However, our fit results are not sensitive to whether we include (as we do) or exclude the 4 most forward data points in the 4 first $p_{\rm T}$ bins.

Finally, the new CMS $8.16\,{\rm TeV}$ data on W$^\pm$ production \cite{Sirunyan:2019dox} are shown in Fig.~\ref{fig:CMS8Wdatat}. In the plot, we also show the EPPS16 uncertainties to visualize the difference between EPPS21 and EPPS16. Particularly towards forward (positive) lepton rapidities the EPPS21 errorbands are clearly smaller than the EPPS16 bands. However, this is mainly due to the D-meson and dijet data which now set more stringent constraints for the gluons. While the W$^\pm$ production at forward direction is directly sensitive to the small-$x$ quark content of the nucleus, the interaction scale, set by the mass of the W$^\pm$ boson, is very high and a significant part of the behaviour of the sea quarks is dictated by the gluons. Just by looking at the plots, the rapidity dependencies of $R_{\rm pPb}$ appear not so perfect but EPPS21 tends to overshoot the data at negative rapidities. At negative rapidities the form of $R_{\rm pPb}$ is largely set by the valence-quark nuclear modifications which are well constrained by the available DIS data -- the backward part of $R_{\rm pPb}$ cannot thus be easily changed without ruining the agreement with the DIS data. If the data were multiplied by a larger normalization factor that matches the negative-rapidity data with the central EPPS21 prediction, the forward part of $R_{\rm pPb}$ would be underestimated by EPPS21. As a result, to better reproduce the overall rapidity dependence, the forward part of EPPS21 should be higher which could be attained by having less gluon shadowing. However, both the dijet and D-meson data prefer a stronger shadowing such that one could be tempted see some indications of a tension between these observables, perhaps due to a coherent energy loss \cite{Arleo:2021bpv} or equivalent. However, $\chi^2/N_{\rm data} \approx 0.94$ for the W$^\pm$ data so statistics-wise these data are well reproduced. We note that the CMS collaboration provides the covariance matrix to be used in calculating the $\chi^2$ values, which implicitly involves solving for the optimal shifts associated with the systematic uncertainties of the data. As only the covariance matrix is given, we cannot solve for these optimal shifts without further input and the data points are plotted at their central values (the data are still multiplied with the optimal normalization factor) -- it is conceivable that the apparent non-prefect reproduction of the rapidity dependence disappears once these shifts are implemented. 

\begin{figure}[htbp!]
\centering
\includegraphics[width=1.00\linewidth]{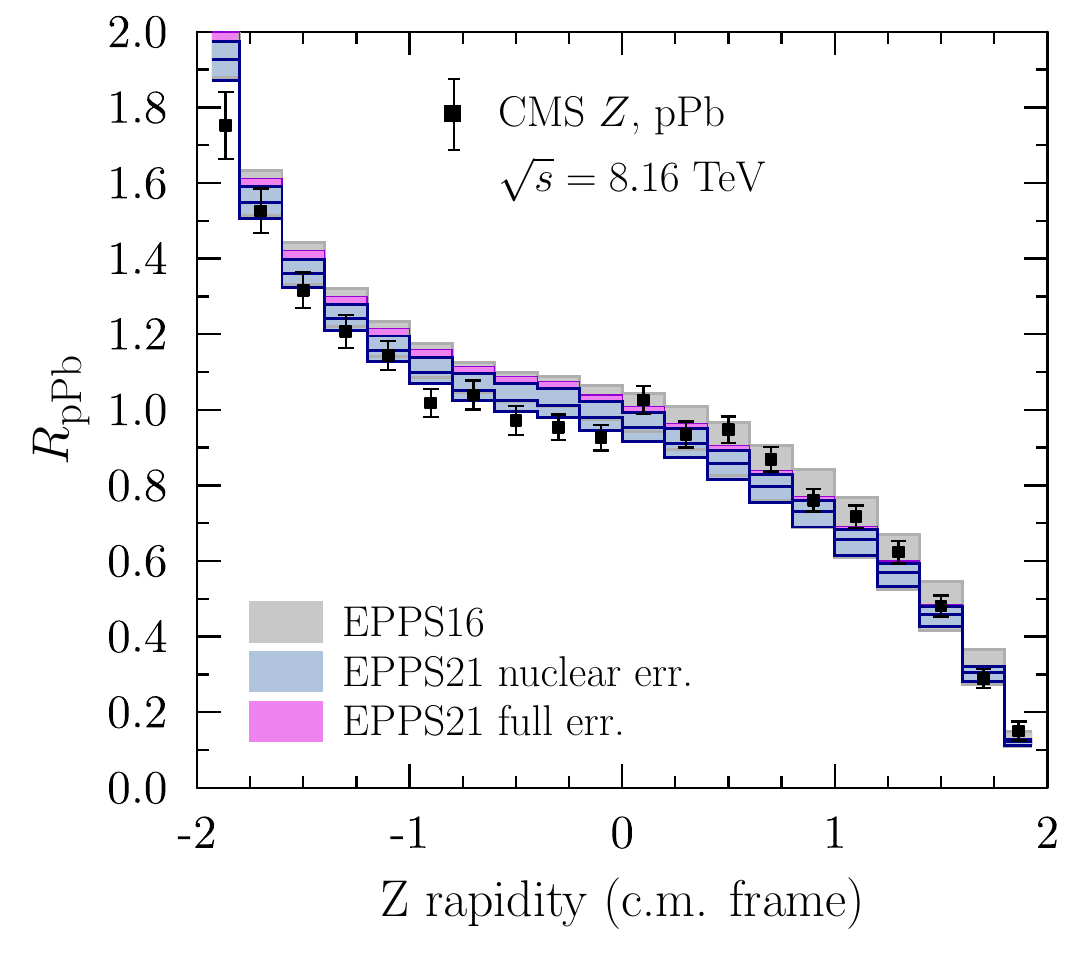}
\caption{
The CMS $8.16\,{\rm TeV}$ Z-boson data \cite{CMS:2021ynu,CMS:2014jea} for nuclear modification $R_{\rm pPb}$ compared with the EPPS21 fit. The solid blue lines show our central results, inner blue bands the nuclear uncertainties, and the purple bands the total uncertainty. The data have been normalized by a factor of 0.951 which is the optimal normalization with EPPS21. The grey bands correspond to the EPPS16 results.
}
\label{fig:CMS8Zdatat}
\end{figure} 

\subsection{Comparison with the CMS 8 TeV Z data}
\label{sec:8TeVZ}

Also the new CMS $8.16\,{\rm TeV}$ data for Z production in pPb collisions are available \cite{CMS:2021ynu}. By using the $8\,{\rm TeV}$ pp reference data \cite{CMS:2014jea} we have constructed the mixed-energy nuclear modification factor as in Eq.~(\ref{eq:mixedR}). A comparison of the experimental results with the EPPS21 predictions is shown in Fig.~\ref{fig:CMS8Zdatat}. Due to the different muon acceptances in pp and pPb collisions, the shape of $R_{\rm pPb}$ differs significantly from unity. While the broad systematics of the data are reproduced by the EPPS21 PDFs, the 90\% EPPS21 uncertainty bands hardly overlap with the data errors. The value of $\chi^2$ is also rather large, $\chi^2/N_{\rm data} \approx 2.1$ (accounting for the correlated normalization uncertainty but adding the other errors in quadrature). Similar difficulties are also visible in the original CMS data paper \cite{CMS:2021ynu} for the absolute cross sections as well as for the forward-to-backward ratio. In particular, the data exhibit large fluctuations around the midrapidity which seem to be impossible to reproduce by any PDF-based calculation. For these reasons, these new Z-boson data are not included in EPPS21.

\section{Summary}

In summary, we have presented a new global analysis of collinearly factorized nuclear PDFs, EPPS21. On the methodological front, we have now set up a framework to account for the free-proton uncertainties in a more consistent way than in previous fits of nuclear PDFs based of Hessian uncertainty analyses. The most significant new experimental ingredients are the LHC p-Pb data for dijets (Run-I), D-mesons (Run-I) and W$^\pm$ bosons (Run-II), which reduce the uncertainty of the nuclear gluons significantly in comparison to our previous EPPS16 fit. Our analysis now confirms the presence of gluon shadowing and antishadowing for large nuclei. We have now also incorporated recent JLab DIS data which probe the valence quarks at low values of $Q^2$, previously only reached in the older SLAC data. The good simultaneous fit to all large-$x$ DIS data indicates that no significant $A$-dependent higher-twist contributions appear in the kinematic region included in the fit.  Overall, the new LHC pPb data are also well described within EPPS21 but some discrepancies persist. In particular, it appears impossible to reproduce the strong suppression in the four most forward dijet data points with any set of PDFs. What causes such a strong effect remains an open question and calls for further investigation. In addition, there are some indications that the Run-II W$^\pm$ data would be better reproduced with a slightly less shadowed gluons in comparison to the central EPPS21. However, this feature may be just a fictitious one as the systematic shifts due to the experimental correlations are not available. The new Run-II Z-boson data from CMS also disagree significantly with EPPS21 and it appears impossible to accurately accommodate the rapidity dependence of these data within a global PDF analysis. From all this we can conclude that the LHC p-Pb program has now reached a level of precision which begins to put the conjecture of the universality of nuclear PDFs into a real test. 

The parametrization of the EPPS21 PDFs will be available from LHAPDF \cite{Buckley:2014ana} as well as from Ref.~\cite{PDFurl}. 

\section*{Acknowledgments}

We acknowledge financial support from the Academy of Finland Projects 297058 \& 330448 (K.J.E.), and 308301 (H.P.), Xunta de Galicia (Centro singular de investigación de Galicia accreditation 2019-2022), European Union ERDF, the “María de Maeztu” Units of Excellence program MDM-2016-0692 and the Spanish Research State Agency, European Research Council project ERC-2018-ADG-835105 YoctoLHC. The Finnish IT Center for Science (CSC) is acknowledged for the computing time through the Project jyy2580. This research was funded as a part of the Center of Excellence in Quark Matter of the Academy of Finland (project 346325).

\appendix
\section{Comparisons with the older data}
\label{sec:app}

The following figures present the comparisons between the EPPS21 PDFs and the data already included in the EPPS16 fit \cite{Eskola:2016oht}.

\begin{figure*}[htb!]
\centering
\includegraphics[width=0.494\linewidth]{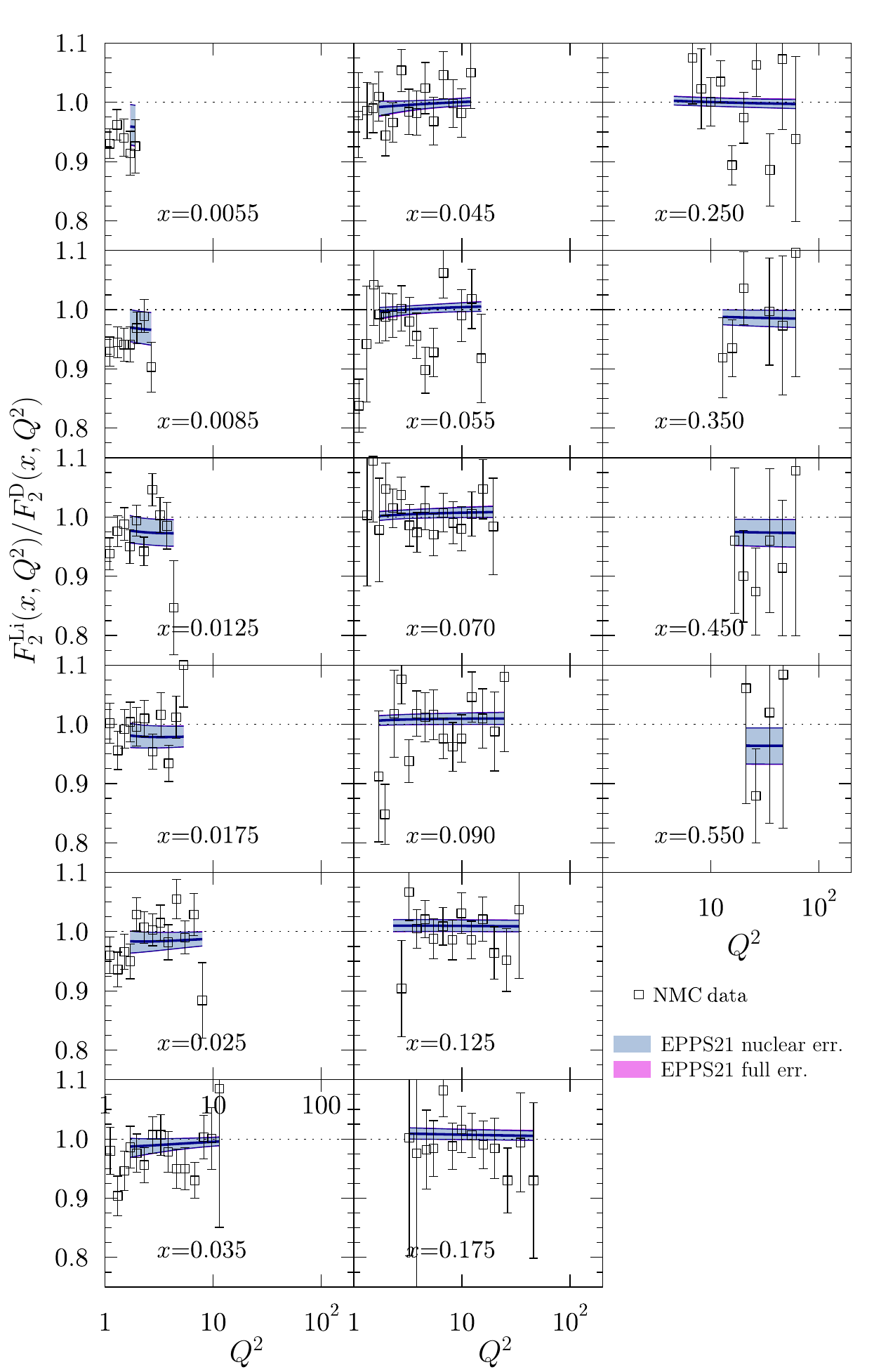} \hspace{-0.0cm}
\includegraphics[width=0.494\linewidth]{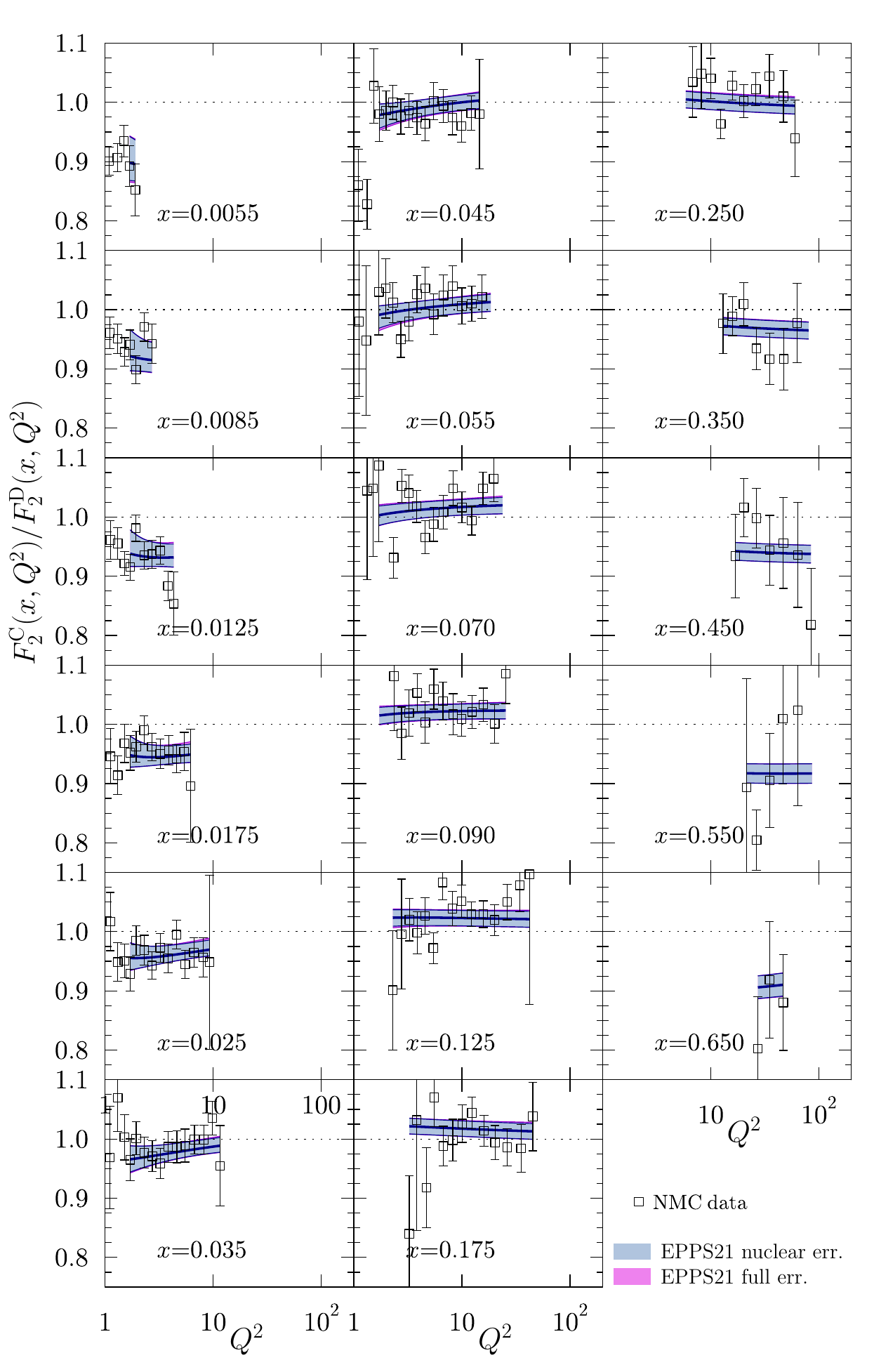}
\caption{The $Q^2$-dependent NMC structure-function ratios \cite{Arneodo:1995cs} compared with the EPPS21 analysis. The solid blue curves show our central results, inner blue bands the nuclear uncertainties, and the purple bands the total uncertainty.}
\label{fig:CLiQ2}
\end{figure*} 
\begin{figure*}[htb!]
\centering
\includegraphics[width=0.890\linewidth]{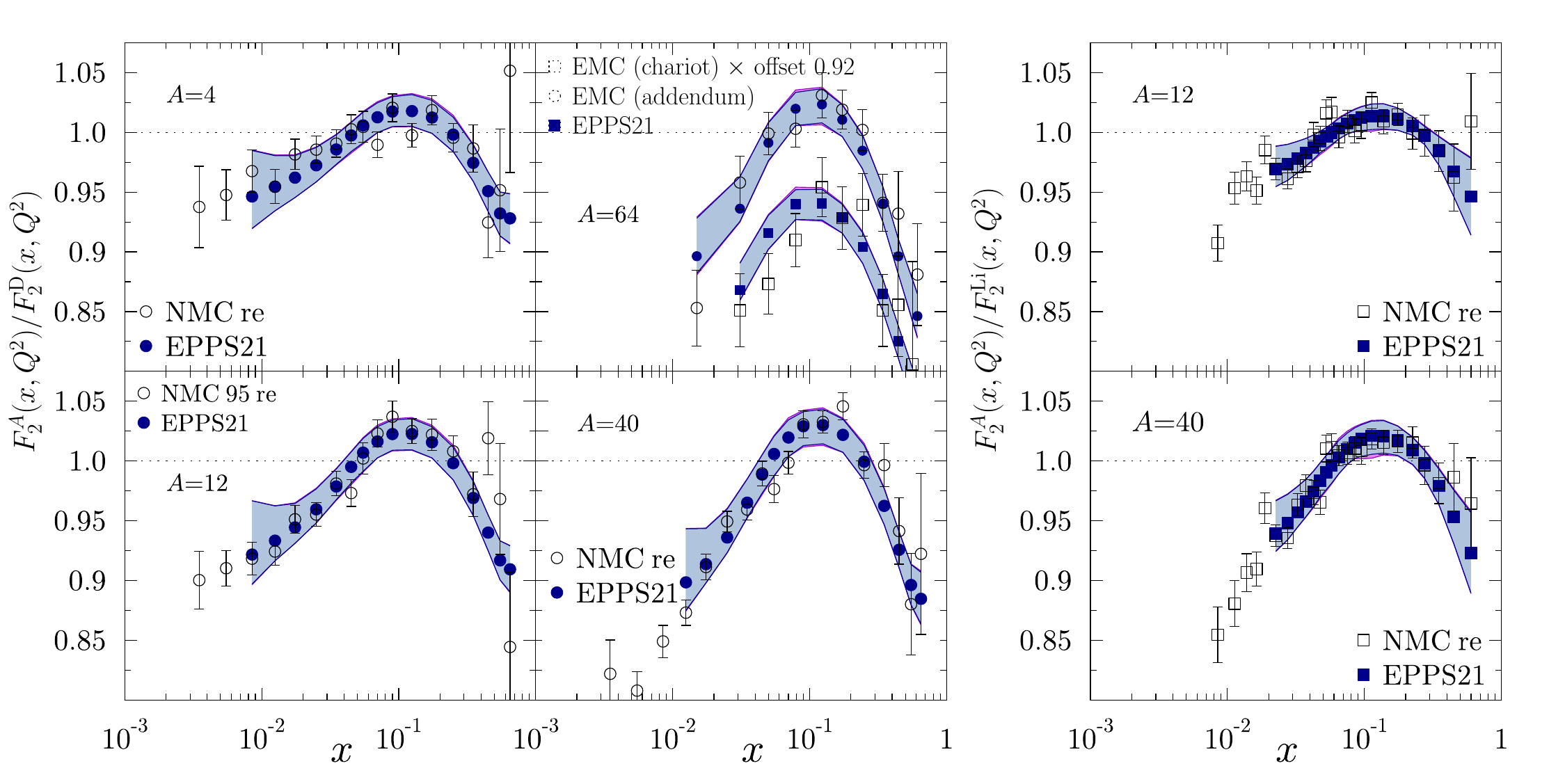}
\caption{
Structure-function ratios from the NMC \cite{Amaudruz:1995tq,Arneodo:1995cs} and EMC \cite{Ashman:1992kv} experiments compared with the EPPS21 analysis. The solid blue points show our central results, inner blue bands the nuclear uncertainties, and the purple bands the total uncertainty.
}
\label{fig:ALiDQ2}
\end{figure*} 

\begin{figure*}[htb!]
\centering
\includegraphics[width=0.95\linewidth]{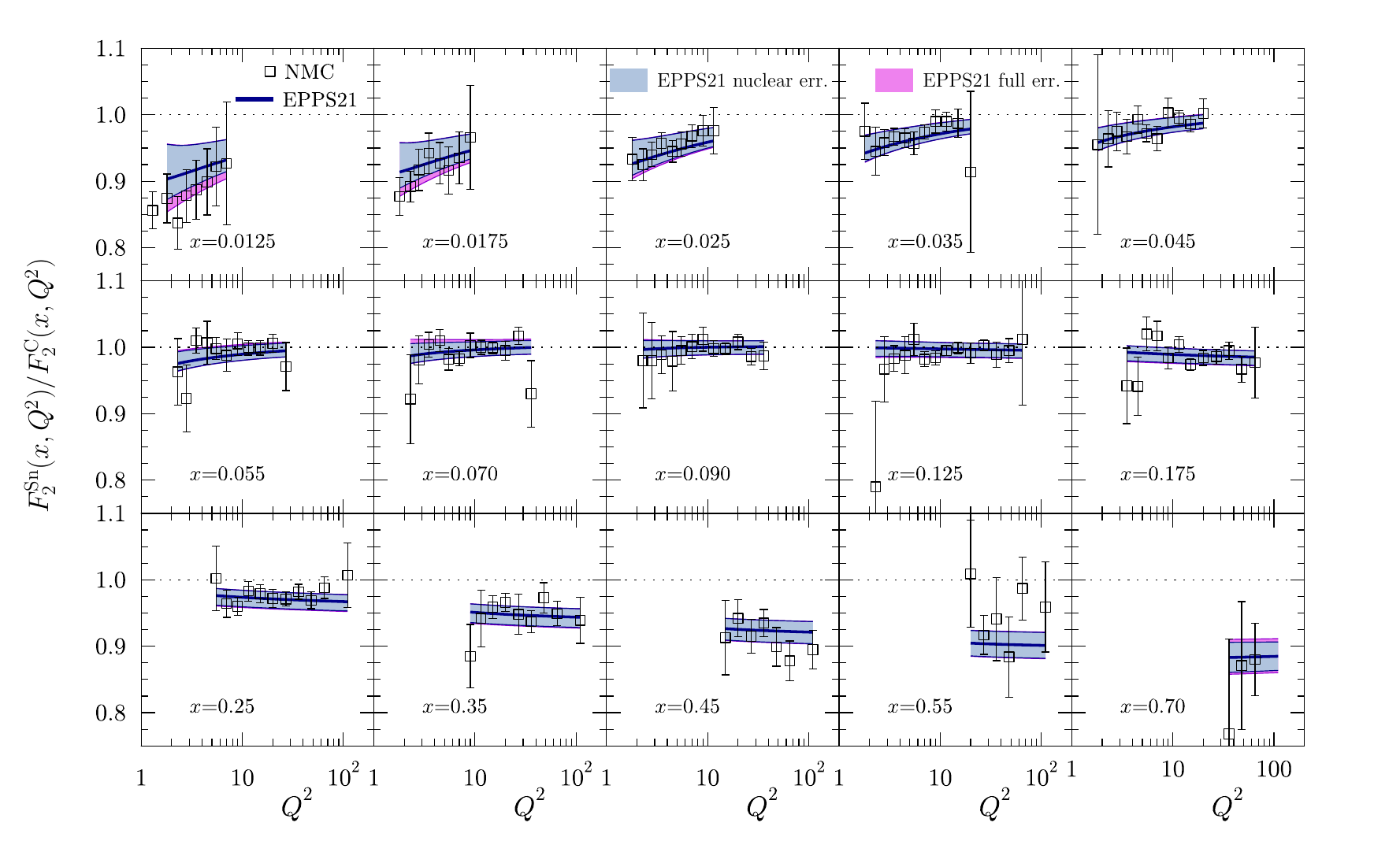}
\caption{
The structure-function ratios $F_2^{\rm Sn}/F_2^{\rm C}$ as a function of $Q^2$ for several fixed values of $x$. The data is from the NMC experiment \cite{Arneodo:1996ru}. The solid blue points show our central results, inner blue bands the nuclear uncertainties, and the purple bands the total uncertainty.
}
\label{fig:RF2SnC}
\end{figure*} 

\begin{figure*}[htb!]
\centering
\includegraphics[width=0.464\linewidth]{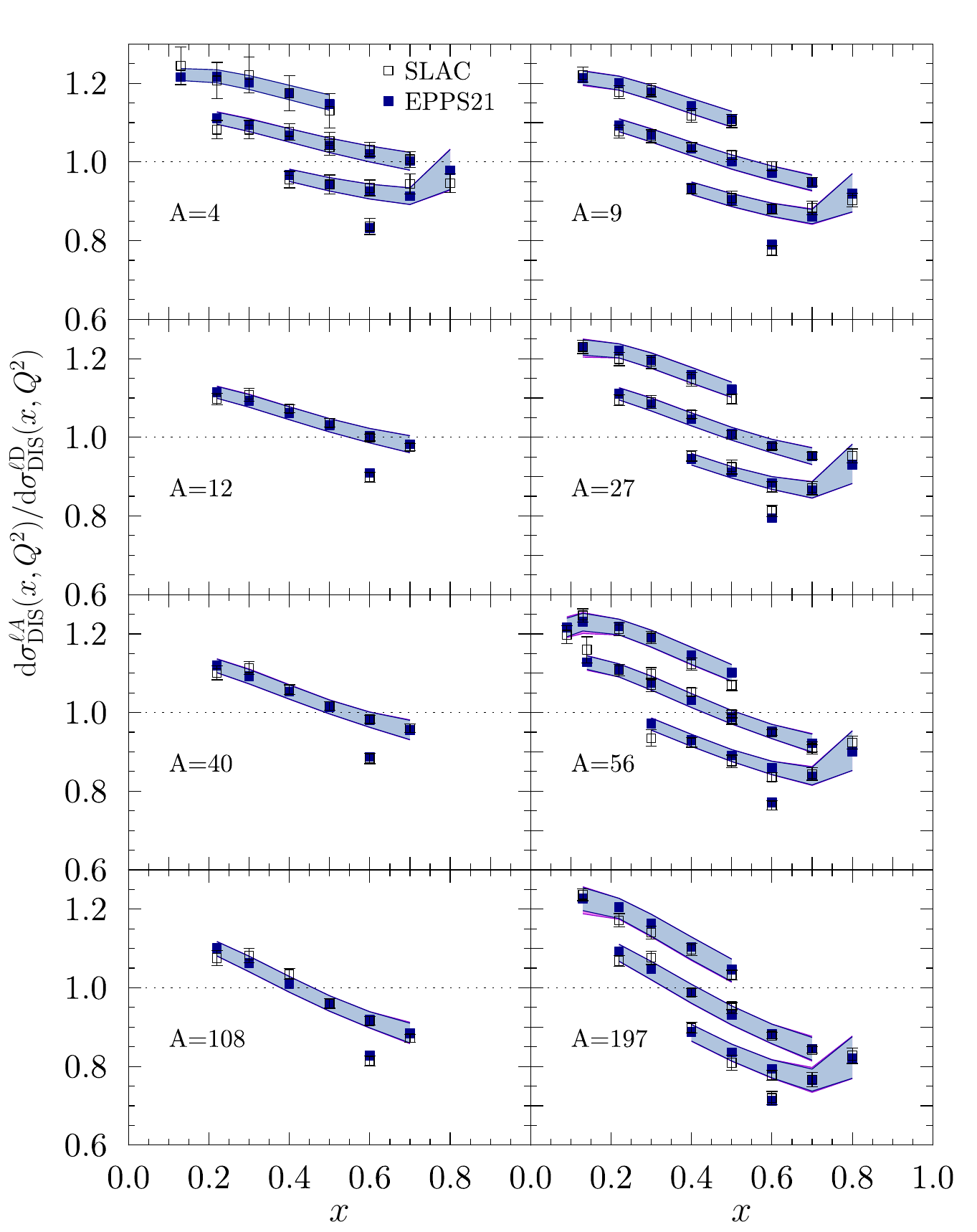} \hspace{-0.0cm}
\includegraphics[width=0.434\linewidth]{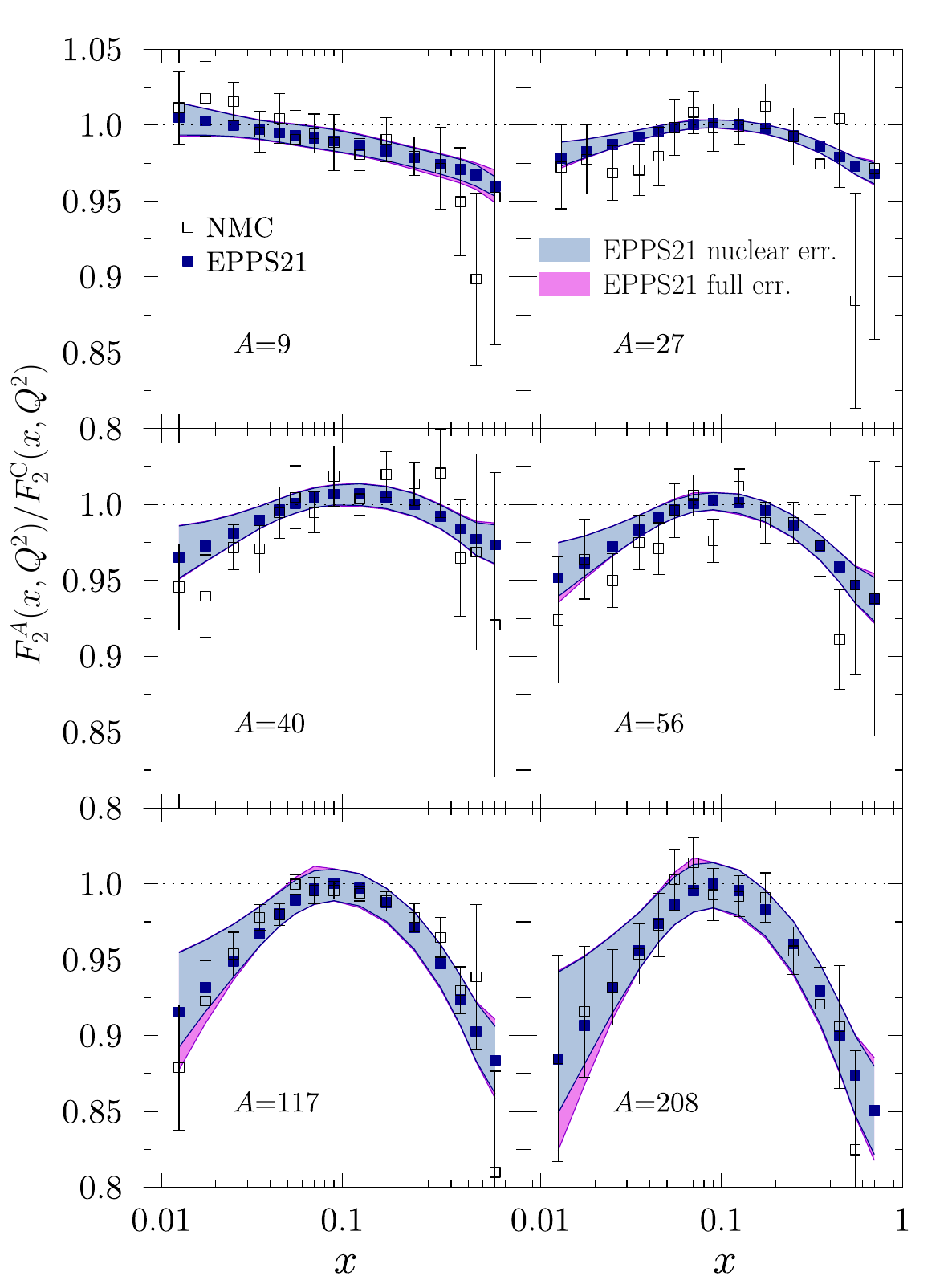}
\caption{
Left: The SLAC data \cite{Gomez:1993ri} for DIS cross-section ratios compared with the EPPS21 analysis. To increase the visibility, the data and theory values have been multiplied by 1.2, 1.1, 1.0, 0.9 for $Q^2=2\,{\rm GeV}^2$, $Q^2=5\,{\rm GeV}^2$, $Q^2=10\,{\rm GeV}^2$, $Q^2=15\,{\rm GeV}^2$. Right: The NMC data \cite{Arneodo:1996rv} for structure-function ratios compared with the EPPS21 analysis. The solid blue points show our central results, inner blue bands the nuclear uncertainties, and the purple bands the total uncertainty.
}
\label{fig:SLAC_NMC}
\end{figure*} 

\begin{figure*}[htb!]
\centering
\vspace{-0.5cm} 
\includegraphics[width=1.0\linewidth]{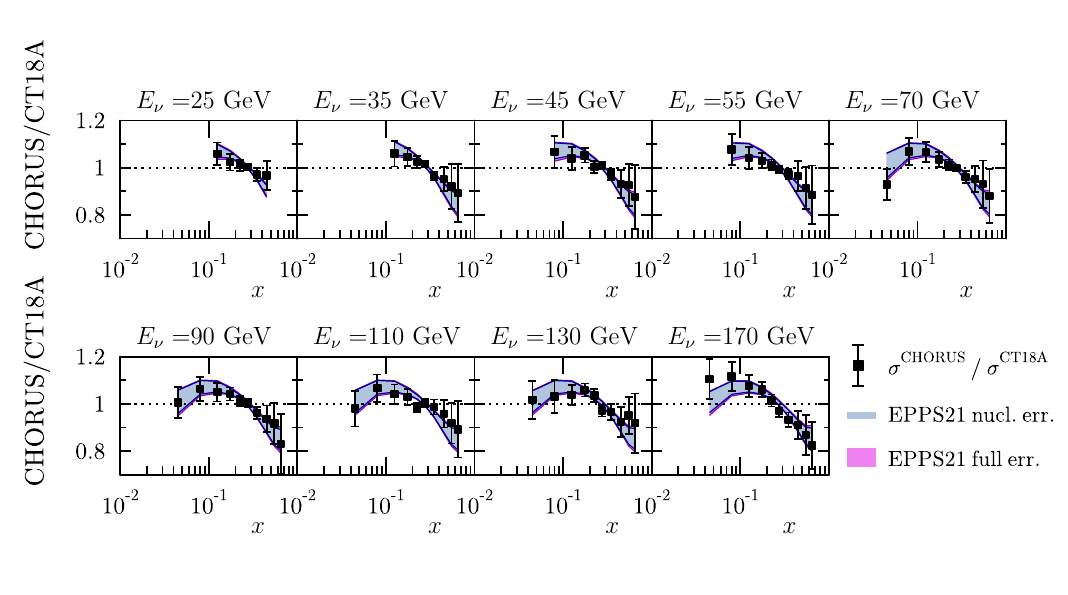} \\
\vspace{-0.5cm} 
\caption{
The charged-current neutrino-nucleus CHORUS data \cite{Onengut:2005kv} compared with the EPPS21 fit. Blue bands show the nuclear uncertainties, and the purple bands the total uncertainty.
}
\label{fig:nudatat}
\end{figure*} 

\begin{figure*}[htb!]
\centering
\vspace{-0.5cm} 
\includegraphics[width=1.0\linewidth]{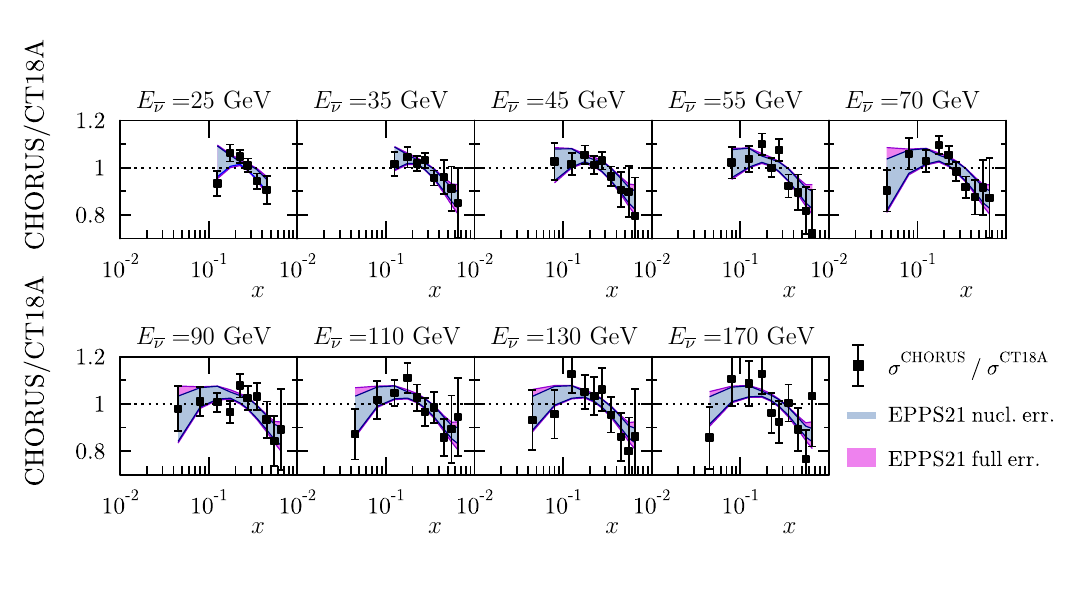}
\vspace{-1cm} 
\caption{As Fig.~\ref{fig:nudatat} but for antineutrino-nucleus interaction.}
\label{fig:anudatat}
\end{figure*}

\begin{figure*}[htb!]
\centering
\includegraphics[width=1.0\linewidth]{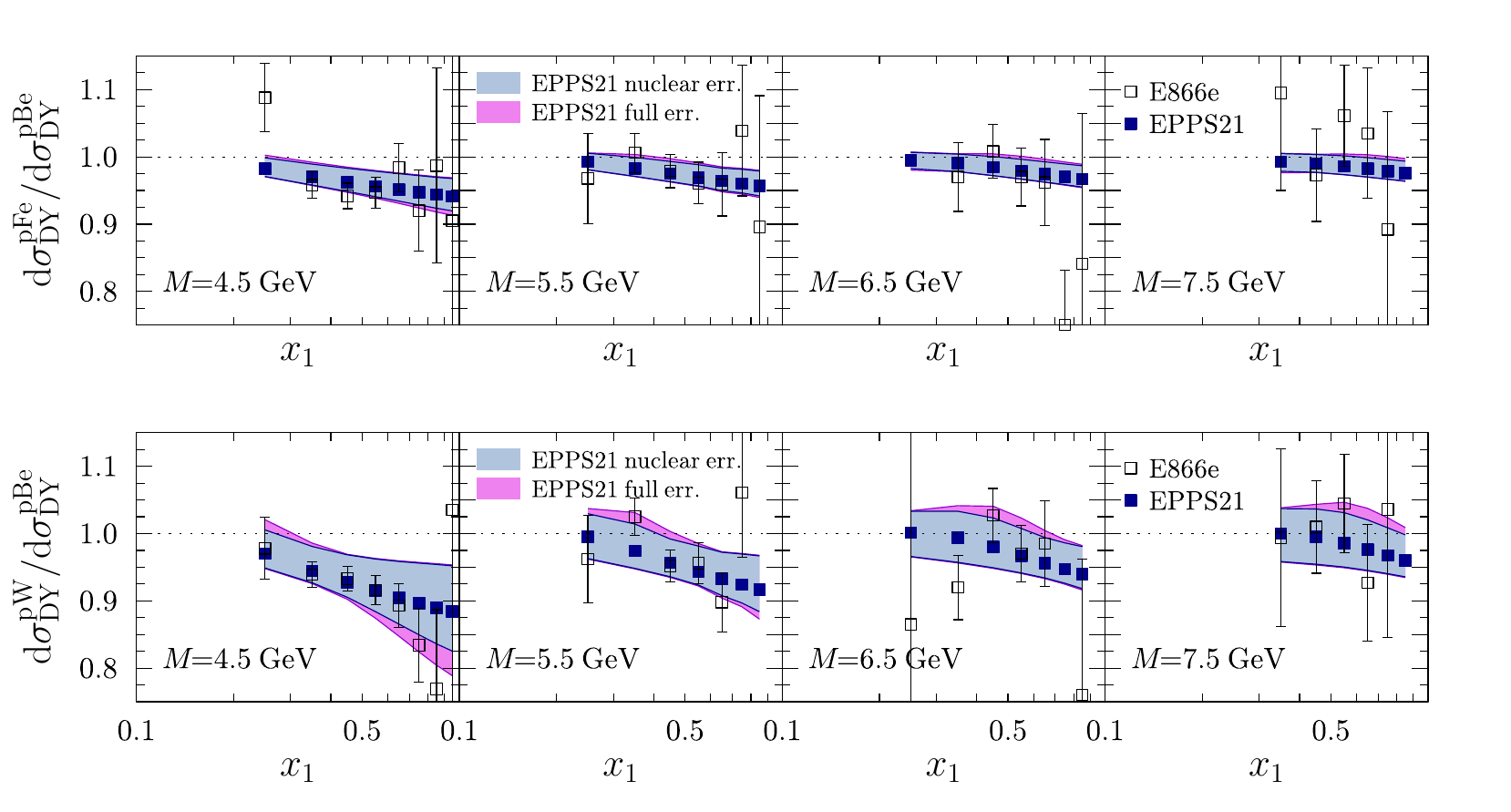}
\caption{
Drell-Yan cross-section ratios $d\sigma^{{\rm p}A} / d\sigma^{\rm pBe}$ measured by the E866 collaboration \cite{Vasilev:1999fa} compared with the EPPS21 analysis. The solid blue points show our central results, inner blue bands the nuclear uncertainties, and the purple bands the total uncertainty.} 
\label{fig:DYdatat}
\end{figure*} 

\begin{figure*}[htb!]
\centering
\includegraphics[width=1.0\linewidth]{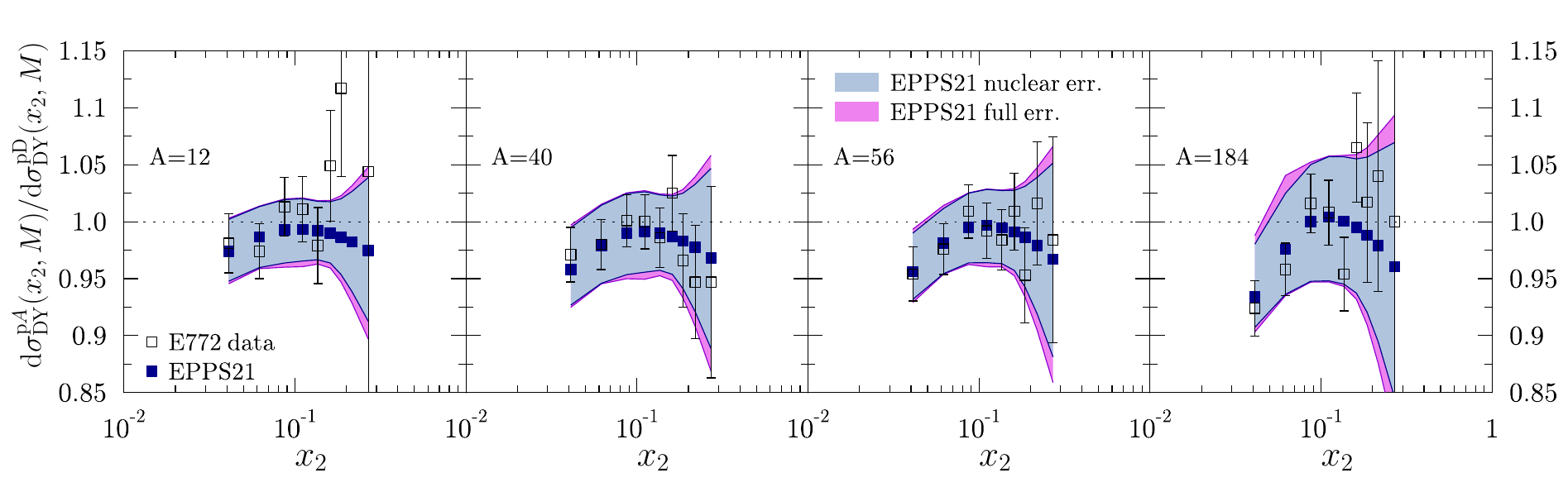}
\caption{
Drell-Yan cross-section ratios of the E772 collaboration \cite{Alde:1990im} compared with the EPPS21 analysis. The solid blue points show our central results, inner blue bands the nuclear uncertainties, and the purple bands the total uncertainty.
}
\label{fig:RDY}
\end{figure*}

\begin{figure*}[htb!]
\centering
\hspace{-0.5cm}
\includegraphics[width=0.40\linewidth]{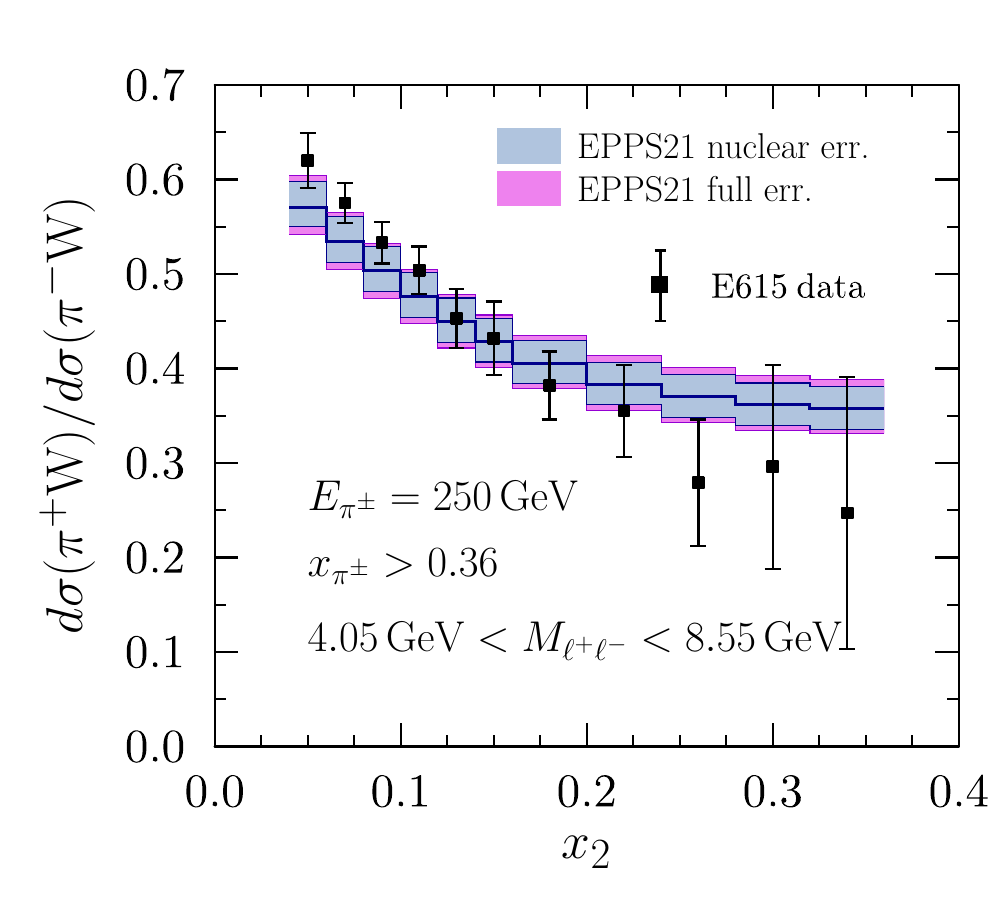} 
\includegraphics[width=0.40\linewidth]{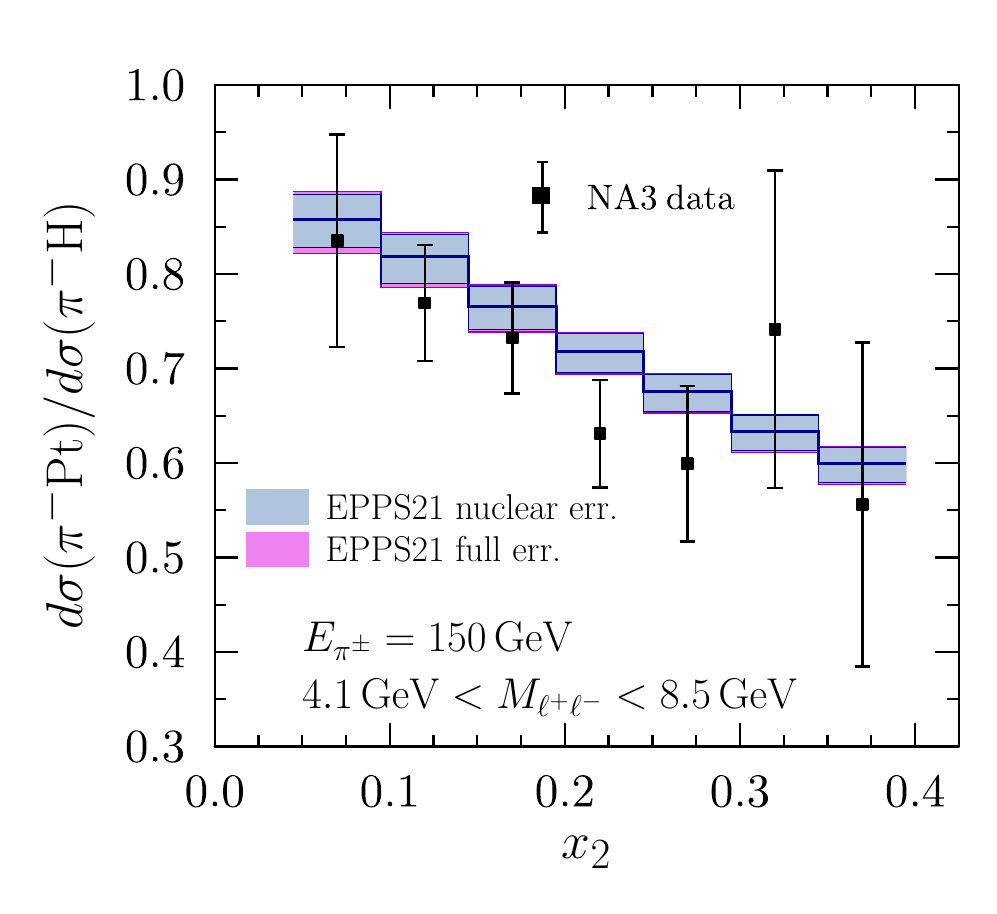}  

\vspace{-0.6cm}
\hspace{-0.8cm}\includegraphics[width=0.8\linewidth]{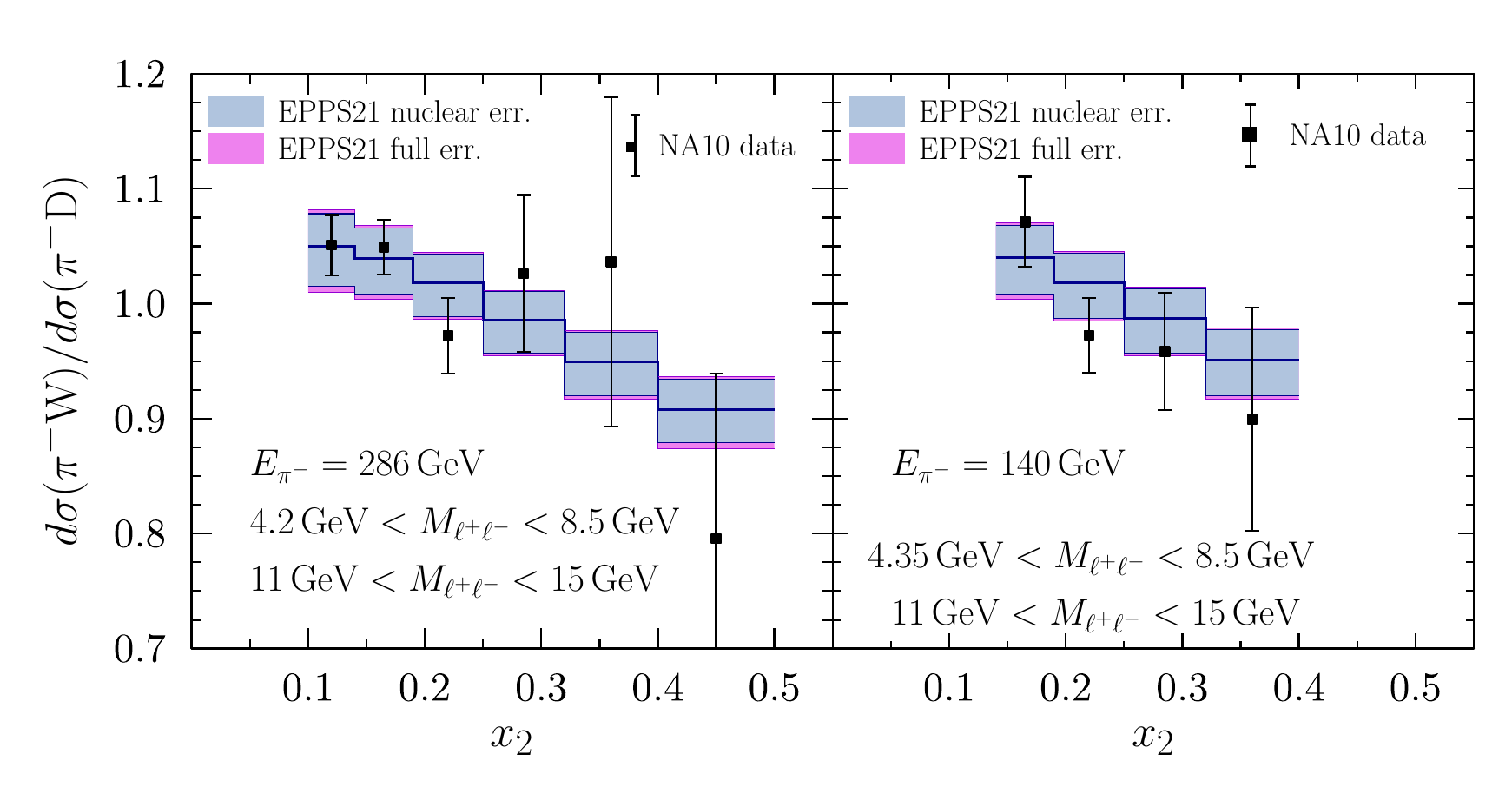} 
\caption{
The Drell-Yan ratios in $\pi^\pm$-$A$ collisions from E615 \cite{Heinrich:1989cp}, NA3 \cite{Badier:1981ci}  and NA10 \cite{Bordalo:1987cs} compared with the EPPS21 analysis. The solid blue points show our central results, inner blue bands the nuclear uncertainties, and the purple bands the total uncertainty.}
\label{fig:piDYdatat}
\end{figure*}

\begin{figure*}[htbp!]
\centering
\includegraphics[width=0.45\linewidth]{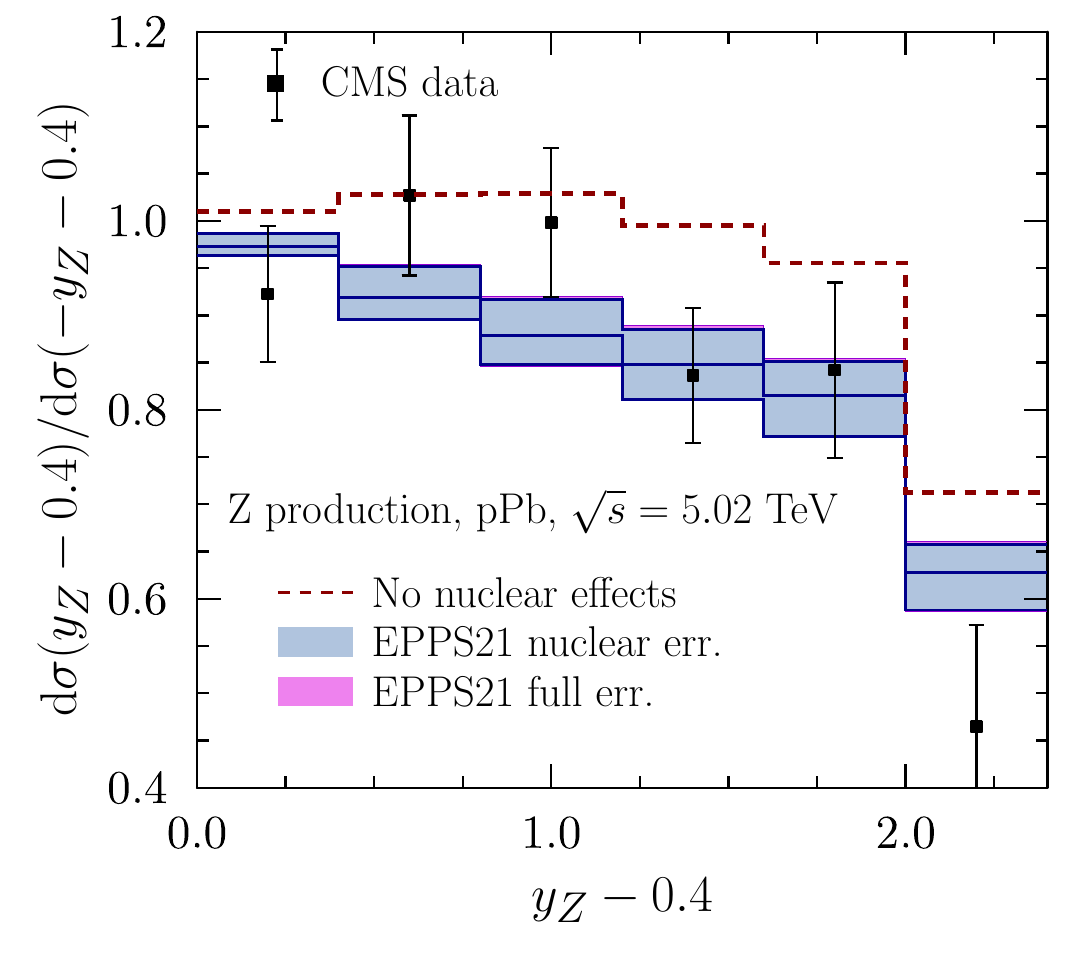}
\includegraphics[width=0.45\linewidth]{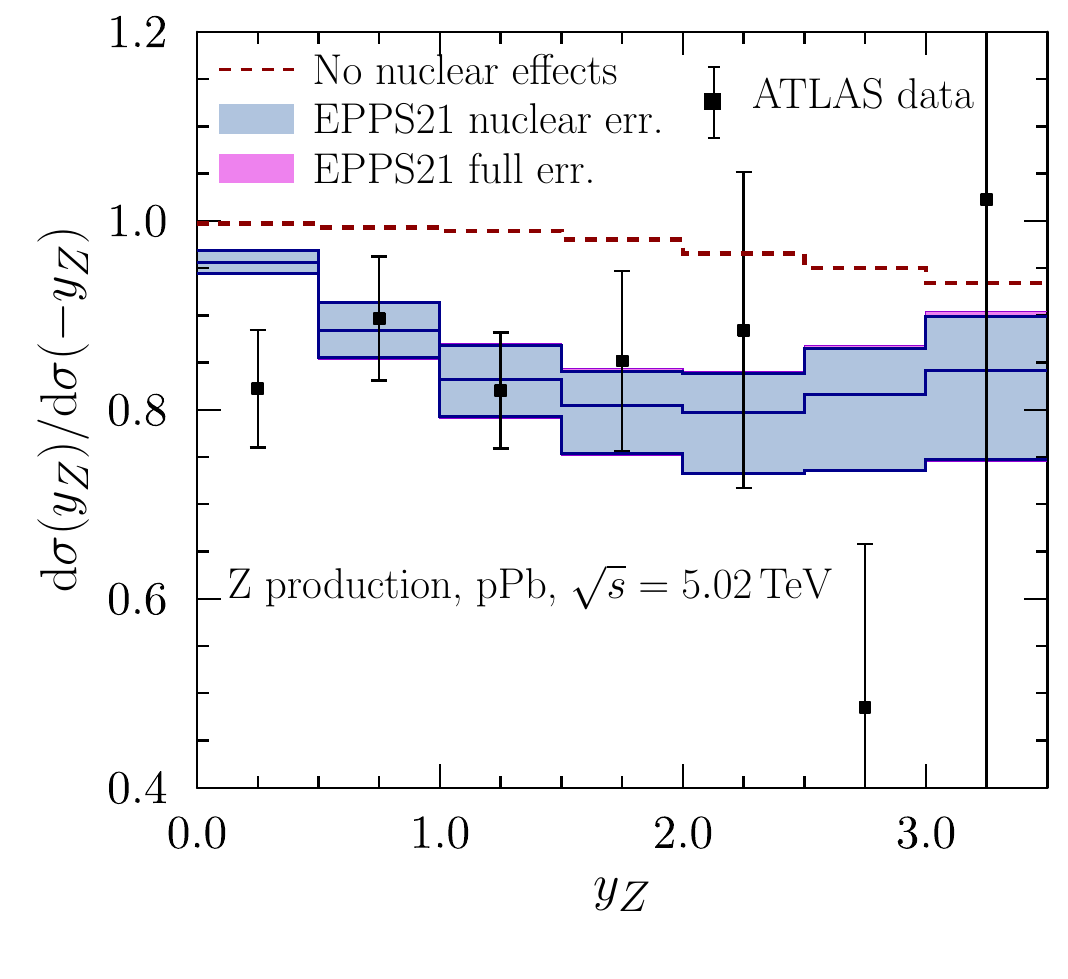}

\includegraphics[width=0.45\linewidth]{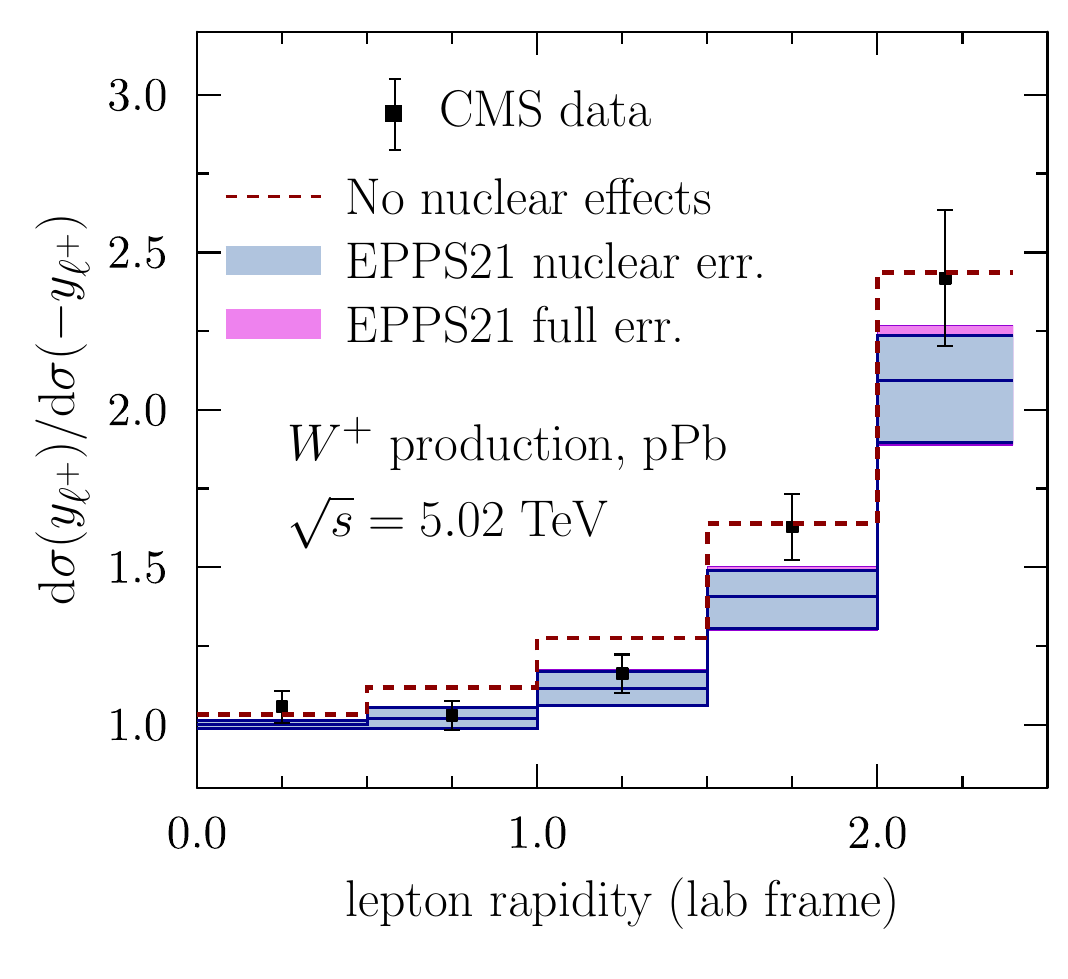}
\includegraphics[width=0.45\linewidth]{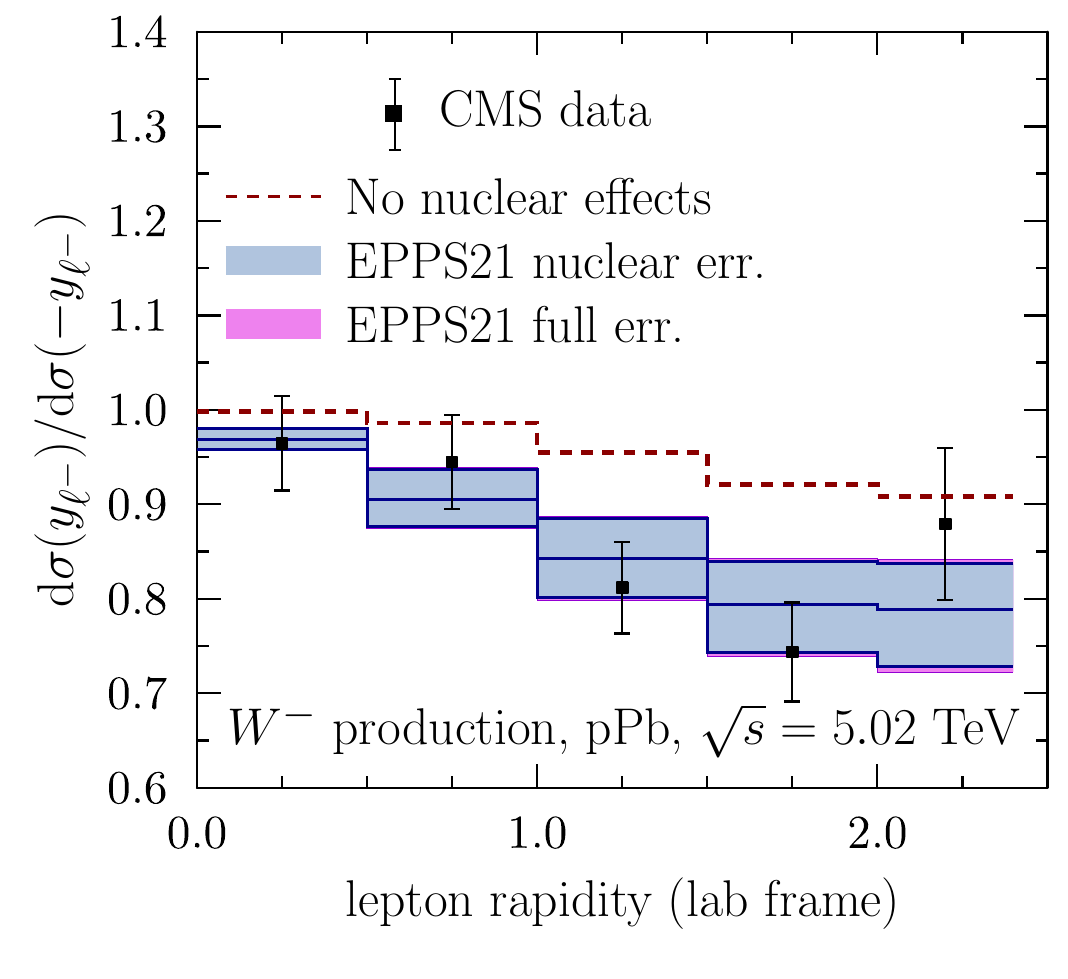}

\includegraphics[width=0.45\linewidth]{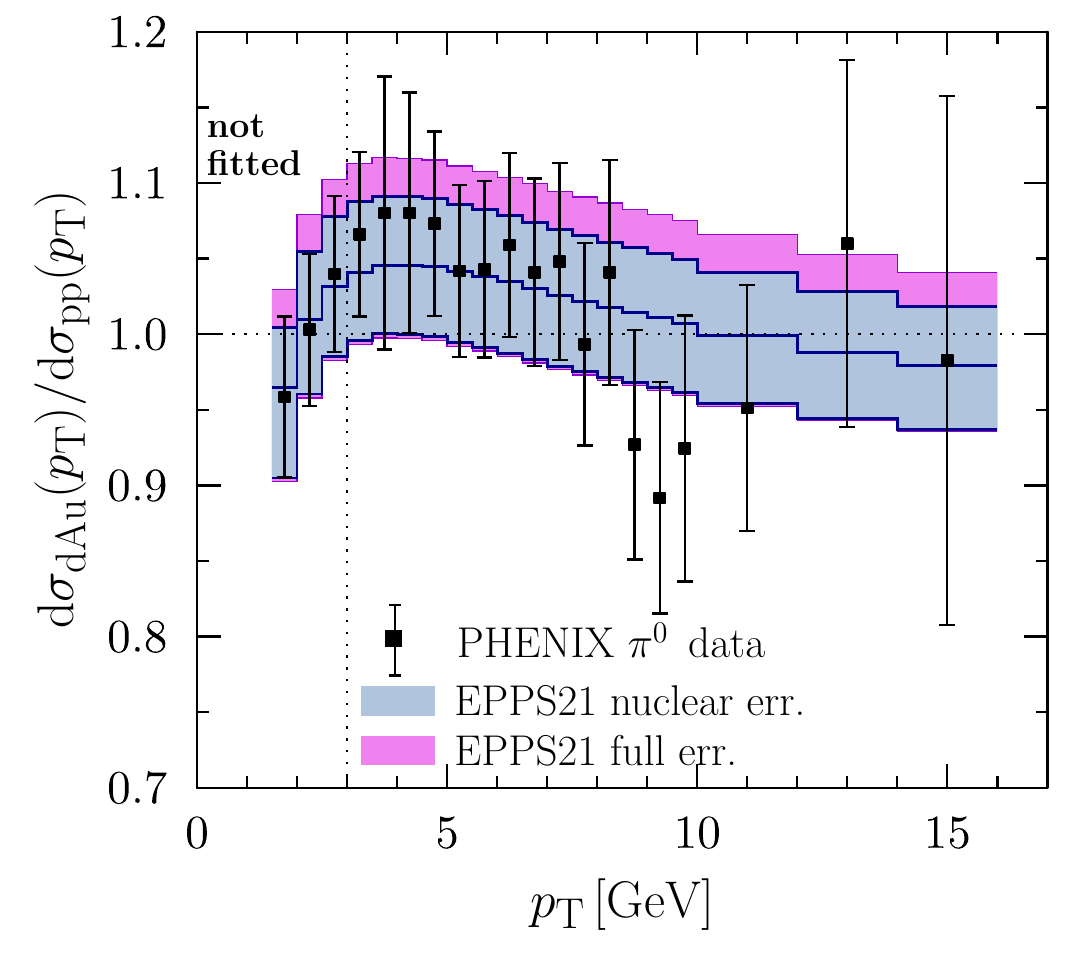}
\caption{The CMS and ATLAS $5\,{\rm TeV}$ pPb data \cite{Khachatryan:2015hha,Khachatryan:2015pzs,Aad:2015gta} for Z (upper panels) and ${\rm W}^\pm$ (middle panels), as well as PHENIX dAu cross-section ratios for inclusive pion production \cite{Adler:2006wg} (lowest panel) compared with the EPPS21 analysis. The solid blue lines show our central results, inner blue bands the nuclear uncertainties, and the purple bands the total uncertainty.
}
\label{fig:LHCdatat}
\end{figure*}

\end{document}